\documentclass[fleqn,usenatbib]{mnras}

\usepackage{newtxtext,newtxmath}

\usepackage[T1]{fontenc}

\usepackage{graphicx}	
\usepackage{amsmath}	

\usepackage[normalem]{ulem}
\usepackage{booktabs}	
\usepackage{mathtools}	
\usepackage{adjustbox} 
\usepackage{tabularx}

\usepackage{color}
\usepackage{acronym}   
\usepackage{xspace}
\usepackage{enumitem} 
\usepackage{calc}
\usepackage[caption=false]{subfig}
\usepackage{ifthen} 
\usepackage{wrapfig} 

\usepackage{outlines} 
\newcommand\rate{\mathcal{R}}
\newcommand\COMPAS{{\sc{COMPAS }}}

\newcommand\bhnsSingle{BHNS\xspace}

\newcommand\PercentageClassicLVK{\ensuremath{86\%}\xspace}
\newcommand\PercentageOnlyStableMTLVK{\ensuremath{4\%}\xspace}
\newcommand\PercentageSCCELVK{\ensuremath{4\%}\xspace}
\newcommand\PercentageDCCELVK{\ensuremath{<1\%}\xspace}
\newcommand\PercentageOtherLVK{\ensuremath{6\%}\xspace}

\newcommand{\monei}{\ensuremath{m_{1,\rm{i}}}\xspace}
\newcommand{\mtwoi}{\ensuremath{m_{2,\rm{i}}}\xspace}
\newcommand{\monef}{\ensuremath{m_{1,\rm{f}}}\xspace}
\newcommand{\mtwof}{\ensuremath{m_{2,\rm{f}}}\xspace}
\newcommand{\ai}{\ensuremath{a_{\rm{i}}}\xspace}
\newcommand{\qi}{\ensuremath{q_{\rm{i}}}\xspace}
\newcommand{\Zi}{\ensuremath{Z_{\rm{i}}}\xspace}
\newcommand{\vk}{\ensuremath{v_{\rm{k}}}\xspace}
\newcommand{\thetak}{\ensuremath{{\theta}_{\rm{k}}}\xspace}

\newcommand{\ei}{\ensuremath{{e}_{\rm{i}}}\xspace}

\newcommand{\Rsun}{\ensuremath{\,\rm{R}_{\odot}}\xspace}
\newcommand{\km}{\ensuremath{\,\rm{km}}\xspace}
\newcommand{\kms}{\ensuremath{\,\rm{km}\,\rm{s}^{-1}}\xspace}
\newcommand{\Msun}{\ensuremath{\,\rm{M}_{\odot}}\xspace}

\newcommand{\Mpc}{\ensuremath{\,\rm{Mpc}}\xspace}
\newcommand{\Zsun}{\ensuremath{\,\rm{Z}_{\odot}}\xspace}

\newcommand{\AU}{\ensuremath{\,\mathrm{AU}}\xspace}
\newcommand{\Myr}{\ensuremath{\,\mathrm{Myr}}\xspace}

\newcommand{\Gyr}{\ensuremath{\,\mathrm{Gyr}}\xspace}
\newcommand{\Gyrs}{\ensuremath{\,\mathrm{Gyr}}\xspace}

\newcommand{\yearmin}{\ensuremath{\,\rm{yr}^{-1}}\xspace}

\newcommand{\GpcminThree}{\ensuremath{\,\rm{Gpc}^{-3}}\xspace}

\newcommand{\Hubble}{\ensuremath{\mathcal{H}_0}\xspace}

\newcommand{\MSFR}{\ensuremath{{M}_{\rm{SFR}}}\xspace}
\newcommand{\SFRD}{\text{SFRD}\ensuremath{(Z_{\rm{i}},z)}\xspace}
\newcommand{\tdelay}{\ensuremath{{t}_{\rm{delay}}}\xspace}
\newcommand{\tDCO}{\ensuremath{{t}_{\rm{DCO}}}\xspace}
\newcommand{\ts}{\ensuremath{{t}_{\rm{s}}}\xspace}
\newcommand{\tevolve}{\ensuremath{{t}_{\rm{evolve}}}\xspace}
\newcommand{\tform}{\ensuremath{{t}_{\rm{form}}}\xspace}
\newcommand{\tmerger}{\ensuremath{{t}_{\rm{m}}}\xspace}
\newcommand{\tinspiral}{\ensuremath{{t}_{\rm{inspiral}}}\xspace}
\newcommand{\thubble}{\ensuremath{{t}_{\mathcal{H}}}\xspace}
\newcommand{\tdet}{\ensuremath{{t}_{\rm{det}}}\xspace}
\newcommand{\Nform}{\ensuremath{{N}_{\rm{form}}}\xspace}
\newcommand{\Ndet}{\ensuremath{{N}_{\rm{det}}}\xspace}
\newcommand{\Nmerger}{\ensuremath{{N}_{\rm{merger}}}\xspace}

\newcommand{\Pdet}{\ensuremath{{P}_{\rm{det}}}\xspace}

\newcommand{\Vc}{\ensuremath{{V}_{\rm{c}}}\xspace}
\newcommand{\DL}{\ensuremath{{D}_{\rm{L}}}\xspace}
\newcommand{\Dc}{\ensuremath{{D}_{\rm{c}}}\xspace}

\newcommand*\diff{\mathop{}\!\mathrm{d}}

\newcommand{\CMP}{C21}
\newcommand{\PII}{Broekgaarden et al. (in prep.)} 

\newcommand{\mnsf}{\ensuremath{m_{\rm{NS}}}\xspace}
\newcommand{\mbhf}{\ensuremath{m_{\rm{BH}}}\xspace}
\newcommand{\mtotf}{\ensuremath{m_{\rm{tot}}}\xspace}
\newcommand{\mchirpf}{\ensuremath{{\mathcal{M}}_{\rm{c}}}\xspace}
\newcommand{\af}{\ensuremath{a_{\rm{f}}}\xspace}
\newcommand{\qf}{\ensuremath{q_{\rm{f}}}\xspace}
\newcommand{\ef}{\ensuremath{{e}_{\rm{f}}}\xspace}
\newcommand{\chibh}{\ensuremath{{\chi}_{\rm{BH}}}\xspace}
\newcommand{\Rns}{\ensuremath{{R}_{\rm{NS}}}\xspace}

\newcommand{\mAzero}{\ensuremath{\rm{A}000}\xspace}

\newcommand{\Nmodels}{\ensuremath{420}\xspace}
\newcommand{\NmodelsBPS}{\ensuremath{15}\xspace}
\newcommand{\NmodelsMSSFR}{\ensuremath{28}\xspace}
\newcommand{\RateIntrinsicZero}{\ensuremath{\mathcal{R}_{\rm{m}}^{0}}\xspace}
\newcommand{\RateObserved}{\ensuremath{\mathcal{R}_{\rm{det}}}\xspace}

\newcommand{\RateIntrinsicAzeroBHNS}{\ensuremath{43}\xspace}  
\newcommand{\RateIntrinsicAzeroBHNSmin}{\ensuremath{4}\xspace}  
\newcommand{\RateIntrinsicAzeroBHNSmax}{\ensuremath{830}\xspace} 

\newcommand{\RateObservedAzeroBHNS}{\ensuremath{11}\xspace} 
\newcommand{\RateObservedAzeroBHNSmax}{\ensuremath{180}\xspace} 
\newcommand{\RateObservedAzeroBHNSmin}{\ensuremath{1}\xspace}

\acrodef{GSMF}{galaxy mass function, the number density of galaxies per logarithmic mass bin}
\acrodef{MZR}{mass-metallicity relation}
\acrodef{SFRD}{metallicity-specific star formation rate density}

\acrodef{NSNS}{binary neutron star}
\acrodef{BHBH}{binary black hole}

\acrodef{DCO}{double compact object}
\acrodef{NS}{neutron star}
\acrodef{BH}{black hole}
\acrodef{BH--NS}{black hole-neutron star}
\acrodef{GRB}{gamma--ray burst}
\acrodef{RLOF}{Roche-lobe overflow}
\acrodef{CE}{common envelope}
\acrodef{GW}{gravitational wave}
\acrodef{SN}{supernova}
\acrodefplural{SN}[SNe]{supernovae}
\acrodef{PISN}{pair-instability SN}
\acrodef{ECSN}{electron-capture SN}
\acrodefplural{ECSN}[ECSNe]{electron-capture SNe}
\acrodef{USSN}{ultra-stripped SN}
\acrodefplural{USSN}[USSNe]{ultra-stripped SN}
\acrodef{CCSN}{core-collapse SN}
\acrodefplural{CCSN}[CCSNe]{core-collapse SN}
\acrodef{COMPAS}{
Compact Object Mergers: Population Astrophysics and Statistics}
\title[Black Hole -- Neutron Star Mergers]{Impact of Massive Binary Star and Cosmic Evolution on Gravitational Wave Observations I: Black Hole -- Neutron Star Mergers}

\author[F. S. Broekgaarden et al.]{Floor S. Broekgaarden,$^{1}$\thanks{E-mail: floor.broekgaarden@cfa.harvard.edu}
Edo Berger,$^{1}$
Coenraad J. Neijssel,$^{2,3,4}$
Alejandro Vigna-G\'{o}mez,$^{5}$
\newauthor
Debatri Chattopadhyay,$^{6,3}$
Simon Stevenson,$^{6,3}$
Martyna Chruslinska,$^{7}$
Stephen Justham,$^{8,9,10,11}$
\newauthor
Selma E. de Mink,$^{11,10,1}$
Ilya Mandel$^{2,3,4}$
\\
$^{1}${Center for Astrophysics \textbar{} Harvard $\&$ Smithsonian,
60 Garden Street, Cambridge, MA 02138, USA}\\
$^{2}${Monash Centre for Astrophysics, School of Physics and Astronomy, Monash University, Clayton, Victoria 3800, Australia}\\
$^{3}${The ARC Center of Excellence for Gravitational Wave Discovery -- OzGrav, Hawthorn VIC 3122, Australia}\\
$^{4}${Birmingham Institute for Gravitational Wave Astronomy and School of Physics and Astronomy, University of Birmingham}\\
Birmingham, B15 2TT, United Kingdom\\
$^{5}${DARK, Niels Bohr Institute, University of Copenhagen, Jagtvej 128, 2200, Copenhagen, Denmark}\\
$^{6}${Center for Astrophysics and Supercomputing, Swinburne University of Technology, Hawthorn VIC 3122, Australia}\\
$^{7}${Institute of Mathematics, Astrophysics and Particle Physics, Radboud University Nijmegen, PO Box 9010, 6500 GL Nijmegen}\\
$^{8}${School of Astronomy $\&$ Space Science, University of the Chinese Academy of Sciences, Beijing 100012, China}\\
$^{9}${National Astronomical Observatories, Chinese Academy of Sciences, Beijing 100012, China} \\
$^{10}${Anton Pannekoek Institute for Astronomy and GRAPPA, University of Amsterdam, Postbus 94249, 1090 GE Amsterdam, The Netherlands }\\
$^{11}${Max-Planck-Institut für Astrophysik, Karl-Schwarzschild-Straße 1, 85741 Garching, Germany}\\
}

\date{Accepted XXX. Received YYY; in original form ZZZ}

\pubyear{2021}

\begin{document}
\label{firstpage}
\pagerange{\pageref{firstpage}--\pageref{lastpage}}
\maketitle

\begin{abstract}
Mergers of black hole-neutron star (BHNS) binaries have now been observed by \ac{GW} detectors with the recent announcement of GW200105 and GW200115. Such observations not only provide confirmation that these systems exist, but will also give unique insights into the death of massive stars, the evolution of binary systems and their possible association with gamma-ray bursts, $r$-process enrichment and kilonovae.  Here we perform binary population synthesis of isolated BHNS systems in order to present their merger rate and characteristics for ground-based GW observatories. We present the results for 420 different model permutations that explore key uncertainties in our assumptions about massive binary star evolution (e.g. mass transfer, common-envelope evolution, supernovae), and the metallicity-specific star formation rate density, and characterize their relative impacts on our predictions. We find intrinsic local \bhnsSingle merger rates spanning $\rate_{\rm{m}}^0 \approx$ \RateIntrinsicAzeroBHNSmin--\RateIntrinsicAzeroBHNSmax \GpcminThree \yearmin for our full range of assumptions. This encompasses the rate inferred from recent BHNS GW detections, and would yield detection rates of $\RateObserved \approx \RateObservedAzeroBHNSmin$--$\RateObservedAzeroBHNSmax$\yearmin for a GW network consisting of LIGO, Virgo and KAGRA at design sensitivity.
We find that the binary evolution and metallicity-specific star formation rate density each impact the predicted merger rates by order $\mathcal{O}(10)$. We also present predictions for the  \ac{GW} detected \bhnsSingle merger properties and find that all 420 model variations predict that $\lesssim 5\%$ of the \bhnsSingle mergers have {BH} masses $\mbhf \gtrsim 18\Msun$, total masses $\mtotf \gtrsim 20\Msun$, chirp masses $\mchirpf \gtrsim 5.5\Msun$, mass ratios $\qf \gtrsim 12$ or $\qf \lesssim 2$. Moreover, we find that massive {NSs} with $\mnsf > 2\Msun$ are expected to be commonly detected in \bhnsSingle mergers in almost all our model variations. Finally, a wide range of $\sim 0\%$--$70\%$ of the \bhnsSingle mergers are predicted to eject mass during the merger.
Our results highlight the importance of considering variations in binary evolution and cosmological models when predicting, and eventually evaluating, populations of BHNS mergers.
\end{abstract}





\section{Introduction}
\label{sec:introduction}
The ground-based \ac{GW} interferometers of the LIGO, Virgo and KAGRA (LVK) network (\citealt{2010JPhCS.228a2012L, 2012CQGra..29l4007S, 2013PhRvD..88d3007A,2015CQGra..32b4001A, 2015CQGra..32g4001L, 2016CQGra..33g5009D}) observed  \acp{GW} from \ac{BHBH} and \ac{NSNS} mergers in their first observing runs \citep{2019PhRvX...9c1040A, 2020arXiv201014527A,2020ApJ...891..123N,2020PhRvD.101h3030V,2019arXiv191009528Z,2019PhRvD.100b3007Z, 2020ApJ...892L...3A,2020arXiv200408342T}. With the recent announcement of the first two observations of mergers between a black hole and neutron star (BHNS), GW200105 and GW200115 \citep{Abbott:2021-first-NSBH}, all three of these \ac{GW} flavors have now been detected. 
In addition, the recent \ac{GW} catalogs (GWTC-2 and GWTC-2.1, \citealt{2020arXiv201014527A,2020arXiv201014533T,GWTC2point1}) presented several \acp{GW} where a \bhnsSingle source has not yet been ruled out,  including  GW190814 \citep{2020ApJ...896L..44A, 2020ApJ...904...39H, 2020arXiv201111934Z}, GW190425 \citep{2020ApJ...892L...3A,2020ApJ...891L...5H,2020ApJ...890L...4K} and the low signal to noise candidates GW190426$\_$152155 and GW190917, but none of these candidates present a confident \bhnsSingle detection. 

The detection of \bhnsSingle mergers is of broad interest as they could be used to measure the Hubble constant and other cosmological parameters to greater distances than \ac{NSNS} mergers \citep{1986Natur.323..310S,2010ApJ...725..496N, 2017PhRvD..95d4024C,2018PhRvL.121b1303V,2020arXiv201206593F}, may help constrain the neutron star equation of state  \citep{2012PhRvD..85d4061L,2010CQGra..27k4106D, 2020ApJ...893..153K} and may help study the rate of heavy element production \citep{2011ApJ...738L..32G,2015MNRAS.448..541J}.
In addition, they are theorized to be 
$r$-process enrichment sites 
\citep[e.g.][]{1974ApJ...192L.145L,1976ApJ...210..549L,1999A&A...341..499R,1999ApJ...525L.121F}
and are possibly accompanied by electromagnetic counterparts such  as kilonovae
\citep[e.g.][]{1998ApJ...507L..59L,2013ApJ...775...18B, 2017LRR....20....3M,2020arXiv201102717Z}, 
 short gamma-ray bursts
 \citep[e.g.][]{1984SvAL...10..177B,1986ApJ...308L..47G, 1986ApJ...308L..43P, 1989Natur.340..126E, 2020ApJ...895...58G},
radio emission \citep{2011Natur.478...82N,2013MNRAS.430.2121P,2015MNRAS.450.1430H,2016ApJ...831..190H} and 
neutrinos \citep{2013ApJ...776...47D,2018PhRvD..97b3009K}. 
By having a plethora of possible observational signatures \citep[see e.g.][]{2012ApJ...746...48M,2014ApJ...791L...7P, 2019MNRAS.486.5289B}, \bhnsSingle mergers provide an interesting class of sources for multi-messenger astronomy. 
On the other hand, despite the possibility of observing the neutron stars in \bhnsSingle as Galactic radio pulsars, no such systems are known at present, which may indicate the relative rarity of systems or selection effects that make their detection unlikely unless the neutron star was formed first and recycled by accretion from the black hole's progenitor \citep{2020MNRAS.494.1587C}.  Meanwhile, some short gamma-ray bursts may have originated from \bhnsSingle mergers \citep[e.g.][]{2008MNRAS.385L..10T, 2020ApJ...895...58G}, but there are no consensus sources at present.

The main formation pathway leading to BHNS mergers  is under debate, but the  favored scenario is that they form from two massive stars that are born in a binary and evolve in isolation, typically involving a \ac{CE} episode that tightens the binary orbit \citep{1976ApJ...207..574S, 1989A&ARv...1..209S}.  This channel can explain the majority of current \ac{BHBH} and \ac{NSNS} mergers detected with \acp{GW}   \citep[e.g.][]{2018MNRAS.479.4391M,2018PhRvD..97d3014W, 2019MNRAS.490.3740N,2020ApJ...898..152S,2020ApJ...901L..39O}.
A popular alternative formation pathway is through dynamical interactions in 
globular clusters 
\citep[e.g.][]{2000ApJ...528L..17P,2003ASPC..302..391S,2010MNRAS.407.1946D, 2013MNRAS.428.3618C,,2014MNRAS.442..207C, 2016PhRvD..93h4029R} or 
young stellar clusters \citep[e.g.][]{2014MNRAS.441.3703Z,2016MNRAS.459.3432M, 2020ApJ...898..152S, 2020MNRAS.497.1563R},
but recent work suggests the predicted \bhnsSingle merger rate from these channels might be low ($\lesssim 10$\GpcminThree\yearmin)   \citep{2014MNRAS.440.2714B,2014MNRAS.441.3703Z, 2016MNRAS.459.3432M, 2018MNRAS.480.4955F,2019ApJ...877..122Y, 2020ApJ...888L..10Y, 2020CmPhy...3...43A,2020MNRAS.tmp.2087B,2020ApJ...901L..16F, 2020arXiv200609744S, 2020ApJ...903....8H}, although  \citet{ArcaSedda:2021,2020MNRAS.497.1563R,2020ApJ...898..152S} predict \bhnsSingle merger rates similar to the rates from isolated binary evolution for dynamical interactions in young stellar clusters. 

Other  formation channels  include  isolated (hierarchical) triple (or quadruple) evolution involving Kozai-Lidov oscillations  \citep{2017ApJ...836...39S,2019MNRAS.486.4443F,2019MNRAS.490.4991F, 2019ApJ...883...23H,2019ApJ...878...58S}, 
 isolated binary evolution where one star evolves chemically homogeneously through efficient rotational mixing  \citep{2016MNRAS.458.2634M,2017A&A...604A..55M}, population III stars \citep{2017MNRAS.471.4702B} and 
 formation in (active) galactic nucleus disks  \citep{2009MNRAS.395.2127O,2019MNRAS.488...47F, 2020MNRAS.498.4088M, 2020ApJ...901L..34Y}. More  exotic channels have also been suggested such as  formation from primordial black holes \citep{2013PhRvD..87l3524C,2014JCAP...06..026P} or mirror dark matter particles \citep{2020PhLB..80435402B}.  
Future  \ac{GW} observations will distinguish between  formation channels
\citep[e.g.][]{2010CQGra..27k4007M,2017MNRAS.471.2801S,2017Natur.548..426F,2017CQGra..34cLT01V,2020arXiv201110057Z}. 
Here we focus on the formation of \bhnsSingle mergers from the isolated binary evolution channel.

The formation of \bhnsSingle mergers through isolated binary evolution has been studied with population synthesis simulations for decades \citep[e.g.][]{1993MNRAS.260..675T, 1999ApJ...526..152F, 2003MNRAS.342.1169V, 2015ApJ...806..263D, 2018MNRAS.480.2011G,2018MNRAS.481.1908K,2019MNRAS.490.3740N,2020A&A...636A.104B}, but their predicted rates are still uncertain to several orders of magnitude   \citep[][]{2010CQGra..27q3001A,MandelBroekgaardenReview:2021}.
This uncertainty has already been shown to come from two main factors. 
First, from uncertain physical processes in massive (binary)-star evolution such as the \ac{CE} phase,  mass transfer efficiency and \ac{SN}  natal kicks \citep[e.g.][]{2018MNRAS.481.1908K,2020arXiv201016333B,2020arXiv201011220B,Belczynski:2021-uncertainStellarEvolution}.  
Second,  from uncertainties in the star formation history and metallicity distribution of star forming gas over cosmic time \citep[e.g.][]{2019MNRAS.488.5300C, 2019MNRAS.490.3740N,2020arXiv200903911S,2020MNRAS.493L...6T}, which we will refer to as the metallicity specific star formation rate density, the  \SFRD, which is a function of birth (initial) metallicity $\Zi$ and redshift $z$.

To make the most of future comparisons between observations and simulations of \bhnsSingle mergers it is crucial to explore the uncertainties from both the assumptions for the massive (binary) evolution and  \SFRD, in order to make predictions for the \bhnsSingle merger rate and characteristics. In turn, the population properties (e.g. the distributions of masses and mass ratios) can be used to make predictions  for, e.g.   the fraction of \bhnsSingle mergers with a possible electromagnetic counterpart.
However, previous studies typically focus on exploring only one of the two uncertainties, making it challenging to understand how the massive (binary) evolution and \SFRD combined impact the results. In addition,  studies often focus on presenting results for \ac{BHBH} or \ac{NSNS} mergers as these \ac{DCO} binaries have been observed longer and are more numerous and because \bhnsSingle mergers are typically a rare outcome in binary population synthesis models, making simulating a statistically significant population of \bhnsSingle systems computationally challenging \citep[e.g. ][]{2017IAUS..325...46B,andrews2017dart_board,  2018arXiv180608365T,2019MNRAS.490.5228B,2019PhRvD.100h3015W}.

In this paper we therefore focus on making predictions for \bhnsSingle mergers and exploring the uncertainties from both varying assumptions for the massive (binary) evolution and \SFRD.  We  increase the efficiency of our simulations for \bhnsSingle mergers by a factor of $\sim 100$ compared with typical simulations, that use sampling from the initial conditions,   using the adaptive importance sampling algorithm  {\sc{STROOPWAFEL}} \citep{2019MNRAS.490.5228B}.  By doing so, we can run simulations with high resolutions in metallicity (using 53 metallicity bins) and create catalogs with many \bhnsSingle sources.

We investigate a total of \Nmodels models, which are combinations of  \NmodelsBPS different  binary population synthesis model settings and \NmodelsMSSFR \SFRD prescriptions, to model these uncertainties.    Using these explorations we address the two main questions: (1) what are the expected properties of \bhnsSingle mergers?  and (2) how do the uncertainties from both massive (binary)  evolution and \SFRD  impact the predicted BHNS merger rates and properties?

 The method is described in Section \ref{sec:method}. We discuss the formation channels leading to \bhnsSingle mergers and their characteristics for both the intrinsic (merging at redshift zero) and \ac{GW} detectable population for our fiducial simulation assumptions in Section~\ref{sec:results-fiducial}.   We discuss how these predictions change for a set of  \Nmodels variations in both population synthesis model assumptions and \SFRD assumptions in Section~\ref{sec:results-variations}.  We end with a discussion in Section~\ref{sec:discussion} and present our conclusions in Section~\ref{sec:conclusions}.

All data produced in this study are publicly available on Zenodo at \url{https://doi.org/10.5281/zenodo.4574727}.
 All code, scripts and files to reproduce all figures and results in this paper are publicly available in the Github repository \url{https://github.com/FloorBroekgaarden/BlackHole-NeutronStar}. We present a comparison of our  \bhnsSingle results to similar predictions for \ac{BHBH} and \ac{NSNS} mergers in an accompanying paper (Broekgaarden et al., in prep.).


\section{Method}
\label{sec:method}
\subsection{Population Synthesis  Model Setup}
\label{subsec:method-BPS-assumptions}
To evolve a population of binary systems we use the rapid binary population synthesis code from the~{\sc{COMPAS}}\footnote{Compact Object Mergers: Population Astrophysics and Statistics,  \url{https://compas.science}} suite \citep{stevenson2017formation, 2018MNRAS.477.4685B, 2018MNRAS.481.4009V, 2019MNRAS.490.3740N, 2019MNRAS.490.5228B}. 
The main methodology of the binary population synthesis code  in COMPAS is built on  algorithms developed by  \citet{1985MNRAS.214..357W, 1987ApJ...321..780D} and \citet{1987SvA....31..228L} and later work by \citet{1997MNRAS.291..732T}. 
For single star evolution (SSE) COMPAS uses the analytic fitting formulae by \citet{2000MNRAS.315..543H,2002MNRAS.329..897H}, which are based on SSE tables presented by \citet{1998IAUS..191P.607P} and earlier work from  \citet{1989ApJ...347..998E} and \citet{1996MNRAS.281..257T}. The stellar evolution and binary interactions are  incorporated through parameterized  and approximate prescriptions of the physical processes. By doing so, COMPAS can typically compute an outcome of a binary system in under a second.  
{\COMPAS} is described in the {\COMPAS method paper} (in prep., from hereon \CMP). 
We describe below the most relevant assumptions and  the settings of our fiducial population synthesis model, which are also summarized in Table~\ref{tab:population-synthesis-settings}.

\begin{table*}
\caption{Initial values and default settings of the population synthesis  simulation with {\sc{COMPAS}} for our fiducial model A. More details can be found in Section~\ref{subsec:method-BPS-physical-assumptions} and in  the  {\sc{COMPAS}}  method paper  \CMP.  Cyan star symbols in front of a row indicate prescriptions that we vary. These variations are listed in  Table~\ref{tab:variations-BPS}.  A (overleaf) latex template for this table that is available for public use can be found at \url{https://github.com/FloorBroekgaarden/templateForTableBPSsettings}.  }
\label{tab:population-synthesis-settings}
\centering
\resizebox{\textwidth}{!}{%
\begin{tabular}{lll}
\hline  \hline
Description and name                                 														& Value/range                       & Note / setting   \\ \hline  \hline
\multicolumn{3}{c}{Initial conditions}                                                                      \\ \hline
Initial mass $\monei$                               															& $[5, 150]$\Msun    & \citet{2001MNRAS.322..231K} IMF  $\propto  {\monei}^{-\alpha}$  with $\alpha_{\rm{IMF}} = 2.3$ for stars above $5$\Msun	  \\
Initial mass ratio $\qi = \mtwoi / \monei $           												& $[0, 1]$                          &       We assume a flat mass ratio distribution  $p(\qi) \propto  1$ with \mtwoi $\geq 0.1\Msun$   \\
Initial separation $\ai$                                            											& $[0.01, 1000]$\AU & Distributed flat-in-log $p(\ai) \propto 1 / {\ai}$ \\ 
Initial metallicity $\Zi$                                           											& $[0.0001, 0.03]$                 & Distributed using a uniform grid in $\log(\Zi)$ with 53 metallicities        \\
Initial orbital eccentricity $\ei$                                 							 				& 0                                & All binaries are assumed to be circular at birth  \\
%
\hline
\multicolumn{3}{c}{Fiducial parameter settings:}                                                            \\ \hline
Stellar winds  for hydrogen rich stars                                   																&      \citet{2010ApJ...714.1217B}    &   Based on {\citet{2000A&A...362..295V,2001A&A...369..574V}}, including  LBV wind mass loss with $f_{\rm{LBV}} = 1.5$.   \\
Stellar winds for hydrogen-poor helium stars &  \citet{2010ApJ...715L.138B} & Based on   {\citet{1998A&A...335.1003H}} and  {\citealt{2005A&A...442..587V}}.  \\

%
Max transfer stability criteria & $\zeta$-prescription & Based on \citet[][]{2018MNRAS.481.4009V} and references therein     \\ 
{\hspace{-.35cm}\Large{\textcolor{cyan}{$\star$}}}{\hspace{+.02cm}} Mass transfer accretion rate & thermal timescale & Limited by thermal timescale for stars  \citet[][]{2018MNRAS.481.4009V,2020MNRAS.498.4705V} \\ 
 & Eddington-limited  & Accretion rate is Eddington-limit for compact objects  \\
Non-conservative mass loss & isotropic re-emission &  {\citet[][]{1975MmSAI..46..217M,1991PhR...203....1B,1997A&A...327..620S}} \\ 
& &  {\citet{2006csxs.book..623T}} \\
{\hspace{-.35cm}\Large{\textcolor{cyan}{$\star$}}}{\hspace{+.02cm}} Case BB mass transfer stability                                														& always stable         &       Based on  \citet{2015MNRAS.451.2123T,2017ApJ...846..170T,2018MNRAS.481.4009V}         \\ 
%
%
CE prescription & $\alpha-\lambda$ & based on  \citet{1984ApJ...277..355W,1990ApJ...358..189D}  \\
{\hspace{-.35cm}\Large{\textcolor{cyan}{$\star$}}}{\hspace{+.02cm}} CE efficiency $\alpha$-parameter                     												& 1.0                               &              \\
CE $\lambda$-parameter                               													& $\lambda_{\rm{Nanjing}}$                             &        Based on \citet{2010ApJ...716..114X,2010ApJ...722.1985X} and  \citet{2012ApJ...759...52D}       \\
{\hspace{-.35cm}\Large{\textcolor{cyan}{$\star$}}}{\hspace{+.02cm}} Hertzsprung gap (HG) donor in \ac{CE}                       														& pessimistic                       &  Defined in \citet{2012ApJ...759...52D}:  HG donors don't survive a \ac{CE}  phase        \\
%
%
\ac{SN} natal kick magnitude \vk                          									& $[0, \infty)$\kms & Drawn from Maxwellian distribution    with standard deviation $\sigma_{\rm{rms}}^{\rm{1D}}$          \\
 \ac{SN} natal kick polar angle $\thetak$          											& $[0, \pi]$                        & $p(\thetak) = \sin(\thetak)/2$ \\
 \ac{SN} natal kick azimuthal angle $\phi_k$                           					  	& $[0, 2\pi]$                        & Uniform $p(\phi) = 1/ (2 \pi)$   \\
 \ac{SN} mean anomaly of the orbit                    											&     $[0, 2\pi]$                             & Uniformly distributed  \\
{\hspace{-.35cm}\Large{\textcolor{cyan}{$\star$}}}{\hspace{+.02cm}} Core-collapse  \ac{SN} remnant mass prescription          									     &  delayed                     &  From \citep{2012ApJ...749...91F}, which  has no lower \ac{BH} mass gap  \\%
{\hspace{-.35cm}\Large{\textcolor{cyan}{$\star$}}}{\hspace{+.02cm}} USSN  remnant mass prescription          									     &  delayed                     &  From \citep{2012ApJ...749...91F}   \\%
ECSN  remnant mass presciption                        												&                                 $m_{\rm{f}} = 1.26\Msun$ &      Based on Equation~8 in \citet{1996ApJ...457..834T}          \\
{\hspace{-.35cm}\Large{\textcolor{cyan}{$\star$}}}{\hspace{+.02cm}} Core-collapse  \ac{SN}  velocity dispersion $\sigma_{\rm{rms}}^{\rm{1D}}$ 			& 265\kms           & 1D rms value based on              \citet{2005MNRAS.360..974H}                          \\
 USSN  and ECSN  velocity dispersion $\sigma_{\rm{rms}}^{\rm{1D}}$ 							 	& 30\kms             &            1D rms value based on e.g.    \citet{2002ApJ...571L..37P,2004ApJ...612.1044P}    \\
{\hspace{-.35cm}\Large{\textcolor{cyan}{$\star$}}}{\hspace{+.02cm}} PISN / PPISN remnant mass prescription               											& \citet{2019ApJ...882...36M}                    &       As implemented in \citet{2019ApJ...882..121S}      \\
{\hspace{-.35cm}\Large{\textcolor{cyan}{$\star$}}}{\hspace{+.02cm}} Maximum NS mass                                      & $\rm{max} \, m_{\rm{NS}} = 2.5$\Msun &             \\
Tides and rotation & & We do not include prescriptions for tides and/or rotation\\
\hline
\multicolumn{3}{c}{Simulation settings}                                                                     \\ \hline
Total number of binaries sampled per metallicity  & $\approx 10^6$                    &      We simulate about a million binaries per \Zi grid point            \\
Sampling method                                      & \sc{STROOPWAFEL} &                Adaptive importance sampling from  \citet{2019MNRAS.490.5228B}.  \\
Binary fraction                                      & $f_{\rm{bin}} = 1$ &       Corrected factor to be consistent with e.g. {\citet[][]{2017IAUS..329..110S}}        \\
Solar metallicity \Zsun                             & \Zsun = 0.0142 & based on {\citet{2009ARA&A..47..481A}} \\
Binary population synthesis code                                      & COMPAS &       \citet{stevenson2017formation, 2018MNRAS.477.4685B, 2018MNRAS.481.4009V, 2019MNRAS.490.3740N} \\
& & \citet{2019MNRAS.490.5228B}.        \\
\hline \hline
\end{tabular}%
}
\end{table*}

\subsubsection{Initial distribution functions and sampling method }
\label{subsec:method-BPS-initial-conditions}
Each binary system in our simulation can be described at birth  (on the zero-age main sequence, ZAMS) by its initial component masses, separation, eccentricity and metallicity. During the simulation  random birth parameter values are drawn for each binary from distributions whose shape is based on observations that are described below. We assume that the initial parameter distributions are independent of each other. Although this might not be valid \citep{2013ARA&A..51..269D, 2017ApJS..230...15M}, this likely only introduces a small uncertainty  \citep{1990ApJS...74..551A, 2018A&A...619A..77K}.

We assume the mass of the initially most massive star in the binary system (the primary) \monei  follows a \citet{2001MNRAS.322..231K} initial mass function (IMF) with distribution function $p(\monei) \propto  {\monei}^{-\alpha}$ with $\alpha = 2.3$, with masses \monei  $\in [5,150]$\Msun, where the lower limit is chosen as stars below this mass typically do not form \acp{NS} and the $150\Msun$ is based on the typical maximum observed mass of stars. The mass of the secondary is chosen by drawing a  mass ratio between the two stars $\qi \equiv \mtwoi / \monei$, which is assumed to follow a flat distribution
on $[0,1]$  \citep[cf.][]{1991MNRAS.250..701T,1992ApJ...401..265M, 1994A&A...282..801G, 2007ApJ...670..747K, 2012Sci...337..444S, 2014ApJS..213...34K}. We set a minimum secondary mass of \mtwoi $\geq 0.1\Msun$, the approximate minimal mass for a main sequence star \citep{HayashiNakano:1963}. The initial separation is assumed to follow a flat in the log distribution $p(\ai) \propto 1 / {\ai}$, with  $\ai \in [0.01, 1000]$\AU \citep{1924PTarO..25f...1O, 1983ARA&A..21..343A, 2013ARA&A..51..269D}, where the lower limit is chosen as stars closer than 0.01\AU typically touch on the ZAMS and the upper limit is chosen as we assume wider binaries are single stars. We reject and resample binaries that are drawn with such small separations that there is mass transfer at birth, as we assume those binaries merge as stars  and are not included in the population of binaries. 
We assume all binaries are circular at birth ($\ei =0$) to reduce the dimensions of our parameter space; \citet{2015ApJ...814...58D} showed that this assumption is likely not critical for predictions of \bhnsSingle mergers. In addition,  close binaries are expected to circularize by the time they have reached their first mass transfer episode  \citep[cf. ][]{1973ApJ...180..307C,1977A&A....57..383Z,1995A&A...296..709V,2008EAS....29...67Z}, although see e.g. \citet{2020PASA...37...38V}. 
Since this study focuses on post-mass transfer binaries we expect that starting with circular orbits does not significantly influence our outcomes.

The birth metallicities of the stars are varied by using a grid of 53 different initial fractional metallicities $\Zi$ in the range  $[0.0001, 0.03]$, which matches the metallicity range of the stellar models by \citet{1998IAUS..191P.607P}. We define the fractional metallicity Z as the mass fraction of metals such that $X+Y+Z =1$ with $X$ and $Y$ the mass fractions of hydrogen and helium, respectively. The \Zi grid points are roughly uniformly distributed in log-$(\Zi)$ space\footnote{See the scatter points in Figure~\ref{fig:BHNS_rate_per_metallicity} for the exact grid of metallicities.}.
For each grid point  $\Zi$ we draw using Monte Carlo $\approx 10^6$ initial binaries using the adaptive importance sampling algorithm {\sc{STROOPWAFEL}} \citep{2019MNRAS.490.5228B}. This  algorithm  improves our efficiency of sampling  the rare astrophysical outcome of  \bhnsSingle binaries (and \ac{NSNS} and \ac{BHBH}) in population synthesis simulations by a factor of about $ 100$ with respect to traditional Monte Carlo sampling from the birth distributions within these initial ranges. 

We assume all stars to initially be non-rotating (see e.g. \citealt{2013ApJ...764..166D} for more details). 

\subsubsection{Physical assumptions in the binary population synthesis model }
\label{subsec:method-BPS-physical-assumptions}
We summarize our most important binary population synthesis model assumptions below and in Table~\ref{tab:population-synthesis-settings}. Our fiducial model has label A, and all \NmodelsBPS different binary population synthesis models studied in this paper are summarized in Table~\ref{tab:variations-BPS}. More details about the modelling in {\COMPAS} are given in \CMP.

For hydrogen-rich stars we implement the mass loss rates for line-driven stellar winds from  \citet{2000A&A...362..295V,2001A&A...369..574V} as implemented by \citet[][see their Equations~6 and 7]{2010ApJ...714.1217B}. This includes applying an additional  wind mass loss of $f_{\rm{LBV}} \cdot  10^{-4}$\Msun\yearmin{}  independent of metallicity to mimic the effect of luminous blue variable (LBV) winds for stars crossing the \citet{1994PASP..106.1025H} limit. 
We adopt the default $f_{\rm{LBV}} = 1.5$ from \citet{2010ApJ...714.1217B}. For hydrogen-poor  stars (which can be observed as Wolf-Rayet stars, \citealt{2007ARA&A..45..177C}) we use the stellar wind prescription from \citet{2010ApJ...715L.138B}.

We distinguish in this paper between three different cases of mass transfer depending on the stellar phase of the donor star \citep[based on][]{1967ZA.....65..251K,1970A&A.....7..150L}.  Case A is when mass transfer is initiated from a main sequence donor (during core hydrogen burning), case B for hydrogen-shell burning or core-helium burning donors and case C  for post core helium burning donors. In addition, we use case BA, case BB and case BC analogues for the A, B and C mass transfer cases from a stripped or helium donor star   \citep[cf.][]{1977Ap&SS..50...75D,1981A&A....96..142D, 1993ARep...37..411T}.

We use the $\zeta$-prescription to determine the stability of mass transfer, which compares the radial response of the donor star with the response of the Roche lobe radius to mass transfer (\citealt[][]{2018MNRAS.481.4009V,2020MNRAS.498.4705V}, \CMP{} and references therein). 
The mass transfer efficiency describes the fraction of the mass lost by the donor that is accreted by the companion star, $\beta = \Delta M_{\rm{acc}} / \Delta M_{\rm{donor}}$, where $ \Delta M_{\rm{donor}}$ and $ \Delta M_{\rm{acc}}$ are the change in mass by the donor and accretor star over time, respectively.  The timescale and amount of donated mass are set by the stellar type of the donor star.  We assume for our fiducial model that the maximum accretion rate for stars is 
$    \Delta M_{\rm{acc}} / \diff t = 10 M_{\rm{acc}} / \tau_{\rm{KH}},$
 similar to \citet{2002MNRAS.329..897H}, with $t$ the time,  $\tau_{\rm{KH}}$ the Kelvin-Helmholtz (thermal) timescale of the star, and  the factor of 10 is added to take into account the expansion of the accretor due to mass transfer \citep{1972AcA....22...73P,2002MNRAS.329..897H,,2015ApJ...805...20S}. For compact objects we assume the maximum mass accretion rate is Eddington-limited, this assumption likely does not impact our result \citep{2020ApJ...897..100V}.
If more mass is transferred from the donor than can be accreted we assume this mass is lost from the vicinity of the accreting star  through  `isotropic re-emission'  \citep[e.g.][]{1975MmSAI..46..217M,1991PhR...203....1B,1997A&A...327..620S, 2006csxs.book..623T} and adopt the specific angular momentum accordingly  \citep[e.g.][]{2008ApJS..174..223B}.  
In  models B, C and D we vary the maximum accretion rate of the accreting star by instead setting $\beta$ to fixed values of $\beta = 0.25, 0.5$ and $ 0.75$, respectively.

We assume for our fiducial model that a mass transfer phase from a stripped post-helium-burning star (case~BB) onto a \ac{NS} or \ac{BH} is always stable as suggested by  \citet{2015MNRAS.451.2123T,2017ApJ...846..170T}. 
\citet{2018MNRAS.481.4009V} show that this assumption leads to a better match of population synthesis models to the observed population of  Galactic \ac{NSNS} binaries.  We vary this in model E, where we assume case BB mass transfer to always be unstable (Table~\ref{tab:variations-BPS}).

We follow the simplified $\alpha$--$\lambda$ prescription from \citet{1984ApJ...277..355W} and \citet{1990ApJ...358..189D} to parameterize the \ac{CE} phase. 
We assume for the $\alpha$ parameter, which regulates the efficiency with which the envelope is ejected, the value $\alpha=1$ in our fiducial model but vary this to $\alpha = 0.5$ and $\alpha=2$ in models F and G, respectively (Table~\ref{tab:variations-BPS}). A suitable value of $\alpha$ for these simulations is uncertain, and challenging to infer from observations and simulations. There may well be no single value of $\alpha$ which accurately describes the physics in the diverse \ac{CE} phases experienced by our compact-object progenitors. Population synthesis predictions have suggested the $\alpha$ value impacts the detectable \ac{GW} merger rates \citep[e.g.][]{2012ApJ...759...52D,2016A&A...596A..58K,2018MNRAS.481.1908K,2021arXiv210205649O}. 
For the $\lambda$ parameter we use the fitting formulas from \citet{2010ApJ...716..114X,2010ApJ...722.1985X},  similar to the $\lambda_{\rm{Nanjing}}$ parameter in \citet{2012ApJ...759...52D}, which includes internal energy ($\lambda_{\rm{b}}$ as in \citealt{2010ApJ...716..114X,2010ApJ...722.1985X}) and the added models up to $100\Msun$ zero-age main sequence masses. Similar to \citet[][see their Section 2.3.2.]{2012ApJ...759...52D} we extrapolate these models up to our maximum mass of $150 \Msun$.  In this method the value of $\lambda$  depends on the stellar evolutionary stages of the stars (see also \citealt{2000A&A...360.1043D,2001A&A...369..170T,2016A&A...596A..58K}). For more details on the $\alpha$ and $\lambda$ see \citet[][]{2013A&ARv..21...59I} and references therein. 

We do not allow hydrogen shell burning (typically Hertzsprung gap) donor stars  that initiate a \ac{CE} event to survive in our fiducial model. These donor stars are not expected to have developed a steep density gradient between core and envelope  \citep{2000ARA&A..38..113T,2004ApJ...601.1058I},  making it challenging to successfully eject the envelope. Instead, it is thought that a merger takes place, and it has been shown for a few cases that such  binaries are unlikely to form a \ac{DCO} that can form a \ac{GW} source \citep{2015MNRAS.449.4415P,2017MNRAS.465.2092P}.  This implementation follows the `pessimistic' \ac{CE} scenario  \citep[cf.][]{2012ApJ...759...52D}.  The `optimistic' \ac{CE} scenario, on the other hand, assumes these systems can survive.   Which of the scenarios more accurately represents observations is still under debate.   Recently, \citet{2017MNRAS.472.2422M} show that the pessimistic scenario slightly better matches the predicted  \ac{BHBH}  rate.   \citet{2018MNRAS.481.4009V}  argue that there is no clear evidence to favor one of the scenarios over the other based on a study of Galactic \ac{NSNS}  binaries.  
We use the pessimistic model for our fiducial assumption, similar to recent  population synthesis studies \citep[e.g.][]{2018MNRAS.480.2011G,2019ApJ...885....1W,2019MNRAS.490.3740N} and use the optimistic assumption in model variation H (Table~\ref{tab:variations-BPS}). We do allow main sequence companion stars in a \ac{CE} event to survive the \ac{CE}. %
Lastly, we remove binaries where the secondary star fills its Roche lobe immediately after a \ac{CE} event, as we treat those events as failed \ac{CE} ejections. 
	More details and discussion of the treatment of \ac{CE} events in {\sc{COMPAS}} are discussed in \citet{2020PASA...37...38V} and \CMP.

We use the `delayed' \ac{SN}  remnant mass prescription \citep{2012ApJ...749...91F}  to map the carbon-oxygen core masses of stars to compact object remnant masses during core-collapse \ac{SN}  events\footnote{Previous COMPAS studies included an extra mass gap in the NS mass distribution around 2.2 \Msun (see, e.g. the middle panel of Figure~7 in \citealt{2018MNRAS.481.4009V} and bottom panels of Figure~7 in \citealt{2019MNRAS.490.5228B}) due to a difference in the assumed relationship between baryonic and gravitational masses for NS and BH remnants. In our corrected treatment, this quirk (see Equations~12 and 13 in \citealt{2012ApJ...749...91F}) leads to a single gap in the NS mass distribution around 1.7\Msun and a small gap between NS and BH gravitational masses even for the `delayed' remnant mass prescription.}. 
We deviate with this choice from most binary population synthesis codes that typically  use instead the \textit{rapid} \ac{SN}  remnant mass prescription for their fiducial model (examples include \citealt[][]{2019IAUS..346..417K, 2020ApJ...898...71B},  but see e.g. also the discussion in \citealt{2020MNRAS.495.2786E} on why this assumption may need to be revisited). We discuss in  Section~\ref{subsec:discussion-delayed-vs-rapid-SN-remnant-mass} why we prefer using the delayed remnant mass prescription. 
The main difference is that the delayed remnant mass prescription does not create, by construction, the remnant mass gap between NSs and BHs that might be apparent from  X-ray binary observations 
\citep{1998ApJ...499..367B, 2010ApJ...725.1918O, 2011APS..APRH11002F}.
 We explore changing our model to the rapid remnant mass prescription in model variation I (Table~\ref{tab:variations-BPS}).

  During the \ac{SN}  event a fraction of the ejected material, $f_{\rm{fb}}$, is assumed to fallback onto the compact object. We use the prescription from Equations~16 and 19 in  \citet{2012ApJ...749...91F} to determine this fraction and adjust the final remnant mass accordingly.

The maximum mass that a NS can have is uncertain and under debate. Observations from pulsars show that most \acp{NS} have masses around $\sim$1.3\Msun \citep[e.g.][]{2011MNRAS.414.1427V,2012ApJ...757...55O,2013ApJ...778...66K,2019mbhe.confE..23L} and have found that the most massive  \acp{NS} to date  have masses of $\sim 2-2.2$\Msun \citep{2013Sci...340..448A,2020NatAs...4...72C, 2020RNAAS...4...65F}, although more massive \acp{NS} might have been observed  (e.g. \citealt[][]{2008ApJ...675..670F, 2011ApJ...728...95V}, but see the discussion in  \citealt{2016ARA&A..54..401O} on why these mass measurements are more uncertain).
From observational and theoretical (population) modelling the possible maximum NS mass  has been predicted to lie in the range 2--3\Msun \citep[e.g.][]{1996ApJ...470L..61K, 2015ApJ...808..186L,2015ApJ...812...24F,2017ApJ...850L..19M,2018MNRAS.478.1377A, 2020CQGra..37d5006A, 2020arXiv200106102S, 2020ApJ...892L...3A}. 
We decide to assume for our fiducial model that \acp{NS} can have a maximum mass of 2.5\Msun. For models J and K we change this to 2\Msun and 3\Msun, respectively. We adopt the \citet{2012ApJ...749...91F} remnant mass prescription accordingly.

\begin{table*}
 \caption{List of the \NmodelsBPS binary population synthesis models studied in this work. $\mu$ and `Label' denote the alphabetical letter and abbreviated name used to label each model, `Changed physics' and `Variation' denote which category of physics and what we changed, respectively. The column $\#$BHNS lists the total number of \bhnsSingle systems that merges in a Hubble time across the 53 metallicity bins comprising the given simulation.
The fiducial model settings are summarized in Table~\ref{tab:population-synthesis-settings}.}
\label{tab:variations-BPS}
\centering

\begin{tabular}{llllr}
\hline \hline
$\mu$ & Label 	& Changed physics & Variation  			&  $\#\rm{BHNS}^{*}$   	\\ \hline \hline 
A      & fiducial		& --        		& --     					& $1\,525\,553$  \\
\hline
%
B     & $\beta =0.25$  		& mass transfer                  				& fixed mass transfer efficiency of $\beta=0.25$  		&$738\,537$ \\
C       & $\beta =0.5$  		& mass transfer                  				& fixed mass transfer efficiency of $\beta=0.5$  		& $148\,043$  \\
D        & $\beta =0.75$ 		& mass transfer                  				& fixed mass transfer efficiency of $\beta=0.75$  		& $118\,921$\\
E       & unstable case BB		& mass transfer 	& case BB mass transfer is assumed to be always unstable 			&   $458\,667$\\
\hline
%
F       &$\alpha=0.5$		& \ac{CE}                    				&  CE efficiency parameter $\alpha = 0.5$ 	& $915\,179$ \\
G        & $\alpha=2$			& \ac{CE}                    				& CE efficiency parameter $\alpha = 2$  	& 	$833\,433$\\
H      	& optimistic CE	&  \ac{CE}       		& HG donor stars initiating a CE survive CE     					& $1\,535\,042$ \\
\hline
%
%
I       & rapid SN		& SN                   				& Fryer rapid \ac{SN} remnant mass prescription   		&  $2\,766\,298$\\
J       & $\rm{max} \, m_{\rm{NS}} = 2\Msun$		& SN                 				& maximum NS mass is fixed to $2\Msun$   		&  $959\,796$  \\
K        & $\rm{max} \, m_{\rm{NS}} = 3\Msun$		& SN                 				& maximum NS mass is fixed to $3\Msun$   		& $1\,990\,330$ \\
L       & no PISN		& SN                   				& we do not implement   PISN and pulsational-PISN  		& $1\,524\,497$ \\
M      	 & $\sigma_{\rm{cc}} =100$ 	& SN                 				& $\sigma_{\rm{rms}}^{\rm{1D}}=100$\kms  	for core-collapse \acp{SN}	& $3\,049\,458$ \\
N        & $\sigma_{\rm{cc}} =30$ 		& SN                  				& $\sigma_{\rm{rms}}^{\rm{1D}}=30$\kms  	for core-collapse \acp{SN}	& $4\,198\,238$ \\
O       & $v_{\rm{k,BH}} = 0$		& SN                   				& we assume \acp{BH} receive no natal kick  		& $5\,068\,628$ \\
\hline \hline
\end{tabular}%
\end{table*}

 We assume stars  with helium core masses in the range  1.6--2.25\Msun \citep{2002MNRAS.329..897H}  lead to \acp{ECSN}\footnote{Lower core masses will lead to the formation of white dwarfs} \citep{1980PASJ...32..303M, 1984ApJ...277..791N,1987ApJ...322..206N, 2008MNRAS.386..553I}. 
If a star undergoes an \ac{ECSN}  we set its remnant mass to 1.26\Msun, as an approximation to the solution of Equation~8 in \citet{1996ApJ...457..834T}.

Case BB mass transfer from a companion star onto a \ac{NS} or \ac{BH} in short period binaries leads to severe stripping (leaving behind an envelope with mass $\lesssim 0.1$\Msun) and we assume the stripped star eventually undergoes an \ac{USSN}  as shown by \citet{2013ApJ...778L..23T,2015MNRAS.451.2123T, 2015MNRAS.454.3073S, 2017MNRAS.466.2085M} and  \citet{2018MNRAS.479.3675M}. This follows the prescription of the fiducial model of \citet{2018MNRAS.481.4009V}, but we also assume case BB mass transfer onto a \ac{BH}  leads to an ultra-stripped star and \ac{USSN}  as suggested by \citet{2013ApJ...778L..23T,2015MNRAS.451.2123T}  and in agreement with the implementation of \ac{USSN}  in other binary population synthesis work (e.g. \citealt[][]{2018MNRAS.481.1908K,2020MNRAS.498.4705V}).   We calculate the remnant mass of an ultra-stripped \ac{SN}  using the delayed \citet{2012ApJ...749...91F} \ac{SN} prescription in our fiducial model.

Stars with helium cores in the range of about 45--150\Msun are thought to become unstable and undergo a \ac{PISN} or pulsational-\ac{PISN}, leading to an absence of \acp{BH} with masses in this range as shown from  theory \citep{1964ApJS....9..201F,1967PhRvL..18..379B,2017ApJ...836..244W, 2019ApJ...887...53F} and observations \citep{2017ApJ...851L..25F,2018ApJ...856..173T,2019PhRvX...9c1040A, 2019PhRvD.100d3012W, 2019MNRAS.484.4216R,2020PhRvD.102h3026G}. \ac{BH} formation is expected again above a helium core mass of $\sim 150$\,M$_\odot$ \citep{2002RvMP...74.1015W,2019ApJ...878...49W,2021arXiv210307933W}. However, this assumption has recently been challenged by the second \ac{GW} catalog that contained \ac{BHBH} merger detections with components with masses $\gtrsim 45\Msun$ \citep{2020arXiv201014527A}, the most massive being GW190521 \citep{2020PhRvL.125j1102A}. The origin of these massive \acp{BH} is still unknown, and one of the current thoughts is that they formed from other channels than the isolated binary evolution channel such as via hierarchical mergers \citep[e.g.][]{2020arXiv201006161A}, stellar mergers \citep[e.g.][]{2019MNRAS.485..889S, 2020MNRAS.497.1043D,2020ApJ...903...45K}, triples \citep[e.g.][]{2021ApJ...907L..19V} or in AGN disks \citep[e.g.][]{2020ApJ...903..133S}. See for a more detailed discussion, for example,  \citet{2020ApJ...900L..13A} and \citet{2020arXiv201105332K}  and references therein. 
We follow the \ac{PISN} and pulsational-\ac{PISN}  prescription from \citet{2019ApJ...882...36M} as implemented in \citet{2019ApJ...882..121S}. 
The mass range of helium cores that undergo \ac{PISN} is shown to be a robust prediction (\citealt{2019ApJ...887...53F,2020MNRAS.493.4333R}) and we do not expect our particular choice for \ac{PISN} and pulsational-\ac{PISN} prescription to drastically influence our results other than somewhat the location of the gap (e.g. \citealt{2017MNRAS.470.4739S,2019ApJ...882..121S, 2020ApJ...902L..36F, 2021MNRAS.501.4514C,2021arXiv210307933W}). In model L we explore a simulation where we do not implement pulsational-\ac{PISN} and \ac{PISN}.

NSs and BHs might receive \ac{SN}  kicks at birth.  
For core-collapse \acp{SN}  we draw these natal kick magnitudes  from a Maxwellian velocity distribution with a one-dimensional root-mean-square velocity dispersion of $\sigma_{\rm{rms}}^{1D} = 265$\kms, based on observations of radio pulsar proper motions  \citep{2005MNRAS.360..974H, 1994Natur.369..127L}. In models M and N we explore the variation of lower kicks by using    $\sigma_{\rm{rms}}^{1D} = 100$\kms and $\sigma_{\rm{rms}}^{1D} = 30$\kms, respectively. 
For \ac{USSN} and \ac{ECSN}  we use instead  a  one-dimensional root-mean-square velocity dispersion of  $\sigma_{\rm{rms}}^{1D} = 30$\kms  following \citet{2002ApJ...571L..37P} and \citet{2004ApJ...612.1044P} as they are thought to have smaller kicks than standard iron core-collapse \ac{SN} \citep[e.g.][]{2015MNRAS.454.3073S,2018ApJ...865...61G,2018MNRAS.479.3675M}. This is in agreement with the subset of Galactic binary \ac{NS} systems and pulsars observed with low  velocities and small eccentricities  \citep{2010ApJ...719..722S,2016MNRAS.456.4089B, 2002ApJ...571..906B,2017ApJ...846..170T, 2017A&A...608A..57V, 2017JApA...38...40V, 2020MNRAS.494.3663I}. 
Combined, the magnitude of  \ac{SN}  kicks will thus represent a broader, bi-modal distribution as supported by \citet{1975Natur.253..698K,2002ApJ...568..289A, 2017A&A...608A..57V,} and \citet{ 2020MNRAS.494.3663I}.

We reduce the natal kick using the fallback fraction by a factor ($1-f_{\rm{fb}}$). This typically reduces the natal kicks of BHs with masses above 11\Msun to zero for the delayed \ac{SN}  prescription. This is in agreement with observations  that show evidence that  BHs might form with lower or no natal kicks, although this is still under debate  \citep[][]{1995MNRAS.277L..35B,1999A&A...352L..87N,2012MNRAS.425.2799R,2013MNRAS.434.1355J,2015MNRAS.453.3341R,2016MNRAS.456..578M}.    Model O explores a binary population synthesis variation where \acp{BH} receive no natal kick (Table~\ref{tab:variations-BPS}).

All natal kicks are assumed to be isotropic in the frame of the collapsing star (e.g. \citealt{2013A&A...552A.126W})  and we sample the kick polar angles $\theta_k$ and kick azimuthal angles $\phi_k$ from a unit sphere. The mean anomaly of the orbit is randomly drawn from $[0, 2\pi]$.

The inspiral timescale, $\tinspiral$,  
as a result from orbital energy lost in \acp{GW}  is based on \citet{1964PhRv..136.1224P}.

\begin{figure}
		\includegraphics[width=\columnwidth]{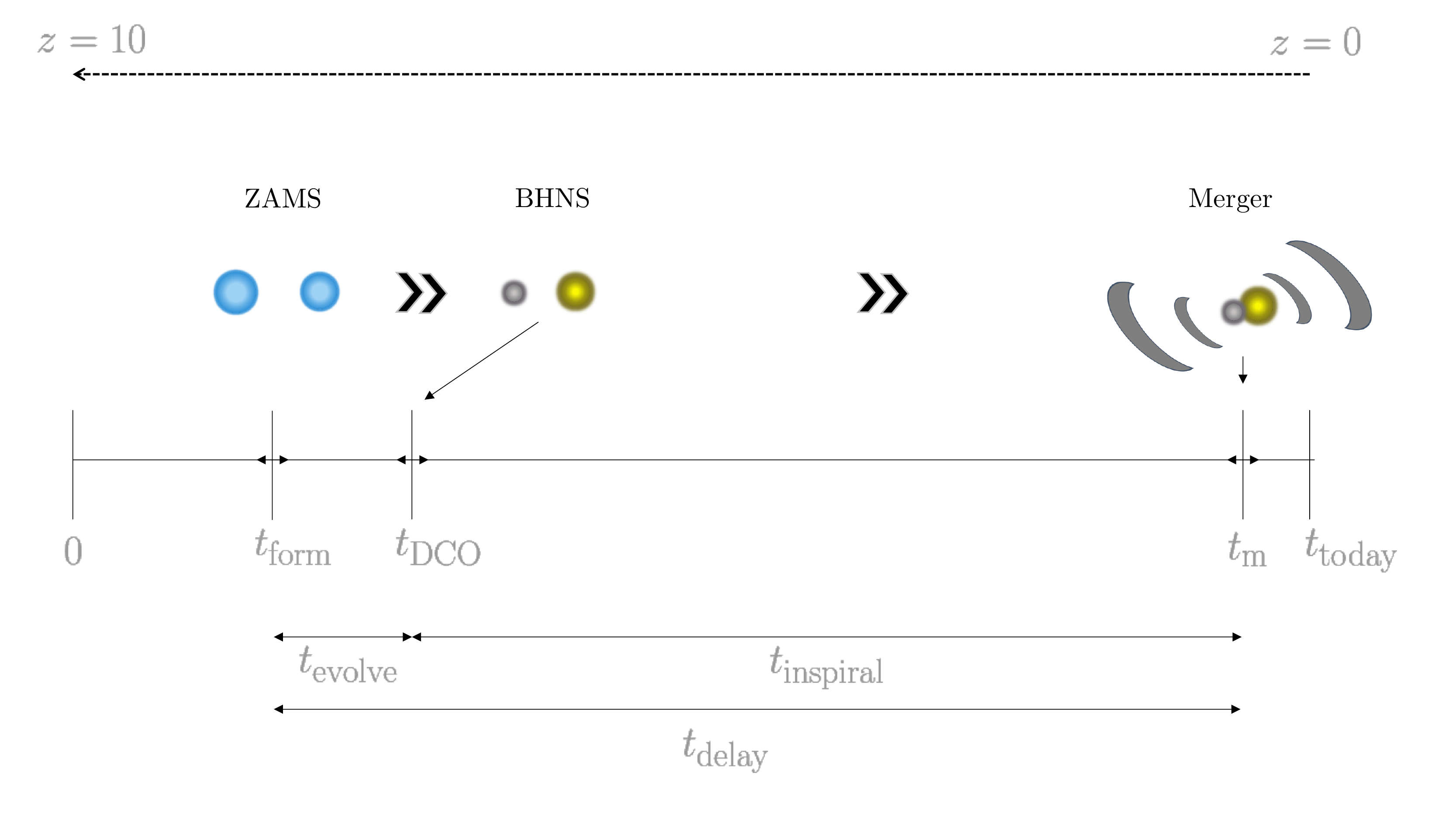} 
    \caption{Schematic display of the different times in the formation and evolution of a binary system that impact the time \tmerger at which a  \bhnsSingle system will merge in the Universe. The relevant timescales are:  the moment the binary is formed at ZAMS from a gas cloud, \tform, the moment of the \ac{DCO}  formation, \tDCO, the time at which the merger takes place, \tmerger, the time it takes the binary to evolve from ZAMS to the \ac{DCO} system, $\tevolve = \tDCO - \tform$, the inspiral time, $\tinspiral = \tmerger-\tDCO$, and the time between binary formation at ZAMS and merger, $\tdelay = \tmerger - \tform$. We assume in our models that star formation commenced at redshift $z = 10$.   } 
   \label{fig:timescalesEvolution}
\end{figure}

\subsubsection{The formation, evolution, inspiral and delay times}	
\label{subsec:method-BPS-different-timescales-binary}
The time at which a \ac{DCO} merges, \tmerger,  is given by  $\tmerger = \tform + \tevolve + \tinspiral$, shown in Figure~\ref{fig:timescalesEvolution}, with $\tform$ the time at which the initial binary forms from a gas cloud and starts hydrogen burning since  the beginning of star formation in the Universe, \tevolve the time it takes the binary from the onset of hydrogen burning at ZAMS to form a \ac{DCO} (i.e., until the second SN) and \tinspiral the time  it takes the \ac{DCO} to coalesce from the moment of the second \ac{SN}. 
The formation and inspiral time together form the delay time $\tdelay = \tevolve+ \tinspiral$.

\subsubsection{Initializing a population of \bhnsSingle mergers}
\label{subsec:selecting-a-population-of-BHNS-mergers}
For each binary population synthesis model  we evolve a population of about  $10^6$ binary systems per metallicity $\Zi$ from birth until they form a \ac{DCO} binary or otherwise merge or disrupt. From the population of \ac{DCO} binaries we select the \bhnsSingle systems. We also select only binaries that merge in a Hubble time, (i.e., that have \tdelay $\leq t_{\mathcal{H}}$, with $t_{\mathcal{H}} =  \mathcal{H}_0^{-1} \approx 14$\Gyr;  cf. the WMAP9--cosmology, \citealt{2013ApJS..208...19H}).
We assume that \bhnsSingle systems with merger times exceeding \thubble will not be detectable by \ac{GW} detectors (in the near-future).

We use {\COMPAS} to calculate and record properties such as the ages, masses, stellar radii, effective temperatures, velocities, eccentricities,  and separations of the binary at important evolutionary stages of the binary such as mass transfer episodes and \acp{SN}. \COMPAS also records \tevolve and \tinspiral for each \bhnsSingle system.

Throughout the rest of the paper we will use the notation \bhnsSingle for binaries containing a \ac{BH} and a NS. The notation BH--NS  (NS--BH) will be used when we explicitly refer to a \bhnsSingle binary where the \ac{BH} (NS) formed in the first \ac{SN}.
The formation order is particularly important for spinning up the \ac{BH} or \ac{NS} and the formation of millisecond pulsars (discussed in Section~\ref{subsec:method-tidal-disruption-BHNS}).

\subsection{Calculating the BHNS formation rate per unit star forming mass}
\label{subsec:method-BPS-merger-rate-per-M-SFR}
We model only a small fraction of the underlying stellar population  by neglecting single stars, not simulating binaries with primary star masses below $5$\Msun and not drawing binaries from their initial birth distributions.
To calculate the \bhnsSingle formation rate, we therefore re-normalize our results to obtain a formation rate of \bhnsSingle mergers for a given metallicity $\Zi$ per unit star forming mass, i.e., $\diff \Nform / \diff \MSFR$, and calculate the formation rate for \bhnsSingle mergers with a given delay time and final compact object masses, \monef and \mtwof,  in {\COMPAS} with
\begin{align}
&\rate_{\rm{form}}(\Zi, \tdelay, \monef, \mtwof) = \notag \\
& \frac{\diff^4 \Nform}{\diff \MSFR \diff \tdelay \diff \monef \diff \mtwof} (\Zi, \tdelay, \monef, \mtwof). 
\label{eq:formation-rate-COMPAS}
\end{align}

We calculate this re-normalized formation rate by incorporating the {\sc{STROOPWAFEL}} weights and assuming a fixed binary fraction of $f_{\rm{bin}} = 1$, which  is  consistent with the observed intrinsic binary fraction for O-stars of $\sim$0.6--0.7 \citep{2011IAUS..272..474S, 2015A&A...580A..93D, 2017A&A...598A..84A, 2017IAUS..329..110S, 2012Sci...337..444S}, when extrapolating for the wider separation range used in this study compared with the observational surveys (cf. \citealt{2015ApJ...814...58D}). Changing  $f_{\rm{bin}}$ to 0.7 did not substantially impact our results.

\subsection{Calculating the cosmological \bhnsSingle merger rate}
\label{subsec:method-MSSFR}
To make predictions for the  \acp{GW} that can be detected with the LVK network today, it is important to consider \bhnsSingle that  formed across a large range of metallicities and redshifts in our Universe. 
This is because  \bhnsSingle systems form from their initial stars in several million years, but their inspiral times can span many Gyr \citep[e.g.][]{1994MNRAS.268..871T,   2002ApJ...572..407B, 2016A&A...589A..64M}. \acp{GW} can, therefore,  originate from binaries with long inspiral times that were formed at high redshifts as well as binaries with shorter inspiral times formed at lower redshifts.  This is especially the case for \ac{GW} observations with the ground-based LVK network that has observation horizons beyond $>100$\Mpc  for \ac{DCO} mergers \citep{2018LRR....21....3A}. 
Moreover, the merger rate density of \bhnsSingle, and more generally \acp{DCO}, can be particularly sensitive to metallicity, which impacts mass loss through stellar winds,  the stellar radii (and radial expansion)  and  thereby the outcome of the (binary) evolution \citep[e.g.][]{1992A&A...264..105M, 2016MNRAS.462.3302E,2018A&A...619A..77K,2018MNRAS.480.2704L,2019MNRAS.482.5012C}. As a result,  \ac{DCO} systems  form sometimes much more efficiently at low metallicities (\Zi$\lesssim 0.1$\Zsun), which  increases the importance of \ac{DCO} formation at high redshifts, where low metallicity stars are more abundantly formed,  when considering the \acp{GW} that can be detected today. It is therefore important to take into account the star formation rate density at different metallicities over the history of our Universe when making predictions for \ac{GW} observations \citep{2019MNRAS.482.5012C}. 

To calculate the \bhnsSingle merger rate density that can be detected with \acp{GW} today we use the method from   \citet{2019MNRAS.490.3740N};  we first integrate the formation rate density from Equation~\ref{eq:formation-rate-COMPAS} over  metallicity and use  the relation $\tform = \tmerger - \tdelay$ (Section~\ref{subsec:method-BPS-different-timescales-binary}) to obtain the merger rate for a binary with masses \monef, \mtwof at any given merger time, \tmerger, using

\begin{figure*}
\includegraphics[width=1\textwidth, trim=0 12cm 0 2.5cm, clip]{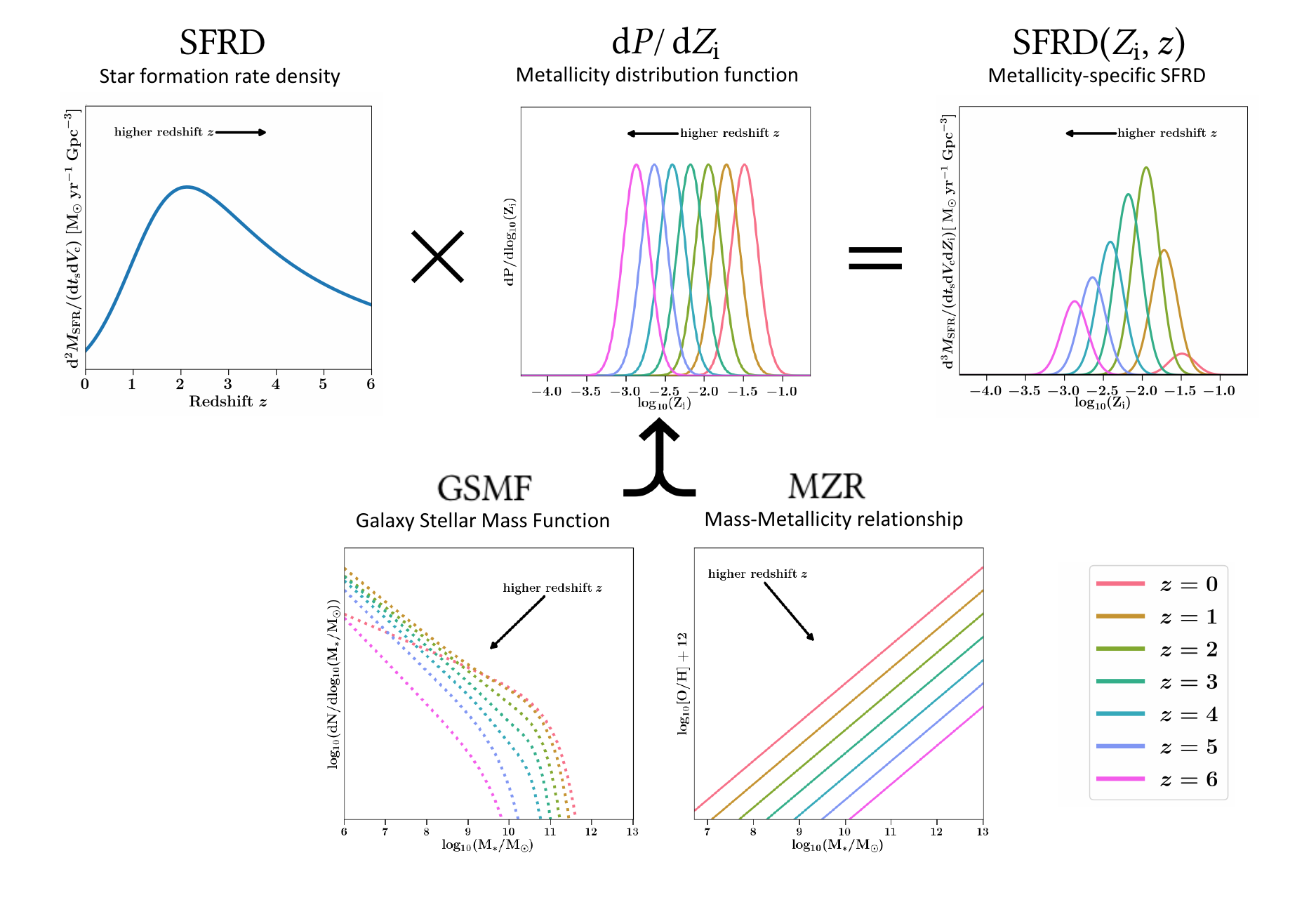}
   \caption{Schematic depiction of how our model for the metallicity specific star formation rate density, \ac{SFRD}{\ensuremath{(Z_{\rm{i}},z)}\xspace},  is created from multiplying a star formation rate density (SFRD) with a metallicity probability distribution function, $\diff P / \diff \Zi$. 
   The metallicity distribution function is constructed by convolving a galaxy stellar mass function with a mass-metallicity relationship. An exception is our fiducial \SFRD model, which uses the `preferred' phenomenological model from \citet{2019MNRAS.490.3740N} for the metallicity distribution function. 
   The arrows in each sub-figure indicate in which direction the distribution moves as redshift increases. }
    \label{fig:MSSFR-sketch}
\end{figure*}
\begin{align}
&\rate_{\rm{m}} (\tmerger,\monef, \mtwof) = \frac{\diff^4 \Nmerger }{\diff \ts \diff \Vc  \diff \monef \diff \mtwof} (\tmerger,\monef, \mtwof)  \notag \\
		&= \int \diff \Zi  \int_0^{\tmerger} \diff \tdelay \, {\rm{SFRD}}(\Zi, \tform=\tmerger-\tdelay) \, \times \notag \\ 
		&\, \hspace{3cm} \rate_{\rm{form}}(\Zi, \tdelay, \monef, \mtwof),  
\label{eq:MSSFR-merger-rate}
\end{align}
where \ts is the time in the source frame of the merger, \Vc is the comoving volume, $\rate_{\rm{form}}$ is obtained using {\sc{COMPAS}} (Section \ref{subsec:method-BPS-merger-rate-per-M-SFR}) and ${\rm{SFRD}}(\Zi, \tform) = {\rm{SFRD}}(\Zi, z(\tform))$.  We obtain the \ac{SFRD}{\ensuremath{(Z_{\rm{i}},z)}\xspace} by 
multiplying a \ac{SFRD}   with a metallicity probability density function
\begin{align}
	&{\rm{SFRD}}(\Zi, z_{\rm{form}}) = 
	\frac{\diff^3 \MSFR}{\diff \ts \diff \Vc \diff \Zi}(z_{\rm{form}})  
	 \notag \\
	& = \underbrace{\frac{\diff^2 \MSFR}{\diff \ts \diff \Vc }(z_{\rm{form}})}_\text{SFRD} 
	 \times 
	 \underbrace{\frac{\diff P }{\diff \Zi}(z_{\rm{form}})}_\text{GSMF $+$ MZR}, 
	 \label{eq:MSSFR-equation}
\end{align}
where we wrote down the equations in  $z$, and used the short hand notation $z_{\rm{form}} = z(\tform)$.  We use for the metallicity density function, ${\diff P }/{\diff \Zi}$, either a direct analytical formula (e.g. model $\mu=000$), or, in most cases, a convolution between a \ac{GSMF} and  \ac{MZR}.  This is discussed in more detail  below and schematically shown in Figure~\ref{fig:MSSFR-sketch}. Throughout our analysis we use the cosmology parameters from the WMAP9 study \citep{2013ApJS..208...19H}\footnote{Obtained from the astropy cosmology module, which has  $\Omega_{\rm{m}}=0.287, \Omega_{\Lambda}^0 = 0.713$ and assumes the flat Lambda-CDM model.}.

This merger rate density, $\rate_{\rm{m}}$, is then converted to a local detection rate by  integrating over the co-moving volume and taking into account the probability, \Pdet,  of detecting a  \ac{GW} source (Section~\ref{subsec:detection-probability}) with 
\begin{align}
\label{eq:rate_detector}
	&\rate_{\rm{det}}(\tdet, \monef, \mtwof) = 
	\frac{\diff^3 \Ndet}{\diff \tdet  \diff \monef \diff \mtwof} 
	\notag  \\
	&= \int 
	\diff \Vc  \,
	 \frac{\diff \ts}{\diff \tdet}  \,  
	 \rate_{\rm{m}}(\tmerger,\monef, \mtwof) \, 
	  \Pdet (\monef, \mtwof, \DL(z)),
\end{align}
where \tdet is the time in the detector (i.e., the observer) frame and \DL is the luminosity distance. See Appendix~\ref{sec:app-calculating-merger-distributionsMSSFR} for more details about the conversion to \tdet, $z$ and \DL. In practise we often marginalize in the remaining sections over the masses and redshifts (or equivalent, delay or merger time) to obtain an overall rate in this Equation as well as Equation~\ref{eq:formation-rate-COMPAS}. We also calculate the detector rate for where $\tdet$ is the current age of our Universe.

In practice, the integral in Equation~\ref{eq:rate_detector} is estimated using  a Riemann sum  over redshift, metallicities and delay time bins, given in Equation~\ref{eq:full-equation-detectable-rate}. This method is similar to previous work including   \citet{2013ApJ...779...72D, 2015ApJ...806..263D,2016ApJ...819..108B,2016MNRAS.458.2634M, 2018MNRAS.477.4685B, 2019MNRAS.482..870E, 2019PhRvD.100f4060B,2020A&A...635A..97B} and \citet{2019MNRAS.482.5012C}. 
Details of our method  are given in \citet{2019MNRAS.490.3740N} and \CMP.

\begin{table*}
\resizebox{\textwidth}{!}{%
\centering

\begin{tabular}{llll}
\hline
\hline
{xyz} index &  \ac{SFRD}  [x]                   & \ac{GSMF}  [y]     & \ac{MZR}  [z]                      \\ \hline \hline
000 (fiducial)   				& \multicolumn{3}{c}{ phenomenological model   \citet{2019MNRAS.490.3740N} }          \\ 
\hline
1              			       	&  {\citet{2014ARA&A..52..415M}} &  {\citet{2004MNRAS.355..764P}} & \citet{2006ApJ...638L..63L}   \\
2              					&  \citet{2004ApJ...613..200S}							& \citet{2015MNRAS.450.4486F} single Schechter   & \citet{2006ApJ...638L..63L} $+$ offset    \\
3              			      	& \citet{2017ApJ...840...39M}     		& \citet{2015MNRAS.450.4486F} double Schechter          &  \citet{2016MNRAS.456.2140M}             \\ \hline
\end{tabular}%
}
\caption{List of the assumptions for the metallicity-specific star formation rate, \SFRD, models  that we explore in this study. 
All  \SFRD models except  our fiducial model are obtained by combining a star formation rate density (SFRD) with a galaxy stellar mass function (GSMF) and mass-metallicity relation (MZR). See  Sections~\ref{subsec:method-MSSFR} and \ref{subsec:MSSFR-variations} for more details. The models are named in the convention $\rm{xyz}$ with $\rm{x, y, z} \in [1,2,3]$ the index numbers for the models used for the  \ac{SFRD}, \ac{GSMF} and \ac{MZR}, respectively.  For example, the combination of using the {\citet{2014ARA&A..52..415M}} \ac{SFRD}  with the  \citet{2004MNRAS.355..764P} \ac{GSMF} and \citet{2016MNRAS.456.2140M}  \ac{MZR}  is labeled ${113}$. The fiducial (phenomenological) model is not a specific combination but a parameterized model that is built to be flexible and is fitted to match the \ac{GW} detected \ac{BHBH} rate and chirp mass distribution from the first two runs of LIGO and Virgo \citep{2019MNRAS.490.3740N}. The simulations using the phenomenological \SFRD model have the label ${000}$ }
\label{tab:MSSFR-variations-labels}
\end{table*}

\subsection{Metallicity-specific star formation rate density  prescriptions}
\label{subsec:MSSFR-variations}
We explore the impact on the \bhnsSingle merger rate and population characteristics from  a total of \NmodelsMSSFR different \ac{SFRD}{\ensuremath{(Z_{\rm{i}},z)}\xspace} models. All our \NmodelsMSSFR \ac{SFRD}{\ensuremath{(Z_{\rm{i}},z)}\xspace} models are constructed by combining analytical, simplified, prescriptions for the star formation rate density and metallicity distribution function  (Equation~\ref{eq:MSSFR-equation}). A schematic depiction is shown in Figure~\ref{fig:MSSFR-sketch}.  Although all models (drastically) simplify the complex behavior of the \ac{SFRD}{\ensuremath{(Z_{\rm{i}},z)}\xspace} as a function of redshift and time,  many of the prescriptions explored in this work are widely used in population synthesis predictions for \ac{DCO} mergers. Our aim in using these models is foremost to explore and study the impact of these uncertainties in current state-of-the-art population synthesis predictions for \bhnsSingle mergers.  In Section~\ref{subsec:disc-MSSFR-assumptions} we discuss in more detail the main limitations and future prospects to this modelling.

Our fiducial \ac{SFRD}{\ensuremath{(Z_{\rm{i}},z)}\xspace} model,  $\rm{xyz} = {{000}}$ (Table~\ref{tab:MSSFR-variations-labels}),  is the  phenomenological  model described in \citet{2019MNRAS.490.3740N} (refered to as the `preferred' model). This  model uses a phenomenological analytical model for the \ac{SFRD}  and the metallicity probability distribution function. The latter is directly constructed using a log-normal metallicity density distribution with a redshift independent standard deviation 
  $\sigma=0.39$  and  redshift dependent  mean $\mu(z)$, which follows the work of  \citet{2006ApJ...638L..63L}.  The values of the free parameters in the phenomenological model are found by combining the  \SFRD prescription with a population synthesis outcome to find the  best fit to the \ac{BHBH} \ac{GW} observations announced in the first two observing runs of LIGO and Virgo   \citep{2019MNRAS.490.3740N}. This work is also done with  {\sc{COMPAS}} using stellar-evolution assumptions that are similar to the ones assumed in this study. We therefore expect the merger rates based on this \ac{SFRD}{\ensuremath{(Z_{\rm{i}},z)}\xspace} choice and our fiducial simulations to be representative for the \ac{GW} observations.  The \ac{SFRD} and metallicity probability distribution function for this prescription are shown in Figure \ref{fig:MSSFR-SFRs} and \ref{fig:MSSFR-Z-PDFs}, respectively.

All our other 27 \ac{SFRD}{\ensuremath{(Z_{\rm{i}},z)}\xspace} models, on the other hand, use one of the commonly used \acp{SFRD}  in combination with a probability density function that is created by combining a \ac{GSMF} with a \ac{MZR}, these are described below\footnote{During this convolution we assume that the SFRD is spread equally among all galaxy stellar mass. Such that a galaxy with twice the amount of mass, has twice the SFRD.}.    The main difference of these models compared to the preferred \citet{2019MNRAS.490.3740N} model is that the former does not assumes the metallicity probability distributions are symmetric in log-metallicity (whilst the preferred model from \citealt{2019MNRAS.490.3740N} does), as can be seen in Figure~\ref{fig:MSSFR-Z-PDFs}. Observational evidence suggests this symmetric behavior is likely not the case \citep[e.g.][]{2006ApJ...638L..63L,2019MNRAS.482.5012C,2020arXiv201202800B}.   Two  combinations of the 27 \SFRD models are shown in Figure~\ref{fig:MSSFR-Z-PDFs}.  By considering all possible combinations of  the three \ac{SFRD},  three \ac{GSMF} and three \ac{MZR} prescriptions, we end up with a total of 27 \ac{SFRD}{\ensuremath{(Z_{\rm{i}},z)}\xspace} models in addition to our fiducial \ac{SFRD}{\ensuremath{(Z_{\rm{i}},z)}\xspace} model based on \citet{2019MNRAS.490.3740N}, resulting in a total of  \NmodelsMSSFR models.

\subsubsection{Star formation rate density \ac{SFRD}}
 The \acp{SFRD} prescriptions are shown in Figure~\ref{fig:MSSFR-SFRs}.
Besides using the phenomenological \ac{SFRD} from \citet{2019MNRAS.490.3740N} for model 000, we follow \citet{2019MNRAS.490.3740N} and vary between three typically used \ac{SFRD} prescriptions described in  Table~\ref{tab:MSSFR-variations-labels}.
First, we use the \ac{SFRD} from  \citet[][Equation~15]{2014ARA&A..52..415M}, which has a slightly earlier peak compared to the other \acp{SFRD} used in this work. Population synthesis studies of \ac{DCO} merger rates that use this \ac{SFRD} prescription include  \citet{2016Natur.534..512B,2018MNRAS.474.2937C, 2019PhRvD.100f4060B} and \citet{2019MNRAS.482..870E}. 
Secondly, we use the \citet{2004ApJ...613..200S} prescription, which  assumes a higher extinction correction, resulting in a higher \ac{SFRD}, particularly at higher redshifts. Population synthesis studies of \ac{DCO} merger rates that use this \ac{SFRD} prescription include  the work by \citet{2013ApJ...779...72D,2015A&A...574A..58K,2017MNRAS.471.4702B, 2018MNRAS.474.4997C} and \citet{2018MNRAS.481.1908K}.
Last, we also use the \ac{SFRD} from \citet[][Equation~1]{2017ApJ...840...39M}, which is an updated version of \citet[][]{2014ARA&A..52..415M} that uses the broken power law initial mass function from \citet{2001MNRAS.322..231K} and  better fits some of the observations between redshifts $4 \lesssim z \lesssim 10$.  \ac{DCO} merger rate studies that use this \ac{SFRD} assumption include \citet{2020A&A...636A.104B, 2020arXiv200906655D, 2020arXiv201103564W} and \citet{2020arXiv201110057Z}.

\subsubsection{Galaxy stellar mass function \ac{GSMF}}
Following \citet{2019MNRAS.490.3740N}, we vary between three different prescriptions for the \ac{GSMF}, which are shown in Figure~\ref{fig:MSSFR-GSMFs}.  Typically,  the \ac{GSMF} is described with a single or double  \citet{1976ApJ...203..297S} function. First, we use a redshift independent single Schechter \ac{GSMF} model as given by \citet{2004MNRAS.355..764P}. This \ac{GSMF} prescription is used to create a metallicity distribution function by  \citet[][]{2006ApJ...638L..63L}, which is used by studies including \citet{2018MNRAS.477.4685B} and \citet{2017A&A...604A..55M}. Second and third we use the redshift dependent single and double Schechter functions based on results from  \citet{2015MNRAS.450.4486F}, respectively. We use the fits by \citet{2019MNRAS.490.3740N} to the tabulated values from  \citet[][table A1]{2015MNRAS.450.4486F}, that extrapolates to a full redshift range of $z\in[0,6.5]$. See  \citet{2019MNRAS.490.3740N} for more details.

\subsubsection{Mass-metallicity relation {MZR}}
We follow \citet{2019MNRAS.490.3740N} for exploring three different \ac{MZR} prescriptions shown in Figure~\ref{fig:MSSFR-MZRs}. The metallicity in this figure is shown as the number density of oxygen over that of hydrogen, which is typically observed. The \ac{MZR} describes the average relation between the typical metallicities found for star forming galaxies at a given redshift.  We use the solar values of $\Zsun = 0.0142$ and $\log_{10}[\rm{O}/\rm{H}] +12 = 8.69$ \citep{2009ARA&A..47..481A} to convert to mass fraction metallicities. For our first two \ac{MZR} prescriptions we use the approximate \ac{MZR} relation that \citet{2006ApJ...638L..63L} construct based on observations from \citet{2005ApJ...635..260S}, which is given by $M_{*}/M_{x} = (\Zi/Z_{\odot})^2$, with $M_x $ as given by \citet{2006ApJ...638L..63L} and $M_{*}$ the galaxy stellar mass. We assume the average metallicity scales with redshift as $\langle \Zi \rangle = Z_{\odot} 10^{-0.3z}$. Following  \citet{2019MNRAS.490.3740N} we create a second prescription based on these two relations, by adding an offset to better match the quadratic fit given in \citet{2005ApJ...635..260S}. As third \ac{MZR} model we use the \ac{MZR} relation given by \citet{2016MNRAS.456.2140M}. See \citet{2019MNRAS.490.3740N} for more details. We do not take into account the observed scatter around the \ac{MZR}, see for more details \citep{2019MNRAS.488.5300C}.

\subsection{Detection probability of a source}
\label{subsec:detection-probability}

Whether a \bhnsSingle merger is detectable by a \ac{GW} interferometer network depends on its distance, orientation, inclination and source characteristics (such as component masses \monef, \mtwof).  
The detectability is described by a source signal-to-noise ratio (SNR). 
We follow the method from \citet{2018MNRAS.477.4685B} to calculate the probability of detecting \ac{GW} sources. We assume a SNR  threshold of $8$ for a single detector \citep{1993PhRvD..47.2198F} as a proxy for detectability by the network. 
The  SNR of the \bhnsSingle mergers are calculated by computing the source waveforms  using a combination of the LAL suite software packages {\sc{IMRPhenomPv2}} \citep{2014PhRvL.113o1101H,2016PhRvD..93d4006H,2016PhRvD..93d4007K} and {\sc{SEOBNRv3}} \citep{2014PhRvD..89h4006P,2017PhRvD..95b4010B}\footnote{See also \citet{lalsuite}.}.  
We marginalize over the sky localization and source orientation of the binary using the antenna pattern function from \citet{1993PhRvD..47.2198F}. The detector sensitivity is assumed to be equal to advanced LIGO in its design configuration \citep{2015CQGra..32g4001L,2016LRR....19....1A, 2018LRR....21....3A}, which is equal to that of a ground-based \ac{GW} detector network  composed of Advanced LIGO, Advanced Virgo and KAGRA (LVK). 
We ignore the effect of the \ac{BH} spin orientation and magnitude on the detectability of \acp{GW}, which is expected to possibly increase  detection rates within a factor 1.5 \citep{2018PhRvD..98h4036G} as binaries with (high) aligned spins are predicted to have larger  horizon distances \citep{2006PhRvD..74d1501C,2015CQGra..32j5009S}.

\subsection{Tidally disrupted BHNS}
\label{subsec:method-tidal-disruption-BHNS}
Simulations show that during a \bhnsSingle merger the \ac{NS} is either tidally disrupted outside of the \ac{BH} innermost stable circular orbit  or instead plunges in,  depending on the mass ratio, \ac{BH} spin and  \ac{NS} equation of state  \citep{2011ApJ...727...95P, 2012PhRvD..86l4007F,2018PhRvD..98h1501F}. If the \ac{NS} is disrupted, part of the disrupted material can form a disk and can eventually power  electromagnetic counterparts such as short gamma-ray bursts and kilonovae \citep{2019MNRAS.486.5289B, 2020EPJA...56....8B,2020arXiv201102717Z}. We estimate the ejected mass  during a \bhnsSingle merger  using Equation~4  from \citet[][]{2018PhRvD..98h1501F} who present a simple formula for the merger outcome,  post-merger remnant mass and ejecta mass based on numerical relativity simulations.  We define a \bhnsSingle merger to `disrupt' the \ac{NS} if the calculated ejecta mass is nonzero. By doing so, we can calculate the fraction of \bhnsSingle mergers that disrupt the \ac{NS} outside of the \ac{BH} innermost-stable orbit, which are interesting candidates for observing an electromagnetic counterpart to their \ac{GW} detection.   Detecting \bhnsSingle mergers  with  electromagnetic counterparts is a golden grail in astronomy as it would confirm the origin of such transients and enables, e.g. measurements of the \ac{NS} equation of state and \bhnsSingle system. There has been a big effort in finding such a counterpart, such as to the possible \bhnsSingle merger GW190814, but so far without a detection \citep[e.g.][]{2019ApJ...884L..55G, 2019ApJ...887L..13D, 2020arXiv200201950A}. 

As the \ac{NS} equation of state is unknown we explore two variations for the \ac{NS} radius: we assume $\Rns=11.5$\km or $\Rns=13$\km  consistent with the APR equation of state \citep{1998PhRvC..58.1804A} and  \ac{GW} observations \citep{2018PhRvL.121p1101A}, and the  NICER observations \citep[e.g.][]{2019ApJ...887L..24M}, respectively.

The spins of the \acp{BH} in \bhnsSingle  mergers, \chibh, are also unknown \citep[e.g.][]{2015PhR...548....1M}. For the  \ac{BH} spin we explore  three models. First, we  assume all black holes to have zero spin, $\chibh=0$. Second,  we assume all \acp{BH} to have half the maximum spin value, $\chibh=0.5$, which explores a scenario where \acp{BH} in \bhnsSingle have more moderate spins. Last, we explore an ad hoc, but physically motivated spin model where we assign spins based on the study by \citet{2018A&A...616A..28Q}.  Here it is assumed that all first formed \acp{BH} in \bhnsSingle binaries have zero spin as a consequence of efficient angular momentum transport \citep{2015ApJ...800...17F,2018A&A...616A..28Q,2019ApJ...881L...1F,2020A&A...636A.104B}.   The helium star progenitors of \acp{BH} that form second in the binary, however,  can spin up through tidal interactions if they are in a close orbit with their companion, leading to \acp{BH} with significant spins \citep[cf.][]{2007Ap&SS.311..177V,2016MNRAS.462..844K,2018A&A...616A..28Q,2020A&A...635A..97B,2020ApJ...895L..28M}. For these NS--BH binaries, we use an approximate prescription to determine the \ac{BH} spin:
\begin{equation}
    \chi_{\mathrm{BH}} = 
                    \begin{cases}
                            0 & \text{for } \log_{10}(P) >  0.3 \\
                            1 & \text{for } \log_{10}(P) < -0.3 \\
                            -5/3 \log_{10}(P) + 0.5 & \text{for } -0.3 < \log_{10}(P) < 0.3,
                    \end{cases}
    \label{eq:aBHfitQin}
\end{equation}    
 where $P$ is the orbital period of the binary in days right before the second \ac{SN}. We follow with this the prescription in  \citet{2020arXiv201113503C}, who create this ad hoc fit from the top middle panel of Figure~6 in \citet{2018A&A...616A..28Q}, which is based on a simulation for solar metallicity.
 Although in reality the spin distribution is more complicated and metallicity dependent, we use this single prescription for simplification. We expect that this does not impact drastically our results as the variation over metallicity are minor compared to the overall behavior of most \acp{BH} having zero spin, and only close NS--BH binaries having high spins. Moreover, we show in Section~\ref{subsec:fraction-BHNS-with-EM-ejecta} that the amount of systems with ejecta is in this prescription dominated by the number of NS--BH binaries, and as this fraction is low, that this prescription is most similar to the assumption where all \chibh are zero.

We assume the \ac{BH} spin to be aligned with the orbit and not vary  from the moment the \bhnsSingle has formed. For each spin model, we vary the two \ac{NS} radii assumptions, resulting in six different combinations of \ac{BH} spin and \ac{NS} radius. 
 See \citet{2019PhRvL.123d1102Z} and \citet{2020arXiv201102717Z} for a further discussion on the effect of different equations of state and  \ac{BH} spins on \bhnsSingle ejecta.

\subsection{Statistical sampling uncertainty}
\label{sec:statistical-uncertainty}
Each of our simulations yield a finite number of \bhnsSingle mergers, as quoted in the last column of Table~\ref{tab:variations-BPS}, which results in a statistical sampling uncertainty.  We calculated this uncertainty on the \bhnsSingle formation rate (Equation~\ref{eq:formation-rate-COMPAS}) using Equation~15 from \citet{2019MNRAS.490.5228B} and found  that this statistical uncertainty is at most $0.06\%$, less than a tenth of a percent. This is negligible compared to the systematic uncertainties from our assumptions in modelling of the massive binary evolution and the \SFRD\footnote{And from our usage of a fixed, finite grid of birth metallicity points \Zi and redshift grid points for the cosmic integration, instead of  (more) continuous distributions.}. We therefore decide to not quote these statistical sampling uncertainties throughout the remaining of the paper, and instead focus on the uncertainty from stellar evolution and \SFRD variations. An interested reader can find more details on the statistical uncertainties in the related script in  \url{https://github.com/FloorBroekgaarden/BlackHole-NeutronStar} and in \citet{2019MNRAS.490.5228B} and references therein.


\section{Fiducial model}
\label{sec:results-fiducial}
\begin{figure*}
\includegraphics[width=1\textwidth]{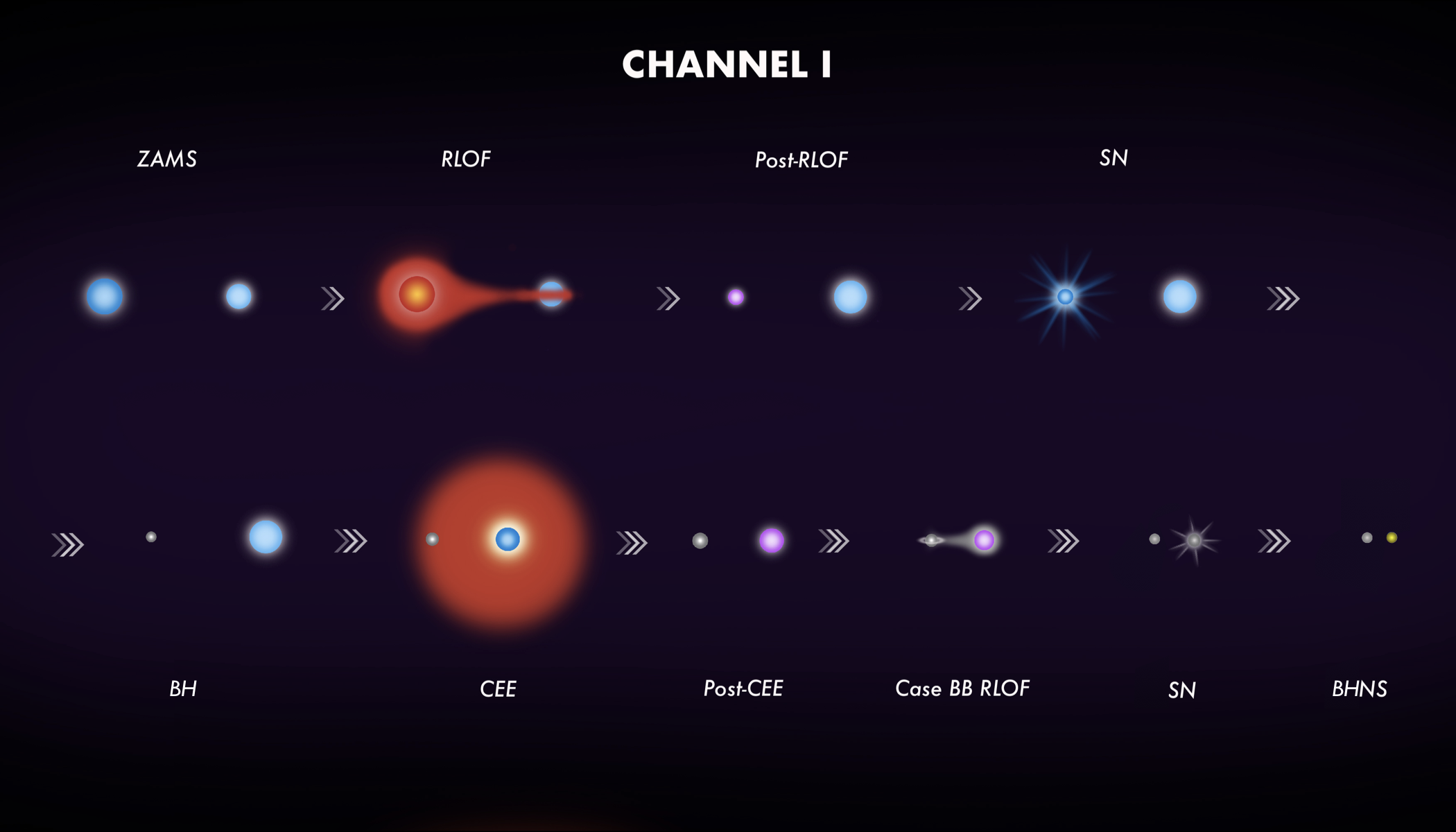}
   \caption{
   Schematic depiction of the classic formation channel as described in  Section~\ref{subsubsec:classic-channel}. The acronyms stand for: zero-age main sequence (ZAMS), Roche lobe overflow (RLOF), supernova (SN), black hole (BH), common-envelope episode (CEE), and black hole-neutron star (BHNS).    The figure is based  on Figure~1 from \citet{2020PASA...37...38V} by T. Rebagliato, which is publicly available at \url{https://zenodo.org/record/3634498\#.XnS9ZC2ZNQI}. Adjustments to the original image were made by S. Vinciguerra. 
   }
\label{fig:formation-channels-sketch}
\end{figure*}

In this section we describe the results of our population synthesis simulation for model \mAzero, which uses both our fiducial assumptions for the massive (binary) star model (described in Section~\ref{subsec:method-BPS-assumptions} and listed in Table~\ref{tab:population-synthesis-settings}) and our fiducial assumptions for the cosmic star formation history model  (described in  Section~\ref{subsec:MSSFR-variations} and listed in Table~\ref{tab:MSSFR-variations-labels}). 
We focus  on this model  to provide insight to a typical output of a \COMPAS simulation. 
This model choice is similar to earlier work done with {\COMPAS} and has been chosen as it, for example, matches  the population of  galactic \ac{NSNS} systems \citep{2018MNRAS.481.4009V, 2020MNRAS.494.1587C} and the rate and chirp mass distributions of the \ac{GW} sources in the first two observing runs of LIGO and Virgo \citep{2019MNRAS.490.3740N}.  In Section~\ref{sec:results-variations} we describe the results for all \Nmodels model variations.

\subsection{Formation channels}
\label{subsec:bhns-BPS-formationChannels}

We identify four main groups of formation channels  described below. 
  The percentages, quoted after each section header, indicate the fraction that each channel contributes to the total number of detected \bhnsSingle mergers. This takes into account the \SFRD weighting using our fiducial model ($\rm{xyz} = 000$) and the detection probability of a \ac{GW} network equivalent to LVK at design sensitivity. These  percentages are calculated using Equation~\ref{eq:rate_detector}, whilst marginalizing over the \bhnsSingle masses.
 The percentage that each formation channel contributes to the detectable merger rate is impacted by variations in the population synthesis and \SFRD models. We present results for our \Nmodels models in Section~\ref{subsec:variations-formation-channels-GWs}.

\subsubsection{(I) Classic channel   \PercentageClassicLVK}
\label{subsubsec:classic-channel}
We find, in agreement with, e.g. \citet[][]{2019MNRAS.490.3740N} for \ac{BHBH} mergers,  that the majority of the binaries  form a \bhnsSingle through the `classic'  formation channel where the binary experiences both a stable mass transfer and an unstable (CE) mass transfer phase. This classic channel is discussed in e.g. \citet[][]{1991PhR...203....1B,1973A&A....25..387V,2006csxs.book..623T} and \citet{2008ApJS..174..223B}, see also \citealt{2018arXiv180605820M} and references therein. We schematically depict this formation channel in Figure~\ref{fig:formation-channels-sketch} and  describe it below in more detail for binaries that form a \bhnsSingle merger.

The binaries in the classic channel are born with a wide range of initial separations  of  about $0.3$--$20$\AU  as shown in  Figure~\ref{fig:BHNS_ZAMSmasses}. The initially more massive star (the primary) eventually expands and fills its Roche lobe initiating a stable mass transfer phase (Roche-lobe overflow, RLOF) onto the initially less massive star (the secondary). This happens in this channel either when the primary is a Hertzsprung gap star or is a core helium burning star (both case B mass transfer), where core helium burning donor stars are in initially wider binaries compared to Hertzsprung gap donor stars. 
 In our fiducial model the companion typically accretes a large fraction of   the mass that is lost from the primary star donor. 
 Mass transfer typically ends in our simulations when the donor has lost all of its hydrogen envelope.  The result for case A or B mass transfer is typically a stripped envelope star that is burning helium, which may be observed as a Wolf-Rayet star \citep[e.g.][]{2007ARA&A..45..177C, 2018A&A...615A..78G}.

Eventually, the stripped primary star ends its life in a core-collapse (\ac{SN})  and the first compact object, a \ac{BH} or NS, is formed. 
The binary needs to stay bound during the \ac{SN} to eventually form a \bhnsSingle. Typically, more than 80$\%$ of the binaries disrupt during the first \ac{SN} \citep[e.g.][]{2019A&A...624A..66R}, where disruption depends on the magnitude and orientation of the \ac{SN} kick, the separation of the binary and the amount of ejected mass (e.g.
\citealt{1975A&A....39...61F,1998A&A...330.1047T}).

 The secondary star later evolves off the main sequence, and expands to fill its Roche lobe. This is the start of a reverse mass-transfer phase from the secondary onto the compact object.  However, for this reverse mass-transfer phase the extreme mass ratio contributes to the mass transfer being dynamically unstable, and the start of CE evolution \citep[e.g.][]{1997A&A...327..620S,2010ApJ...717..724G,2015ApJ...812...40G,2018MNRAS.481.4009V}.  During the CE event the separation of the binary decreases as orbital angular momentum and energy are transferred to the CE. If the CE is ejected successfully, the result is a close binary system consisting of a BH or NS and a massive helium star, an example of a possible observed system in this phase is Cyg X-3 \citep{2013ApJ...764...96B,2013MNRAS.429L.104Z}. Otherwise the system results in a merger of the star with the compact object, which can possibly form a  Thorne--{\.Z}ytkow object \citep{1977ApJ...212..832T} or lead to peculiar \acp{SN} \citep[e.g.][]{2012ApJ...752L...2C,2016MNRAS.455.4351P,2020ApJ...892...13S}.

In a subset of the binaries there is a stable case BB  mass transfer phase after the \ac{CE} from the helium star onto the primary compact object \citep[cf.][]{2003MNRAS.344..629D}.  This typically occurs when the secondary is a relatively low mass helium star as they expand to larger radii compared to more massive helium stars in the single star prescriptions of \citet{2000MNRAS.315..543H} implemented in \COMPAS \citep[cf.][]{2003MNRAS.344..629D}. However,   \citet{2020A&A...637A...6L}  point out that these prescriptions might underestimate the expansion of helium stars, especially at metallicities $\Zi \approx 0.001$.  \bhnsSingle systems undergoing case BB mass transfer typically have the shortest semi-major axis in Figure~\ref{fig:BHNS_DCOmasses}. 

The helium star eventually forms a \ac{NS} or BH in the second \ac{SN}.   If the binary remains bound, a \bhnsSingle  binary is formed with masses as in Figure~\ref{fig:BHNS_DCOmasses} that gradually decays in separation due to the emission of \acp{GW}.  If the separation is small enough (typically $\lesssim 10$\Rsun, see Figure~\ref{fig:BHNS_DCOmasses}) the \bhnsSingle will merge within \thubble.  

%
\subsubsection{(II) Only stable mass transfer channel \PercentageOnlyStableMTLVK}

In  about \PercentageOnlyStableMTLVK{}  of all  detectable mergers the binary forms  similar to the classic channel (I) but does not experience an unstable mass transfer phase leading to a \ac{CE}. 
This channel thus only has  stable mass transfer phases    (cf. \citealt[][]{Heuvel:2017sfe, 2019MNRAS.490.3740N}, see also, e.g. \citealt[][]{2017MNRAS.465.2092P}). 
To form  \bhnsSingle systems with semi-major axis $\lesssim 10$\Rsun that can merge in \thubble, these  binaries typically experience a second stable mass transfer phase  from the secondary star onto the compact object after the first  \ac{SN} that decreases the separation of the binary. This formation channel typically leads to \bhnsSingle mergers with final mass ratios $\qf \equiv \mbhf / \mnsf  \approx 3$ as shown in Figure~\ref{fig:BHNS_DCOmasses} and~\ref{fig:BHNS_DCO_observed}.

\subsubsection{(III) Single-core \ac{CE} as first mass transfer channel \PercentageSCCELVK }
%
In the window of initial separations between $\sim 5$--$40$\AU as shown in Figure~\ref{fig:BHNS_ZAMSmasses},  the first mass transfer phase leads to an unstable single-core \ac{CE} event,  with only the donor star having a clear core-envelope structure and the secondary star still being a main sequence star. This is in agreement with other studies on mass transfer stability \citep[see e.g.  Figure~19 and 20 in][]{2015ApJ...805...20S}.  The donor star that initialized the \ac{CE} is typically a core helium burning star or a star on the giant branch. If the \ac{CE} is successfully ejected this leads to a tight binary star system. The primary star will eventually form a \ac{BH} in a \ac{SN}. Eventually the secondary star also evolves off the main sequence, leading to either a stable mass transfer phase or a second unstable \ac{CE} event. The latter occurs for the widest binaries in Figure~\ref{fig:BHNS_ZAMSmasses}. The secondary eventually forms a \ac{NS}. In this formation channel the \ac{BH} always forms first in our simulations. This formation channel typically leads to \bhnsSingle mergers with final \bhnsSingle mass ratios $\qf \approx 3$ as shown in Figure~\ref{fig:BHNS_DCOmasses} and~\ref{fig:BHNS_DCO_observed}.


%
\subsubsection{(IV) Double-core CE  as first mass transfer channel  \PercentageDCCELVK }
\begin{figure*}
\includegraphics[width=\textwidth]{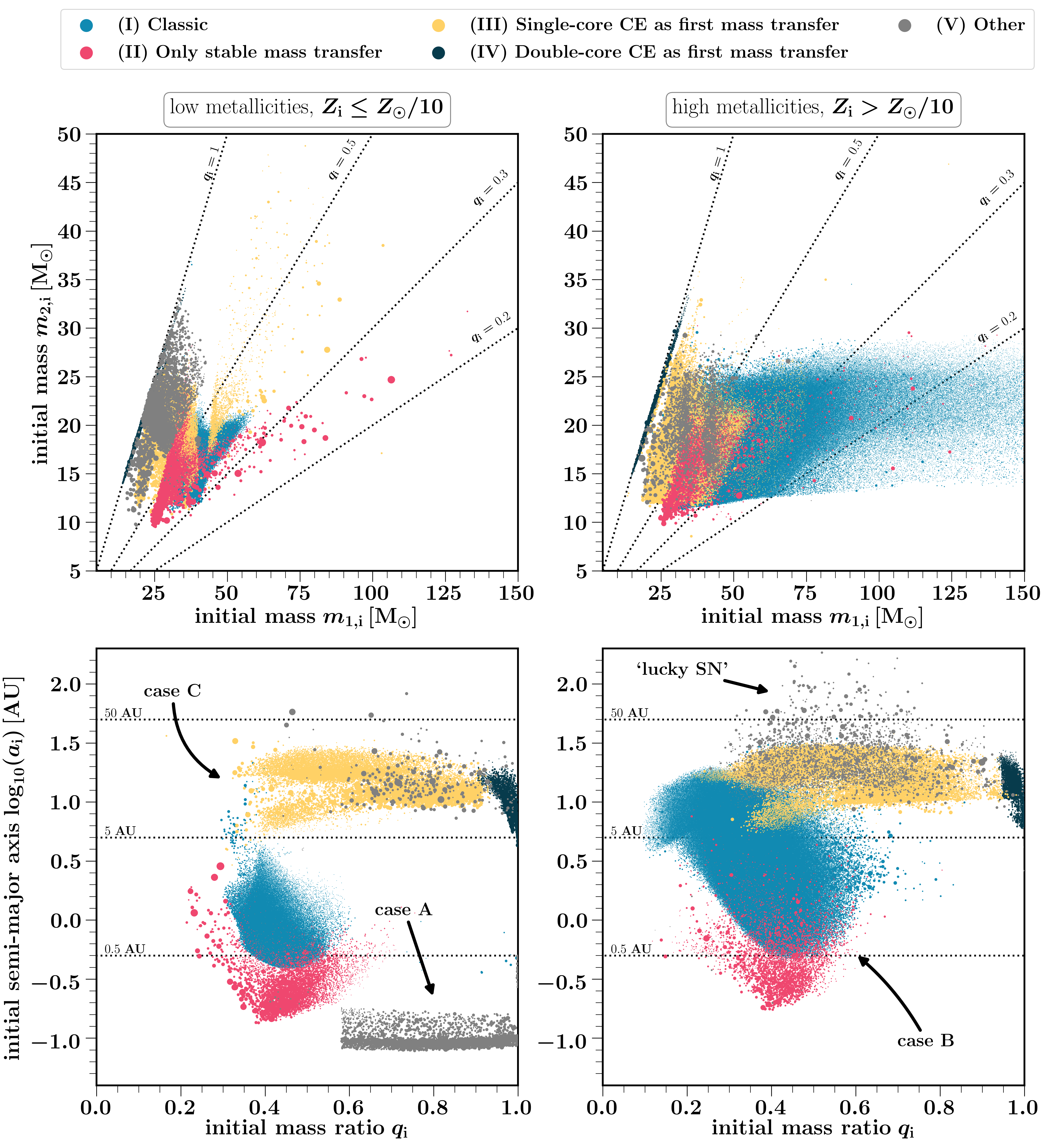}
   \caption{Initial parameters of binaries forming \bhnsSingle systems  that merge in \thubble for  our fiducial model \mAzero  (Section~\ref{sec:method}) for `lower' metallicities $\Zi\leq \Zsun/10$ (left panels) and `higher' metallicities $\Zi>\Zsun/10$ (right panels), where $\Zsun=0.0142$. Colors represent different formation channels, described in Section~\ref{subsec:bhns-BPS-formationChannels}. Each point represents one simulated binary system that leads to a \bhnsSingle merger. The areas of the points represent the statistical weight $w_{\rm{i}}$ (i.e., probability of occurrence)  of that binary in our simulation.  Dotted lines indicate some values of the parameters to guide the reader. We define the initial mass ratio $\qi = \mtwoi / \monei$. Arrows indicate regions where the first mass transfer is case A, B or C, as well as `lucky \ac{SN}' systems that do not experience mass transfer before the first \ac{SN}.  Figures and  videos showing how these distributions change over our model variations are available at \url{https://github.com/FloorBroekgaarden/BlackHole-NeutronStar}.}
    \label{fig:BHNS_ZAMSmasses}
\end{figure*}

In this channel the primary star fills its Roche lobe when both stars are on the giant branch and both stars  have  a core-envelope structure. The mass transfer is  unstable, leading to a double-core CE.  Both stars need to evolve on a similar timescale and, therefore,  have similar initial masses (i.e., $0.9 \lesssim \qi \lesssim 1$) as can be seen in the bottom panel of Figure~\ref{fig:BHNS_ZAMSmasses}. The further evolution of the binary proceeds similar to channel III, except that in most cases there is never a case BB mass transfer phase. A \bhnsSingle system can be formed if the stars have carbon-oxygen core masses close to the boundary between \ac{NS} and \ac{BH} formation in our remnant mass prescription. This is visible in Figure~\ref{fig:BHNS_DCOmasses} where it can be seen that the \bhnsSingle in this channel have \ac{BH} and \ac{NS} masses that are very equal compared to the other formation channels.  This formation channel typically also leads to \bhnsSingle mergers with the smallest final mass ratios $\qf \approx 1.5$ as shown in Figure~\ref{fig:BHNS_DCOmasses} and~\ref{fig:BHNS_DCO_observed}. These low \bhnsSingle masses cause \ac{GW} detectors to be less sensitive to finding binaries from this channel.  This results in \PercentageDCCELVK of the detections being predicted from this channel. Which is different from the expected contribution of this channel to e.g. binary neutron star mergers  \citep{2018MNRAS.481.4009V}
This channel is similar to the one described by \citet{1995ApJ...440..270B,1998ApJ...506..780B} and  later on by, e.g.   \citet{Dewi:2006bx,2011MNRAS.410..984J} and \citet{2018MNRAS.481.4009V}. \\
\subsubsection{ (V) Other channel  \PercentageOtherLVK }
We classify all other \bhnsSingle under the `other' channel. The majority of contributions comes from two formation pathways. First,  a large fraction of the other channel consists of binaries born with low metallicities and initial separations between about $0.01$--$0.20$\AU (gray scatter points in bottom right of the bottom left panel of Figure~\ref{fig:BHNS_DCOmasses}). These binaries undergo mass transfer when the donor star is a main sequence star (case A), which typically results in the secondary star accreting a large amount of mass from the primary. A lot of binaries from this formation pathway form the \ac{NS} first. Second, most of the remaining binaries in the `other' channel form by having a `lucky \ac{SN} natal kick'. The first event in this pathway is the primary star undergoing a \ac{SN}, and in a tiny fraction of those binaries the natal kick has the right magnitude and direction so that the binary stays bound.

%


\subsection{Initial properties leading to \bhnsSingle mergers}
\label{subsec:bhns-BPS-ZAMSm1m2}
The locations  in the initial parameter space of the binaries (e.g.  initial masses, initial mass ratio, and initial semi-major axis)  leading to the formation of a \bhnsSingle system that merges in \thubble are shown in Figure~\ref{fig:BHNS_ZAMSmasses}  for metallicities $\Zi\leq \Zsun/10$ and $\Zi>\Zsun/10$, which represent respectively lower and more solar like  metallicity environments. We assume in our simulations $\Zsun = 0.0142$ \citep{2009ARA&A..47..481A}. We chose the boundary $\Zi=\Zsun/10$ somewhat arbitrarily as it is about half way in our \Zi grid (Figure~\ref{fig:BHNS_rate_per_metallicity}). 

\subsubsection{Initial masses}
As can be seen in the top panels of Figure~\ref{fig:BHNS_ZAMSmasses},  the majority of \bhnsSingle mergers at low $\Zi$ originate from binaries with  masses  $10 \lesssim $ \monei/\Msun $\lesssim 60$ and   $10 \lesssim $\mtwoi/\Msun $\lesssim 30$, whilst at higher metallicities this shifts to  $10 \lesssim $ \monei/\Msun $\lesssim 150$ and $10 \lesssim $\mtwoi/\Msun $\lesssim 30$ respectively. A  difference between the two panels is that, at higher metallicities, there are \bhnsSingle systems formed from binaries with initial mass ratios $\lesssim 0.3$, whereas these mostly lack at lower metallicities,   as can also be seen in the lower panels of Figure~\ref{fig:BHNS_ZAMSmasses}. This is because higher metallicities correspond to stronger line-driven stellar winds, leading to more mass loss, which equalizes the more extreme mass ratio before the onset of mass transfer, making the mass transfer more stable and the system more likely to survive to form a \ac{GW} progenitor \citep[cf.][]{2010ApJ...715L.138B,2018MNRAS.480.2011G,2019MNRAS.490.3740N}.  At lower metallicities, there is fewer mass loss and so the mass ratio stays more extreme at the moment of mass transfer, making it often unstable and the stars merge. 
Moreover, on average the total initial mass of binaries is higher at higher metallicities, as at those metallicities stellar winds strip more mass from the system  compared to lower metallicities \citep[][]{2010ApJ...715L.138B}. This stripping leads to lower mass carbon-oxygen cores compared to those of stars born with the same masses at lower metallicities. So where at higher metallicities \bhnsSingle form, the same systems may form \ac{BHBH} binaries at lower metallicities in our simulations.

\subsubsection{Initial semi-major axis and mass ratio}
The initial semi-major axis of the binaries forming \bhnsSingle mergers in Figure~\ref{fig:BHNS_ZAMSmasses} spans the range of about $0.1 \lesssim \ai   \lesssim 50$\AU  with higher metallicities favoring slightly larger \ai.  
The latter seems counter-intuitive since stellar winds typically widen the binary, but comes from  subtle and indirect effects of the wind loss being stronger at higher metallicities.
As discussed above, the binaries at higher metallicities originate from initially more massive primary stars in the range $25$--$150$\Msun.  The radii of these stars typically expand more during the Hertzsprung-gap phase  in the stellar evolution tracks implemented in {\sc{COMPAS}} \citep{2000MNRAS.315..543H}. 
Since this expansion, in combination with the initial separation, determines when mass transfer happens, the binary needs to have a much larger \ai at higher metallicities to form through the same channel as a binary  at lower metallicities.
In addition,  the secondary star is typically more massive at the onset of a \ac{CE} for higher metallicity binaries, causing the binary to shrink more compared to lower metallicity binaries \citep[cf.][]{2019MNRAS.490.3740N}.

\begin{figure}
    \centering
    {\includegraphics[width=1.0\columnwidth]{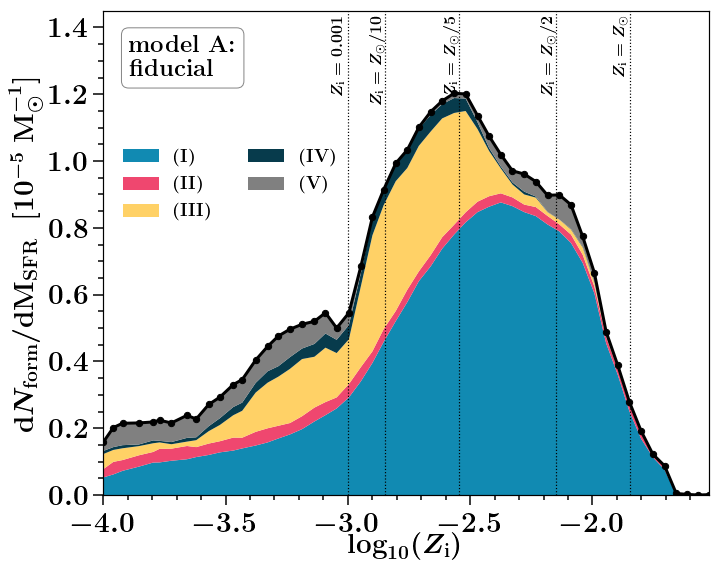} }%
   \caption{Number of \bhnsSingle mergers with $\tdelay \leq \thubble$  that form per unit star forming mass  (\MSFR) as a function of initial metallicity $\Zi$ for our fiducial model. This rate is obtained from Equation~\ref{eq:formation-rate-COMPAS} by marginalizing over \tdelay, \monef and \mtwof. The colors represent the contribution from each formation channel color coded as in  Figure~\ref{fig:BHNS_ZAMSmasses} and described in Section~\ref{subsec:bhns-BPS-formationChannels}. The main formation channels are the (I) `classic' channel (blue) and the (III) `single-core CE as first mass transfer channel' (yellow). The 53 black scatter points denote the simulated \Zi grid points. Dotted lines indicate some \Zi values to guide the reader.
   }
  \label{fig:BHNS_rate_per_metallicity}
\end{figure}

 \subsection{Yield of \bhnsSingle mergers as a function of birth metallicity}

 Figure~\ref{fig:BHNS_rate_per_metallicity} shows the contribution of the five formation channels to the yield of \bhnsSingle  mergers as a function of metallicity.   The  yield of  \bhnsSingle mergers peaks around metallicities $\Zi\approx \Zsun/5$  ($\Zi \approx 0.003$) and is lowest around $\Zi\gtrsim  \Zsun$, in broad agreement with e.g.  \citet[][]{2018MNRAS.480.2011G, 2019MNRAS.487....2M} and \citet{2019MNRAS.490.3740N}, but there are some variations between these models in single stellar evolution, winds, mass transfer, and supernova prescriptions.
The classic formation channel (channel I) dominates the yield. At  metallicities  $\log(\Zi) \lesssim \Zsun/2$ the other formation channels also contribute, particularly the single-core CE channel.   For $\Zi>\Zsun/2$ almost all \bhnsSingle mergers form through just two formation channels: the classic channel and a fraction forms through the   only stable mass transfer channel. This is in  agreement with \citet[][see their Table~C1]{2018MNRAS.481.1908K} and is a result from a combination of the metallicity-dependent effects described in the paragraphs above. 

That the \bhnsSingle yield peaks around $\Zi\approx 0.003$ is due to  line-driven stellar winds scaling positively with metallicity in our simulations (Section~\ref{subsec:method-BPS-assumptions}).
First, at higher metallicities,  higher  wind-loss rates strip more mass from the star leading to lower compact object masses in our {SSE} and  \ac{SN} remnant prescription. Lower mass \acp{BH}  receive larger natal kicks, have less mass fallback (leading to larger Blaauw kicks) and  smaller total system masses making them more likely to disrupt during the \ac{SN}.
Second, at higher metallicities,  binaries typically have wider separations after the second SN and  therefore  longer \tinspiral that may exceed \thubble. This is both because stellar winds widen the binary and because they result in stars with less massive envelopes, which reduces the amount of orbital hardening in mass transfer events (i.e., \ac{CE} and stable Roche-lobe overflow). 
At metallicities $\Zi \lesssim 0.001$, on the other hand, the formation rate of \bhnsSingle mergers is suppressed as the reduced stellar winds lead to massive enough carbon-oxygen cores that many systems instead form a \ac{BHBH} merger. A second effect comes from that although overall the radius expansion of stars increases with increasing metallicity, particularly between  $ -3 \lesssim \log(\Zi) \lesssim -2$ the radius extension of Hertzsprung-gap stars decreases for primary star masses that lie in the range to form \bhnsSingle mergers. This decrease in radial extension decreases the number of systems that merge as stars during mass transfer, \citep[cf.][]{2018MNRAS.480.2011G}, which increases the rate of \bhnsSingle formation.

\begin{figure*}
\includegraphics[width=\textwidth]{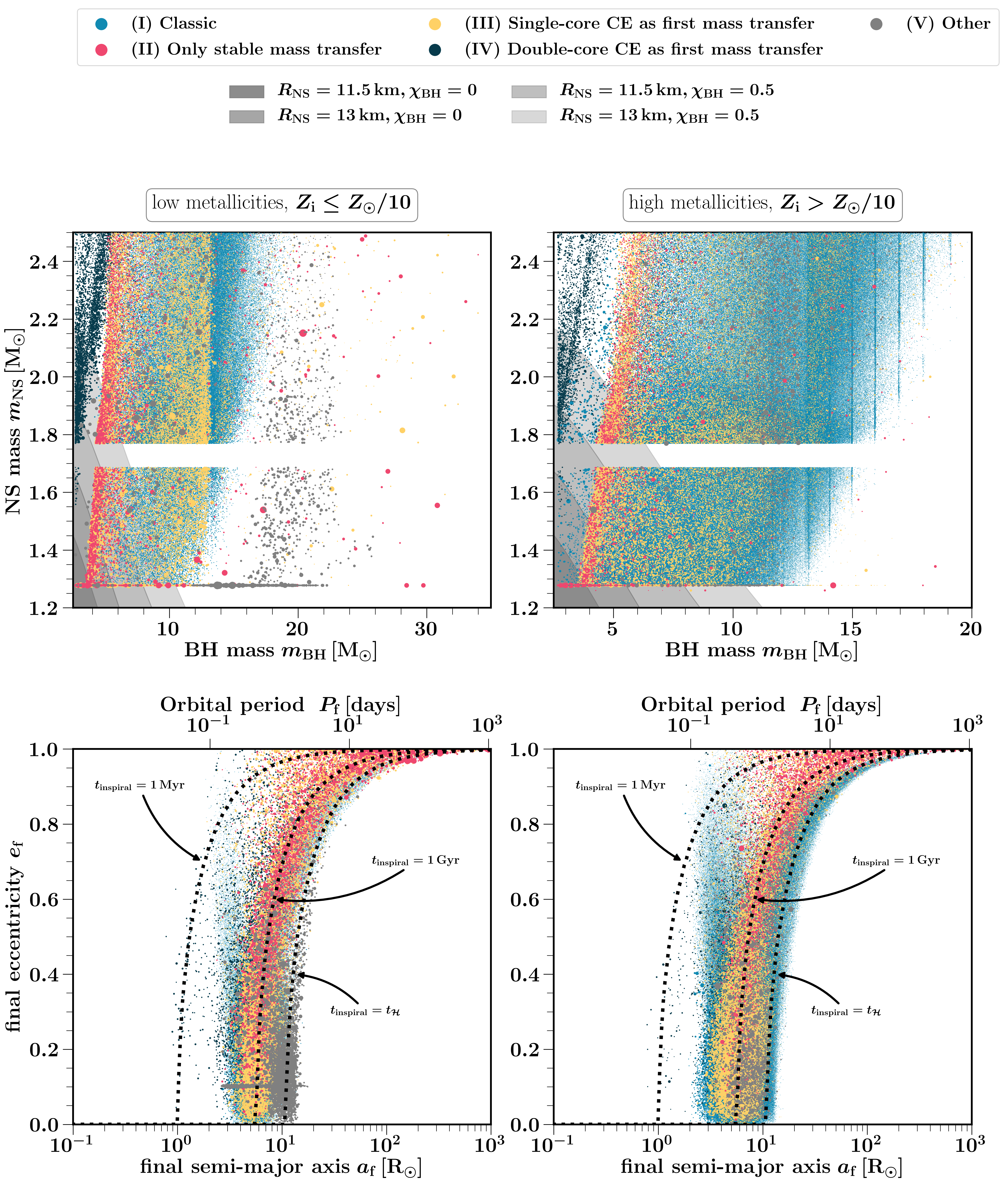}
   \caption{Same as Figure~\ref{fig:BHNS_rate_per_metallicity} for the \bhnsSingle properties  at  \ac{DCO} formation, \tDCO, for our Fiducial model \mAzero. 
   The gray areas  in the top panels denote the \ac{DCO}s that have final masses such that the \ac{NS} is tidally disrupted  outside the  \ac{BH} innermost stable circular orbit based on \citet[][]{2018PhRvD..98h1501F}  when assuming a \ac{NS} radius $\Rns \in\{ 11.5\km,  13\km\}$ and  \ac{BH} spins   $\chibh \in \{0, 0.5\}$ for all BHs. Vertical lines, visible in the top right panel, are artificially caused by our model that maps for a specific \Zi a small range of stars with different initial masses to the exact same \ac{BH} mass as described in Section~\ref{subsec:BHandNSremnantMasses-FiducialModel}. 
   The dotted black lines in the bottom panels show lines of constant $\tinspiral \in \{1\Myr, 1\Gyr,  \thubble \}$. 
   For the orbital period on the second x-axis and the  lines of constant inspiral time we assumed a fixed $\mbhf=10\Msun$ and  $\mnsf=1.4\Msun$.  Figures and videos showing how these distributions change over our model variations are available at \url{https://github.com/FloorBroekgaarden/BlackHole-NeutronStar}. } 
  \label{fig:BHNS_DCOmasses}
\end{figure*}

\subsection{Final properties of  \bhnsSingle mergers  }
\label{subsec:bhns-BPS-DCOmasses}
The final characteristics of the \bhnsSingle systems at  \tDCO (after the formation of the second compact object) are shown in Figure~\ref{fig:BHNS_DCOmasses}. 
In addition, Figure~\ref{fig:BHNS_ObservableDistributions_per_metallicity} shows the predicted distributions functions of the \bhnsSingle merger yield for five different simulated metallicities \Zi  for typical \bhnsSingle characteristic at \tDCO.  These characteristics are the \ac{BH} mass \mbhf, the \ac{NS} mass  \mnsf, mass ratio $\qf = \mbhf / \mnsf$,  eccentricity \ef, total mass $\mtotf = \mbhf + \mnsf$, chirp mass \mchirpf, which is a binary characteristic that is well measured by  ground-based \ac{GW} obervatories and is given by 
\begin{align}
    \mchirpf = \frac{(\mbhf \mnsf)^{3/5}}{(\mbhf + \mnsf)^{1/5}}, 
    \label{eq:chirp-mass}
\end{align}
the inspiral time \tinspiral and the semi-major axis, \af, at \tDCO. 


\subsubsection{BH and NS remnant masses}
\label{subsec:BHandNSremnantMasses-FiducialModel}
\begin{figure*}
\includegraphics[width=1\textwidth]{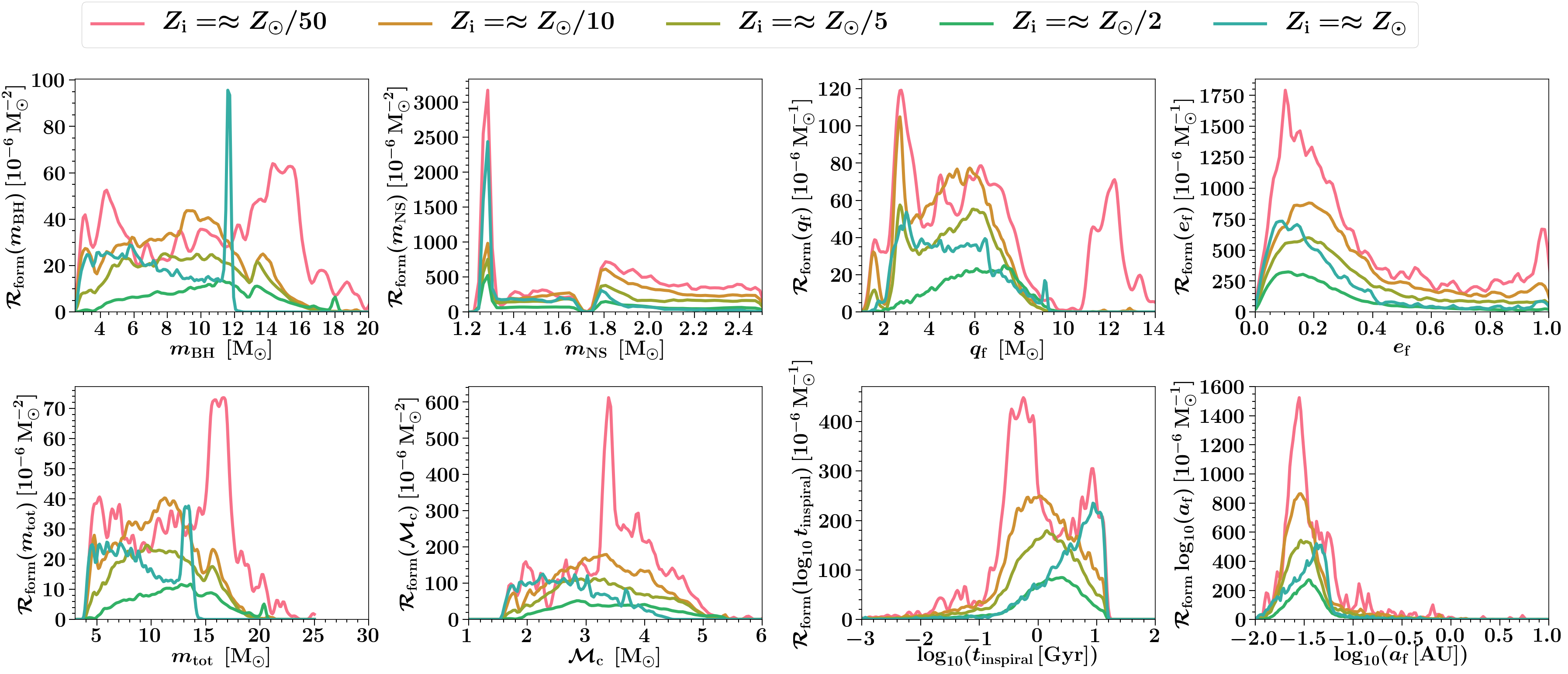}
   \caption{Distributions of the characteristics of \bhnsSingle systems that merge in \thubble at \tDCO for five different metallicities $\Zi$.  The yield is calculated using Equation~\ref{eq:formation-rate-COMPAS} and plotted using a weighted kernel density estimator with a dimensionless kernel bandwidth of 0.04, see the publicly available code for more details.  We use the short hand notation $\mathcal{R}_{\rm{form}}(x)  \equiv  {\rm{d}}N /({\rm{d}}M_{\rm{SFR}} {\rm{d}}x) $ for variable $x$. The variables are described in Section~\ref{subsec:bhns-BPS-DCOmasses}. We warn the reader that interpertating the distributions of \tinspiral and \af can be misleading as they are shown in log to best demonstrate the characteristics over the wide parameter range.  Plots showing more intuitive (not log) distributions of \tinspiral and \af are  given in \url{https://github.com/FloorBroekgaarden/BlackHole-NeutronStar}.}
  \label{fig:BHNS_ObservableDistributions_per_metallicity}
\end{figure*}

The top panels in Figure~\ref{fig:BHNS_DCOmasses} show that  the simulated \bhnsSingle mergers have \acp{NS} with masses in the range $1.25 \lesssim \mnsf/\Msun \leq 2.5$ (where the maximum \ac{NS} mass is set to 2.5\Msun in our fiducial model). 
The discontinuity in the \ac{NS} remnant mass around 1.7\Msun in Figure~\ref{fig:BHNS_DCOmasses} and \ref{fig:BHNS_ObservableDistributions_per_metallicity}  results from the discontinuity in the proto-compact object mass equation at carbon-oxygen cores of $3.5$\Msun  in the delayed \ac{SN} remnant mass prescription (Equation~18 in  \citealt{2012ApJ...749...91F}. 
The over-density of remnant masses around $1.3\Msun$ comes from two effects. First, the  \ac{ECSN}  prescription map stars with different masses to a \ac{NS} mass of 1.26\Msun, as described in Section~\ref{subsec:method-BPS-assumptions} and \citet{2018MNRAS.481.4009V}.  Second, all \ac{NS} progenitors with carbon-oxygen core masses below $2.5$\Msun are in the delayed  \citealt{2012ApJ...749...91F} remnant mass prescription mapped to \acp{NS} with with fixed masses  $\approx 1.28\Msun$\footnote{For the rapid \citealt{2012ApJ...749...91F} remnant mass prescription this maps to a \ac{NS} mass of $\approx 1.1\Msun$.}.

The majority of \acp{BH} in the \bhnsSingle binaries have masses in the range  $2.5 \leq \mbhf/\Msun \lesssim 20$, but this can extend to \mbhf $\approx 30$\Msun for very low values of \Zi, as shown in the top left panel of Figure~\ref{fig:BHNS_DCOmasses}. 
We find that \bhnsSingle binaries with \mbhf $\gtrsim 20$\Msun are rare, in agreement with e.g. \citet{2020MNRAS.497.1563R}. 
At lower metallicities \bhnsSingle mergers are formed with more massive \acp{BH} compared to higher metallicities, as can be seen in the distributions of \mbhf, \qf,  \mtotf and  \mchirpf in Figure~\ref{fig:BHNS_ObservableDistributions_per_metallicity} that are more extended to higher masses for lower metallicities. 
This is because stars at lower metallicities lose less mass during their lives through  line-driven stellar winds leading to larger remnant masses.  
The delayed  \ac{SN} remnant mass prescription does not lead to a \ac{BH} mass gap between $2.5$--$5\Msun$ and we therefore find \acp{BH} with masses close to the maximum \ac{NS} mass of $2.5\Msun$, as can be seen in Figures~\ref{fig:BHNS_DCOmasses}  and \ref{fig:BHNS_ObservableDistributions_per_metallicity}. 
The over-density in $\mbhf$ in straight vertical lines, particularly visible in the right top panel of Figure~\ref{fig:BHNS_DCOmasses}, is due to our prescription of LBV wind mass loss that maps a broad range of initial ZAMS masses to the same carbon-oxygen core masses and hence the same \ac{BH} remnant masses (see for a discussion Appendix B of  \citealt{2019MNRAS.490.3740N}). 
This  results for some of our  \Zi in peaks in the \ac{BH} mass distribution around the highest \ac{BH} mass for that metallicity as can be seen in the top right panel of Figure~\ref{fig:BHNS_DCOmasses} and the top left panel of Figure~\ref{fig:BHNS_ObservableDistributions_per_metallicity}.  

A subset of the \bhnsSingle mergers in our fiducial simulation have  final masses such that the \ac{NS} is disrupted outside of the  \ac{BH} innermost stable circular orbit during the merger. 
This is shown  in Figure~\ref{fig:BHNS_DCOmasses} with the shaded areas for four  different models for the NS radii and \ac{BH} spins. Typically only the \bhnsSingle with low mass \acp{BH} and  \ac{NS} result in a tidal disruption of the \ac{NS}. The fraction of \bhnsSingle mergers that disrupt the \ac{NS} outside of the \ac{BH} innermost stable orbit is strongly dependent on the assumed \ac{BH} spin and \acp{NS} radii, with higher spins and larger \ac{NS} radii leading to more tidally disrupted \acp{NS}.  We discuss this in more detail in Section~\ref{sec:results-NS-disruption-all-models}.

%
\subsubsection{Eccentricity and semi-major axis at \tDCO}
Figure~\ref{fig:BHNS_DCOmasses} shows the semi-major axis and eccentricity for the \bhnsSingle mergers in our Fiducial model \mAzero. 
 \bhnsSingle binaries that merge in \thubble typically have merger times in the range $1\Myr \leq \tinspiral \leq \thubble$ at the moment of the \bhnsSingle formation, corresponding to a semi-major axis between $1 \lesssim \af / \Rsun  \lesssim 100$,  as can be seen in the bottom panels of Figure~\ref{fig:BHNS_DCOmasses}. 
 Systems with larger semi-major axis still merge in \thubble if the binary is more eccentric as this decreases \tinspiral. A subset of the binaries from the classic formation channel have the shortest semi-major axis at \tDCO, which is because these system undergo a case BB mass transfer phase tightening the binary further after the \ac{CE} phase. 

Figure~\ref{fig:BHNS_DCOmasses} shows that the  \bhnsSingle eccentricities densely populate the full range 0--1, although smaller eccentricities are slightly favored as can be seen in Figure~\ref{fig:BHNS_ObservableDistributions_per_metallicity}. 
We do not find clear subpopulations of \bhnsSingle systems with distinguishable eccentricity as discussed for \ac{NSNS} systems by \citet{2019ApJ...880L...8A}.

\subsection{Predicted GW detectable \bhnsSingle distributions }
\label{subsec:distributions-FIducial-GW-observable}

\begin{figure*}
\includegraphics[width=.95\textwidth]{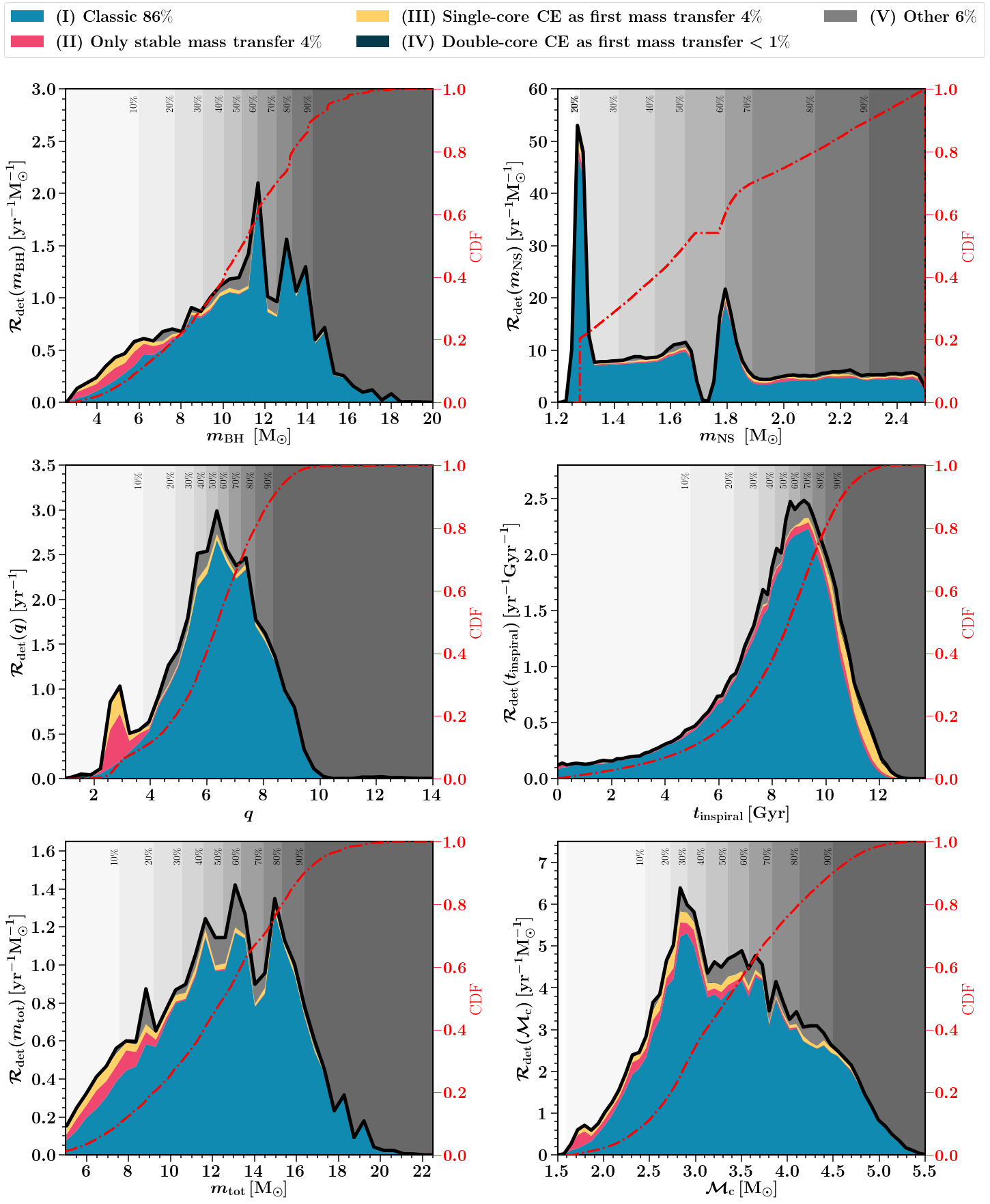}
  \caption{ Predicted distributions of the  detectable \bhnsSingle merger characteristics for our fiducial binary population synthesis and \SFRD model \mAzero for a ground-based LVK detector network at design sensitivity.   The detection rate is calculated using Equation~\ref{eq:rate_detector}. The colors in each graph, and the numbers quoted in the legend, represent the percentage that each formation channel (presented in Section~\ref{subsec:bhns-BPS-formationChannels}) contributes to the total \ac{GW} detected yield.    The gap in the \ac{NS} remnant mass distribution is caused by a discontinuity in the remnant mass prescription as discussed in Section~\ref{subsec:method-BPS-assumptions}. The red dashed-dotted lines show the cumulative distribution function (CDF). Gray areas indicate values for the CDF.
  The distributions are produced using a kernel density estimator with a bandwidth factor of 0.04, see our publicly available code for more details.
    }
  \label{fig:BHNS_DCO_observed}
\end{figure*}

The predicted \ac{GW} detectable \bhnsSingle distributions for a LVK  network at design sensitivity are shown in Figure~\ref{fig:BHNS_DCO_observed}. We show the distribution for the \ac{BH} mass, \ac{NS} mass, mass ratio, inspiral times, total mass and chirp mass (Equation~\ref{eq:chirp-mass}). Except for the inspiral times, \tinspiral, all other \bhnsSingle parameters are typically obtained from \ac{GW} observations \citep[e.g.][]{2017AnP...52900209A}. 
The distributions and yields are determined using Equation~\ref{eq:rate_detector}, taking into account the \SFRD and detector probability (\Pdet)  weighting. 
In this section we show the predicted distributions for our fiducial  population synthesis and \SFRD model,  given by \mAzero. In Section~\ref{sec:results-variations} we present the predicted \bhnsSingle distributions for all our \Nmodels model variations.  
The predicted distributions and characteristics for \ac{NSNS} and \ac{BHBH} mergers are given in a companion paper.

 Our fiducial model \mAzero presents an intrinsic  \bhnsSingle merger rate at redshift zero of $\RateIntrinsicZero \approx \RateIntrinsicAzeroBHNS $\GpcminThree \yearmin  consistent with the   inferred local \bhnsSingle merger rate density from \citet{Abbott:2021-first-NSBH} of $\RateIntrinsicZero = {45}_{-33}^{+75}$\GpcminThree \yearmin when assuming that GW200105 and GW200115 are representative of the entire \bhnsSingle population, whilst slightly below the inferred  $\RateIntrinsicZero = {130}_{-69}^{+112}$\GpcminThree \yearmin when \citet{Abbott:2021-first-NSBH}  assume a broader (and likely optimistic) distribution of component masses. 
 When weighting for the sensitivity  of a ground based \ac{GW} network we find a detection rate of $\RateObserved \approx \RateObservedAzeroBHNS $\yearmin.

\subsubsection{Formation channels} 
\label{subsec:fiducial-formation-channels-GWs}
Our fiducial model predicts that about  \PercentageClassicLVK of the \bhnsSingle mergers  detected by a LVK network at design sensitivity form through the classic formation channel (I,  Section~\ref{subsubsec:classic-channel} and Figure~\ref{fig:formation-channels-sketch}). 
This percentage is higher compared to the average percentage that the classic channel contributes for each \Zi in our simulation (without the \SFRD and \ac{GW} observation weighting). This results from two main effects.  First, our fiducial \SFRD model convolved with the typical short delay times (Figure~\ref{fig:BHNS_ObservableDistributions_per_metallicity}) for \bhnsSingle systems biases the detectable \bhnsSingle systems to originate from binary systems with initial metallicities $\Zi \gtrsim \Zsun / 2$.
At these metallicities the contribution from other channels is relatively low (Figure~\ref{fig:BHNS_rate_per_metallicity}).
Second,  the classic formation channel produces overall more massive \bhnsSingle systems compared to the other channels as shown in Figure~\ref{fig:BHNS_DCOmasses}, which are observed to larger distances with  \ac{GW} detectors. 
This further increases the contribution of the classic formation channel to eventually form the \PercentageClassicLVK. 

The dominant channel for \ac{BHBH} and \ac{NSNS} observations is different compared to our findings for \bhnsSingle mergers.  For \ac{BHBH} mergers the majority of  \ac{GW} detected  mergers are predicted to form through the only stable mass transfer channel (II, \citealp[e.g.][]{2019MNRAS.490.3740N}).  For \ac{NSNS} mergers about $60-70\%$ of the detected systems are predicted to form through the double-core \ac{CE} channel (IV, \citealp[cf.][]{2018MNRAS.481.4009V}).   
\bhnsSingle mergers might thus provide a good probe to study the formation of \ac{GW} sources that form through the classic formation channel compared to \ac{BHBH} and \ac{NSNS} mergers as there is less contamination from other channels. However, the contribution of each channel can drastically change as we vary population synthesis models, in particularly for changes in the \ac{CE} parameter and mass transfer efficiency, which we discuss in Section~\ref{subsec:variations-formation-channels-GWs}.

\subsubsection{BH mass} 
The fiducial model predicts that the \bhnsSingle mergers  will typically have $2.5 \lesssim \mbhf/ \Msun \lesssim 16$, with less than $5\%$ of \ac{GW} detected \bhnsSingle mergers having $\mbhf \gtrsim  15$\Msun (cf.  \citealt[][]{2020MNRAS.497.1563R}), as can be seen in Figure~\ref{fig:BHNS_DCO_observed}. 
This is different for  \ac{BHBH} mergers where the majority of predicted \acp{BH} is typically predicted to exceed  $15$\Msun \citep[e.g.][]{2019MNRAS.490.3740N}. Some of the sharp peaks in the \ac{BH} mass, such as the pile up around $\mbhf \approx 12\Msun$, are caused by our grid of metallicity sampling in combination with our LBV wind prescription (see also Figure~\ref{fig:BHNS_ObservableDistributions_per_metallicity} and  Section~\ref{subsec:BHandNSremnantMasses-FiducialModel}).

\subsubsection{\ac{NS} mass} 
 Figure~\ref{fig:BHNS_DCO_observed} shows that our fiducial model predicts  a somewhat flat distribution in \ac{NS} mass with two peaks around $1.3$\Msun and 1.8\Msun caused by our choice of \ac{SN} remnant mass prescription. 
A large fraction of \acp{NS} in \bhnsSingle mergers are massive: almost $ 60\% $  of the \bhnsSingle  observations are predicted to have $\mnsf \gtrsim 1.5$\Msun  \citep[cf.][]{2018MNRAS.474.2959G}. This is a much larger fraction compared to the typical fraction of NSNS mergers with a $\mnsf \gtrsim 1.5$\Msun \citep[e.g.][]{2018MNRAS.481.4009V}.
This mainly results from the \acp{NS} in  \bhnsSingle systems originating from more massive stars compared to \acp{NS} in \ac{NSNS} mergers. These more massive stars result in more equal mass ratio stellar binaries (since the binary also contains the  BH  progenitor), which is typically more likely to avoid a stellar merger and disruption during the \ac{SN}  in the binary evolution. This favors more massive \acp{NS} in \bhnsSingle mergers.
Massive \acp{BH} in \bhnsSingle mergers typically have more massive \ac{NS} as can be seen in Figure~\ref{fig:BHNS_DCOmasses}, in agreement with e.g. \citet[][Figure~7]{2018MNRAS.481.1908K}. 
All in all, \bhnsSingle mergers can be  important for the study of massive \acp{NS}.

\subsubsection{Mass ratio} 
  Figure~\ref{fig:BHNS_DCO_observed} shows that the predicted \bhnsSingle merger mass ratio is typically in the range $2 \lesssim \qf \lesssim 10$ where we now define $\qf =\mbhf / \mnsf$ and peaks around  $\qf \approx 5$--$8$  (cf. \citealt{2018MNRAS.474.2959G}) due to many \acp{BH} with \mbhf $\approx 10$--$12\Msun$ and the \ac{NS} mass peaks around $1.3\Msun$ and $1.8\Msun$. 
There is also a small second peak around $\qf \approx 3$ from contributions from channels II and IV (Section~\ref{subsec:bhns-BPS-formationChannels}) that produce \bhnsSingle mergers with low \ac{BH} masses resulting in small \qf.

\subsubsection{Inspiral time} 
The \bhnsSingle inspiral times typically span \tinspiral $\approx 2$--$12$\Gyrs.   
Although most \bhnsSingle that merge in \thubble are formed with   $\tinspiral \lesssim 6$\Gyr (Figure~\ref{fig:BHNS_ObservableDistributions_per_metallicity}), selection effects favour the detection of systems that merge in the local Universe, and so have longer delay times in order to match the higher rate of star formation and increased yields at higher redshifts. 
The \tinspiral distribution is  sensitive to the assumed \SFRD model.
The delay time distribution of these mergers could be constrained in the future from observations \citep[e.g.][]{2018ApJ...863L..41F,2019ApJ...878L..14S,2021ApJ...914L..30F} and might, therefore, help distinguishing  binary population synthesis and \SFRD  models.

\subsubsection{Total mass and chirp mass}
The left bottom panel of Figure~\ref{fig:BHNS_DCO_observed} shows that the \bhnsSingle total masses are predicted to lie in the range $5 \lesssim \mtotf / \Msun \lesssim 20$. 
 The shape of the total mass distribution follows the \mbhf distribution as the \ac{BH} mass dominates the total mass for \bhnsSingle mergers. 
The right  bottom panel of Figure~\ref{fig:BHNS_DCO_observed} shows  that the predicted chirp masses of the  \bhnsSingle mergers lie in the range $1.5 \lesssim \mchirpf/ \Msun \lesssim  5.5$ with the majority of systems having chirp masses  between about $2$--$5$\Msun.

\begin{table}
\centering
\begin{tabular}{l|c|c|c}
 \hline
 \hline
        $f_{\rm{EM}}$             & $\chibh	=0$	 	        & $\chibh \sim$ Qin18             & $\chibh	=0.5$           \\ \hline %
$\Rns = 11.5$\km    & $0.013 $	 	    & $0.015$               & $0.13$           \\
$\Rns = 13$\km      & $0.043$	 	    & $0.045 $               & $0.28$            \\ \hline \hline
\end{tabular}%
\caption{Fraction, $f_{\rm{EM}}$, of the detectable \bhnsSingle mergers that are predicted to disrupt the \ac{NS} outside of the \ac{BH} innermost stable orbit.  The different columns correspond to the different spin models assuming: (i) all \ac{BH} spins are zero, (ii) an ad hoc, but physically motivated, model based on \citet{2018A&A...616A..28Q} where the \ac{BH} can have a moderate spin if it is formed second and (iii) all  \ac{BH} spins are half the maximum spin. The rows correspond to assuming a $11.5$\km and $13$\km NS radius, respectively. The fractions are  calculated using \citet[][]{2018PhRvD..98h1501F} and Equation~\ref{eq:rate_detector}, taking into account the \SFRD weighting from our fiducial model and detector sensitivity for a ground-based LVK detector network at design sensitivity. }
\label{tab:fiducial_BHNS_EM_ejecta_observed}
\end{table}

\subsubsection{NS disruption} 
\label{sec:results-NS-disruption-all-models}
We predict that  a fraction between $\approx 1\%$--$28\%$ of the \ac{GW} detectable \bhnsSingle mergers will disrupt the \ac{NS} outside of the \ac{BH} innermost stable orbit as shown in Table~\ref{tab:fiducial_BHNS_EM_ejecta_observed} for our fiducial model \mAzero. 
These systems are predicted to have nonzero ejecta mass and can potentially have an electromagnetic counterpart such as a short gamma-ray burst or kilonova. A subset of these  systems  could potentially be observed as electromagnetic counterpart.  The highest fractions of \bhnsSingle mergers with nonzero ejecta masses is when assuming that  systems have large \ac{NS} radii and large BH spins, where especially the spins of the \ac{BH} are dominant. Our models, except those assuming $\chibh = 0.5$, result in percentages on the order of 1$\%$ consistent with, e.g.  \citet{2020arXiv200906655D}.


\begin{figure*}
    \centering
\includegraphics[width=1.0\textwidth]{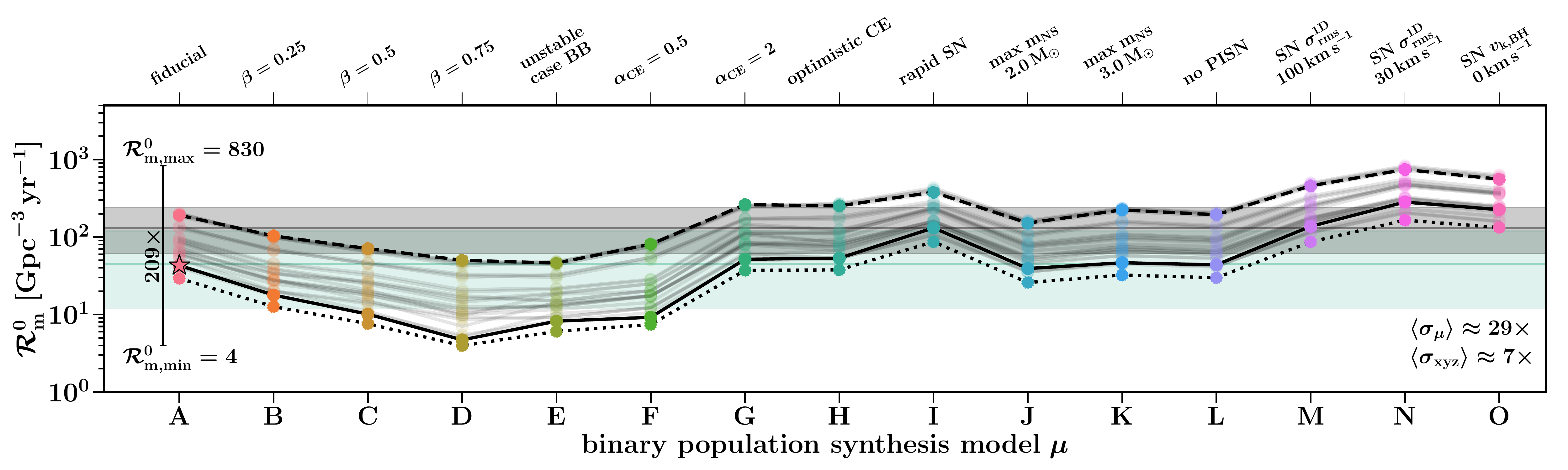} %
    \caption{Predicted intrinsic \bhnsSingle merger rates for our \Nmodels model variations.
    The rates are for mergers at $z=0$ without applying \ac{GW} selection effects. 
    We show for each of the \NmodelsBPS binary population synthesis model variations (given in Table~\ref{tab:variations-BPS}) the \bhnsSingle merger rates for the \NmodelsMSSFR variations in \SFRD (given in Table~\ref{tab:MSSFR-variations-labels}). We connect predictions that use the same \SFRD model with a line for visualisation reasons only (it is not an interpolation between models). 
    Three \SFRD variations,  \rm{xyz}$=$000 (solid), \rm{xyz}$=$231 (dashed) and \rm{xyz}$=$312 (dotted)  are highlighted,  corresponding to our fiducial \SFRD model and the models resulting in one of the highest and lowest rate predictions. The shaded horizontal bars mark the corresponding \ac{GW}-inferred $90\%$ credible intervals for the merger rate densities from \citet[][]{Abbott:2021-first-NSBH}: $\rate_{\rm{m}}^0 = {45}_{-33}^{+75}$\GpcminThree \yearmin when assuming that GW200105 and GW200115 are representative of the entire \bhnsSingle population (teal) and  $\rate_{\rm{m}}^0 = {130}_{-69}^{+112}$\GpcminThree \yearmin when assuming a broader distribution of component masses (grey). Our fiducial model  (\mAzero) estimate is shown with a star symbol. On the right of the  panel  $\langle \sigma_{\mu}\rangle$ and $\langle \sigma_{\rm{xyz}}\rangle$ represent the mean scatter in rates due to varying our assumptions in binary population synthesis and \SFRD prescriptions, respectively. These are calculated using Equation~\ref{eq:sigma-mu} and~\ref{eq:sigma-xyz}. The minimum and maximum rates and the ratio between those are quoted with an error bar on the left.  We use the short-hand notation $\rate_{\rm{m}}^0 \equiv (\diff \Ndet^2 / \diff \ts \diff \Vc)(\tmerger(z=0))$. 
    }%
    \label{fig:IntrinsicRates}
\end{figure*}
\begin{figure*}
    \centering
\includegraphics[width=1.0\textwidth]{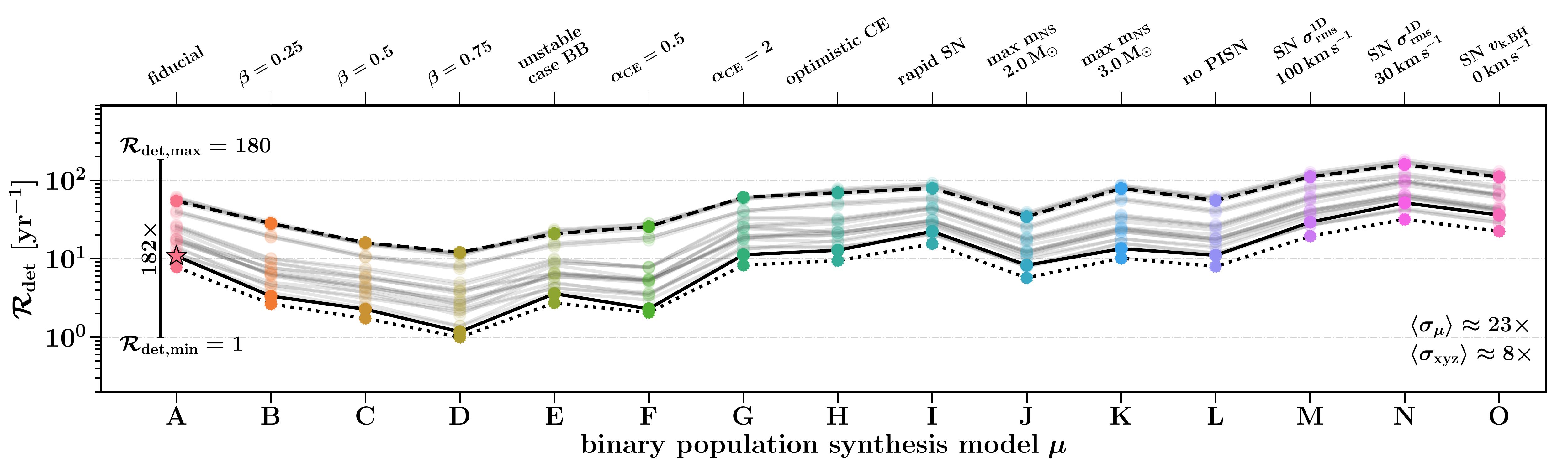} %
    \caption{Same as Figure~\ref{fig:IntrinsicRates} but for the predicted  detectable \bhnsSingle rates for a \ac{GW} network at design sensitivity. We use the short-hand notation $\rate_{\rm{det}} \equiv \diff \Ndet / \diff \tdet$.}%
    \label{fig:ObservedRates}
\end{figure*}

\section{Varying model assumptions}
\label{sec:results-variations}

The predicted rates and characteristics of  \bhnsSingle mergers, presented in Section~\ref{sec:results-fiducial} for our fiducial model  \mAzero,  are sensitive to uncertainties in binary population synthesis and \ac{SFRD}{\ensuremath{(Z_{\rm{i}},z)}\xspace}  model assumptions. 
We compare in this section the predicted \bhnsSingle merger rates and characteristics for our total of \Nmodels  combinations of the \NmodelsBPS binary population synthesis and  \NmodelsMSSFR \SFRD models. 
These model variations are summarized in Table~\ref{tab:variations-BPS} and~\ref{tab:MSSFR-variations-labels}.  In most figures throughout this section we highlight three of our \SFRD models: the preferred phenomenological model from \citet{2019MNRAS.490.3740N},  $\rm{xyz} = 000$, and the \SFRD prescriptions $\rm{xyz} = 312$ and $\rm{xyz} = 231$, which we show correspond to the lowest and highest \bhnsSingle merger rates, respectively.

\subsection{Predicted \bhnsSingle merger rate for varying model assumptions}
\label{subsec:results-variations-rates}

\subsubsection{Intrinsic \bhnsSingle merger rates}
\label{subsec:results-variations-rates-intrinsic}
The predicted intrinsic (at redshift $z=0$) rates for \bhnsSingle mergers are shown for our \Nmodels models in Figure~\ref{fig:IntrinsicRates}. 
These rates are calculated using Equation~\ref{eq:MSSFR-merger-rate}  and do not yet take into account the \ac{GW} detector selection effects.  
We find that the intrinsic \bhnsSingle merger rates are predicted to lie in the range $\rate_{\rm{m}}^0 \approx$ \RateIntrinsicAzeroBHNSmin--\RateIntrinsicAzeroBHNSmax \GpcminThree \yearmin. Our fiducial model \mAzero predicts $\RateIntrinsicZero \approx \RateIntrinsicAzeroBHNS $\GpcminThree \yearmin. We discuss in more detail the variations in rates over the models in Section~\ref{subsec:effect-variations-on-merger-rates}.
Almost all \Nmodels models predict intrinsic \bhnsSingle  merger rates that are consistent with one of the  \ac{GW} inferred $90\%$ confidence interval  for the \bhnsSingle  merger rates from \citet{Abbott:2021-first-NSBH}.

\subsubsection{GW detectable \bhnsSingle merger rates}
\label{subsec:results-variations-rates-observed}
The predicted \ac{GW} detectable  rates for  \bhnsSingle  mergers  are shown in Figure~\ref{fig:ObservedRates}. 
These rates are calculated using Equation~\ref{eq:rate_detector} for a ground-based \ac{GW} detector  equivalent to the LVK network at design sensitivity for a full year observing run. 
Our fiducial model predicts a \bhnsSingle merger rate of $\rate_{\rm{det}}  \approx \RateObservedAzeroBHNS  \yearmin.$ Considering the \Nmodels model variations we find predicted  rates in the range $\rate_{\rm{det}} \approx \RateObservedAzeroBHNSmin$--$\RateObservedAzeroBHNSmax \yearmin$. The \ac{GW} detectable rates in Figure~\ref{fig:ObservedRates} behave similar under our \Nmodels model variations as the intrinsic rates presented in Figure~\ref{fig:IntrinsicRates}. This is because  the  impact of the redshift evolution of the \bhnsSingle merger properties in our simulations is minor compared to changes over our model variations,  within the \ac{GW} horizon distance of \bhnsSingle mergers.

\subsubsection{Formation channel rates}
\label{subsec:variations-formation-channels-GWs}
\begin{figure*}
    \centering
\includegraphics[width=1.0\textwidth]{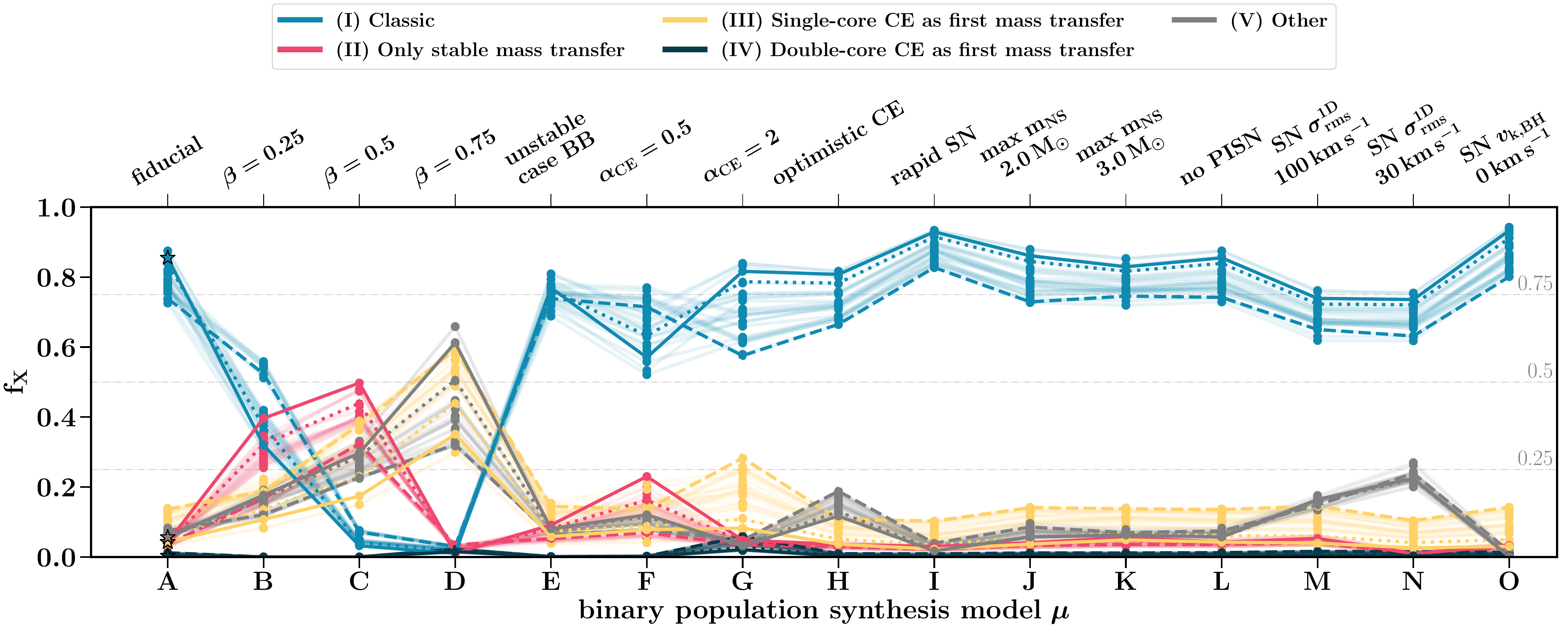} %
    \caption{Fraction, $\rm{f}_{\rm{X}}$,    that each formation channel  contributes to the total predicted  \ac{GW} detected \bhnsSingle merger rate $\rate_{\rm{det}}$ for our \Nmodels model variations (where the subscript X is a placeholder for the formation channel label). Labels and lines are as in Figure~\ref{fig:ObservedRates}, but are colored based on the five formation channels we classify as described in Section~\ref{subsec:bhns-BPS-formationChannels}. 
    }%
    \label{fig:ObservedRates-formation-channels}
\end{figure*}
\begin{table}
\centering
\begin{tabular}{l|l|l|l|l}
 \hline
 \hline
        $\rm{f}_{\rm{I}}$            & $\rm{f}_{\rm{II}}$                  & $\rm{f}_{\rm{III}}$                    & $\rm{f}_{\rm{IV}}$              & $\rm{f}_{\rm{V}}$     \\ \hline %
$0.01$--$0.94$             & $0.008$--$0.50$ 	 	    & $0.021$--$0.60$             & $0$--$0.59$       & $<0.1$--$0.66$  \\ \hline \hline
\end{tabular}%
\caption{Range of the minimum and maximum predicted percentage that each \bhnsSingle formation channel contributes when considering all \Nmodels model variations studied in this work. From left to right the columns represent the formation channels labeled: (I) classic, (II) only stable mass transfer, (III) Single-core CE as first mass transfer, (IV) Double-core CE as first mass transfer and (V) other. The formation channels are described in Section~\ref{subsec:bhns-BPS-formationChannels}. }
\label{tab:formation-channel-variations-min-and-max}
\end{table}

The predicted percentages that each formation channel contributes to the predicted total  \ac{GW} detected \bhnsSingle merger rate  is affected by variations in \SFRD and binary population synthesis models and shown in Figure~\ref{fig:ObservedRates-formation-channels} and Table~\ref{tab:formation-channel-variations-min-and-max} for our five formation channels as  described in Section~\ref{subsec:bhns-BPS-formationChannels}.
The minimum and maximum predicted percentages that each formation channel from Figure~\ref{fig:ObservedRates-formation-channels}  can contribute to the  \ac{GW} detected \bhnsSingle rate is quoted in Table~\ref{tab:formation-channel-variations-min-and-max}.  We find that the fraction each formation channel contributes is dominated by variations in the binary population synthesis model over variations in \SFRD. All models, except those involving binary population synthesis models B, C and D,  predict the classic formation channel (I) dominates the \bhnsSingle rate, with a percentage  $\rm{f}_{\rm{I}} > 50\%$ of the  \ac{GW} detectable \bhnsSingle coming from this channel. Particularly in models B, C and D, that assume a fixed mass transfer efficiency of $\beta = 0.25,\beta=0.5$ and $\beta=0.75$ for non compact objects, respectively, this percentage drastically decreases, as in those models binaries going through the classic channel merge before they form a \bhnsSingle system. This is particularly visible in the variations for  Figures~\ref{fig:BHNS_ZAMSmasses}, \ref{fig:BHNS_rate_per_metallicity} and \ref{fig:BHNS_DCOmasses}  for models B and C, which are given in our online supplementary material. All in all, we find that the formation channel contributions for \bhnsSingle mergers is particularly sensitive to mass transfer, and so we expect that also varying binary population synthesis prescriptions for the stability criteria, the angular momentum loss and the \ac{CE} prescription might further impact our results, as also suggested by e.g. \citet[][see Section 4.3]{2020A&A...638A..55K}.

\begin{figure*}
    \centering
\includegraphics[width=1.0\textwidth]{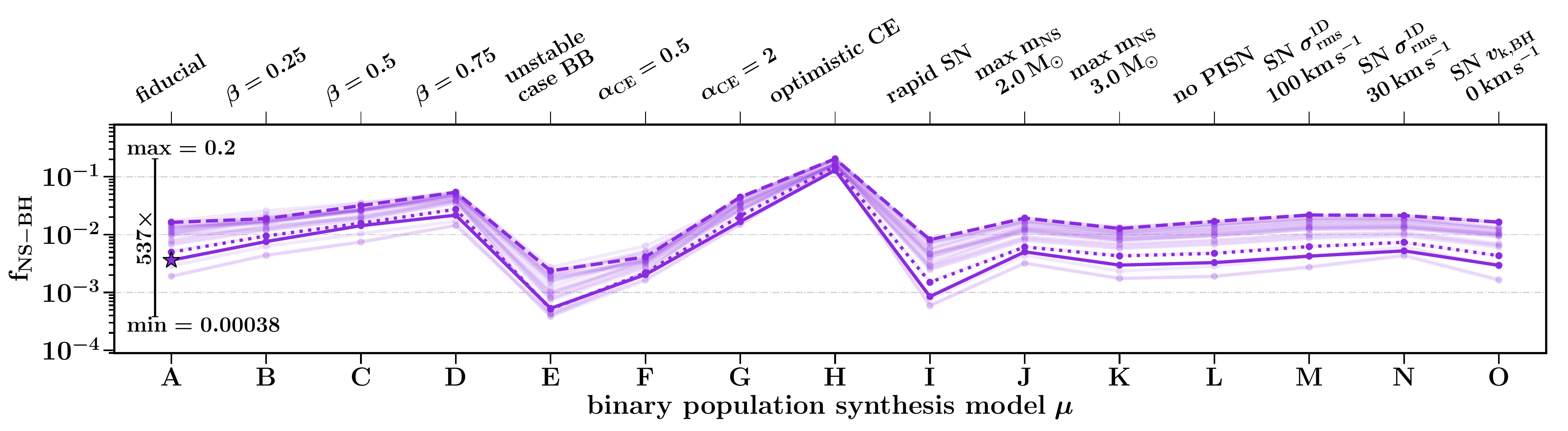} 
    \caption{The predicted  fractions of detectable  \bhnsSingle systems where the \ac{NS} forms first (NS--BH systems), $\rm{f}_{\text{NS--BH}}$,     for a \ac{GW} network at design sensitivity. The labels and lines are as in  Figure~\ref{fig:IntrinsicRates}. } 
    \label{fig:ObservedRatesNSBH}
\end{figure*}

\subsubsection{GW Detectable NS--BH merger rates}
\label{subsec:fraction-BHNS-with-NSBH}

NS--BH systems where the \ac{NS} forms first are interesting astrophysical sources. First,  because the first formed \ac{NS} may spin up during mass transfer episodes and eventually form a millisecond-pulsar \ac{BH} binary that might be observable with radio telescopes \citep[e.g.][]{2005ApJ...628..343P,2020arXiv201113503C}. Detecting  such a system would provide a unique laboratory to test general relativity and alternative theories of gravity \citep{Wex:1998wt,KRAMER2004993, 2014arXiv1402.5594W} and will enable high precision  measurements of the  properties of \acp{BH}    \citep{1975ApJ...198L..27B,1975SvAL....1....2B}. However, to date, no pulsar--BH system has been observed through radio observations. This might not be entirely surprising as several studies estimate that the fraction of   pulsar--\ac{BH}  over \ac{NSNS} binaries in our Milky Way is small \citep{2005ApJ...628..343P, 2020arXiv201113503C}. Currently there are about 20 \ac{NSNS} known \citep[e.g.][]{tauris2017formation,2019ApJ...876...18F}. Second, in NS--BH systems the second formed \ac{BH} may obtain a high \ac{BH} spin if its progenitor helium star spun up due to tidal interactions with the \ac{NS} companion, whereas a first formed \ac{BH} in \bhnsSingle mergers may always have negligible spin  \citep{2018A&A...616A..28Q}. NS--BH mergers might, therefore, be distinguishable in \ac{GW} observations by measuring a high $\chi_{\rm{eff}}$ from the high \ac{NS} or \ac{BH} spins compared to BH--NS mergers that will always have low $\chi_{\rm{eff}}$ \citep[e.g.][]{2020arXiv201113503C}.

\begin{figure*}
    \centering
\includegraphics[width=1.0\textwidth]{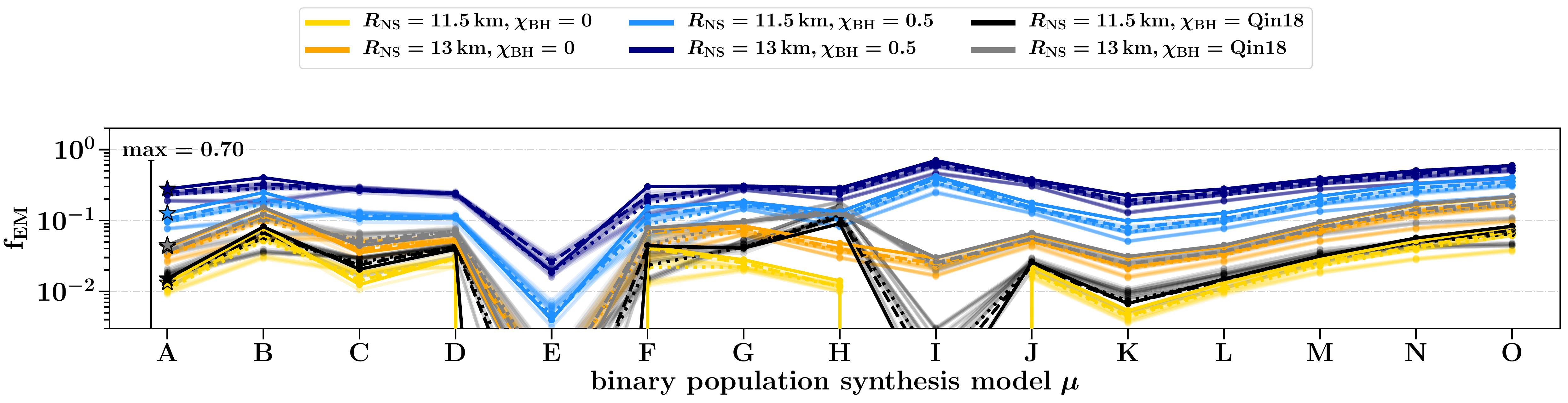} 
    \caption{The fraction of \ac{GW} detected \bhnsSingle mergers in which the \ac{NS} is disrupted outside of the \ac{BH} innermost stable circular orbit (instead of  plunging into the \ac{BH}), $\rm{f}_{\rm{EM}}$. Different colors correspond to different assumptions for the \ac{NS} radii and \ac{BH} spins. The labels and lines are as in  Figure~\ref{fig:IntrinsicRates}.}
    \label{fig:ObservedRatesNSdisrupted}
\end{figure*}

Figure~\ref{fig:ObservedRatesNSBH} shows  the predicted fraction of detectable \bhnsSingle mergers where the \ac{NS} formed first (NS--BH systems).  We predict fractions of NS--BH mergers  in  the range $\rm{f}_{\text{NS--BH}} \approx 0.00038$--$0.2$ for a \ac{GW} network equivalent to LVK at design sensitivity. 
The highest fraction of NS--BH mergers is in the optimistic-\ac{CE} model (H). This is because most NS--BH systems form from binaries where the first mass transfer occurs relatively early on (case A or early case B mass transfer). In our model this phase of mass transfer is typically highly conservative and the stars can therefore exchange a large fraction of mass and reverse in masses, making the initial primary star the least massive star in the system and eventually form the \ac{NS} first. Such systems also typically undergo a \ac{CE} phase initiated by a donor star on the Hertzsprung gap, which are assumed to lead to a stellar merger in the pessimistic \ac{CE} models.  From Figure~\ref{fig:ObservedRatesNSBH} it can be seen that the fraction of  \ac{GW} detectable NS--BH mergers is particularly sensitive to assumptions in the mass transfer and  \ac{CE} prescriptions, varied in binary population synthesis models B, C, D, E, F, G and H. \ac{GW} observations of $\chi_{\rm{eff}}$ of \bhnsSingle mergers could therefore be particularly insightful to constraining the mass transfer and \ac{CE} prescriptions, given that the spin properties of NS--BH are  distinct from those of BH--NS mergers.

\subsubsection{GW detectable merger rates of \bhnsSingle systems that disrupt the NS outside of the BH innermost-stable orbit}
\label{subsec:fraction-BHNS-with-EM-ejecta}
Figure~\ref{fig:ObservedRatesNSdisrupted} shows the  predictions for the fraction of \ac{GW} detectable  \bhnsSingle mergers for which the \ac{NS} is disrupted outside the \ac{BH} innermost stable circular orbit for six combinations of assumed \ac{NS} radii and \ac{BH} spins (described in Section~\ref{subsec:method-tidal-disruption-BHNS}).  
Our \Nmodels model variations predict fractions in the range $\rm{f}_{\rm{EM}} \approx 0$--$0.70$.  
The predictions that assume the highest \ac{BH} spin values predict the highest fraction of \bhnsSingle mergers that disrupt the \ac{NS} outside of the \ac{BH} innermost stable orbit. Larger \ac{NS} radii assumptions also lead to higher fractions as this makes it easier to disrupt the \ac{NS}. 
 The unstable case BB model (E) and rapid \ac{SN} remnant mass prescription model (I)  give the lowest fractions, where for the 11.5\km \ac{NS} radius and zero or \citet{2018A&A...616A..28Q} \ac{BH} spin assumptions none of the \bhnsSingle systems disrupt the \ac{NS} outside of the \ac{BH}. This is because only \bhnsSingle mergers with  small \ac{NS} and \ac{BH} masses\footnote{Typically $\mbhf \lesssim 6\Msun$ and $\mnsf \lesssim 1.7\Msun$, see Figure~\ref{fig:BHNS_DCOmasses}.} disrupt the \ac{NS} outside of the \ac{BH} innermost stable circular orbit for these most pessimistic assumptions for the \ac{NS} radii and \ac{BH} spins (see dark gray shaded area in top panels of Figure~\ref{fig:BHNS_DCOmasses}). These low mass remnants form from relatively low mass helium stars, that in model E undergo unstable case BB mass transfer and merge before forming a \bhnsSingle system. Model I   does not produce, by construction,  any \acp{BH}  with   $ \mbhf \lesssim6$\Msun, which are the  \ac{BH} masses that can disrupt the \ac{NS} outside of the \ac{BH}  for the most pessimistic \ac{NS} radius and \ac{BH} spin assumption.  At the same time, model I also predicts the highest fraction of \ac{NS} disruptions outside the BH innermost-stable orbit, as high as $\rm{f}_{\rm{EM}} \approx 0.70$ for the most optimistic assumptions of 13\km \ac{NS} radii and \ac{BH} spins of $\chibh=0.5$. This is because the rapid \ac{SN} remnant prescription produces a peak of \acp{BH} with masses around $\sim 8\Msun$ that disrupt their \ac{NS} under these assumptions.

Figure~\ref{fig:ObservedRatesNSdisrupted} shows that the predicted rates of \bhnsSingle mergers that disrupt the \ac{NS} outside of the \ac{BH} innermost-stable orbit that assume our \citet{2018A&A...616A..28Q} spin model (black and gray lines) are closest to our models that assume all \acp{BH} have zero spins (yellow and orange lines). This is because in most \bhnsSingle mergers the \acp{BH} forms first, resulting in zero \ac{BH} spins in our  \citet{2018A&A...616A..28Q} spin model. The only binary population synthesis model where the  \citet{2018A&A...616A..28Q} spin assumption predicted rates drastically deviate from the zero \ac{BH} spin predicted rates  is in the optimistic-\ac{CE} model (H). This is because in this model the fraction of NS--BH mergers, where the \ac{NS} formed first,  can be as high as $\rm{f}_{\text{NS--BH}} \approx 20\%$ as shown in Figure~\ref{fig:ObservedRatesNSBH}. As a result, a significant fraction of these NS--BH can achieve high spins in our \citet{2018A&A...616A..28Q} model and disrupt the \ac{NS} outside of the \ac{BH} innermost-stable orbit. 

The disruption of the \ac{NS} outside of the \ac{BH} innermost stable circular orbit can result in the production of electromagnetic transients such  as short gamma-ray bursts, neutrinos and kilonovae \citep[e.g.][]{2019MNRAS.486.5289B, 2020EPJA...56....8B, 2021arXiv210303378D}. In a (small) fraction of such events the electromagnetic counterpart might be detectable, allowing a multi-messenger observation \citep[e.g. a fraction 0.1--0.5 for kilonoavae cf.][]{2019MNRAS.486.5289B,2020arXiv201102717Z}. This depends among other things on the distance to the \bhnsSingle merger, the luminosity of  the electromagnetic transient, its orientation and sky location.

\subsubsection{Effect of varying binary population synthesis and \ac{SFRD}{\ensuremath{(Z_{\rm{i}},z)}\xspace} assumptions  on the predicted  \bhnsSingle rates}
\label{subsec:effect-variations-on-merger-rates}
To quantify the scatter in the predicted rates, i.e. the impact from varying our model assumptions, we calculate  the mean of the ratios between the maximum and minimum predicted rates given by
\begin{equation}
\label{eq:sigma-mu}
\langle \sigma_{\rm{\mu}} \rangle = \frac{1}{15} \sum_{\mu=A}^{\mu=O}
 \frac{\rm{max}(\rate_{\rm{m, \mu000}}^0,... ,\rate_{\rm{m, \mu333}}^0 )}{\rm{min}(\rate_{\rm{m, \mu000}}^0,... ,\rate_{\rm{m, \mu333}}^0 ) }, 
\end{equation} 

and 
\begin{equation}
\label{eq:sigma-xyz}
\langle \sigma_{\rm{xyz}}\rangle = \frac{1}{28} \sum_{\rm{xyz}=000}^{\rm{xyz}=333}
 \frac{\rm{max}(\rate_{\rm{m, Axyz}}^0,... ,\rate_{\rm{m, Oxyz}}^0 )}{\rm{min}(\rate_{\rm{m, Axyz}}^0,... ,\rate_{\rm{m, Oxyz}}^0)}.
\end{equation} 
The values for the scatter caused by our binary population synthesis and \SFRD assumptions are quoted for our \bhnsSingle mergers   in  Figure~\ref{fig:IntrinsicRates} and \ref{fig:ObservedRates}.  
We find that for both the predicted intrinsic and detected \bhnsSingle merger rates  $\langle \sigma_{\rm{\mu}}\rangle \approx 20$--$30$ and $\langle \sigma_{\rm{xyz}}\rangle  \approx 8$. Overall, we thus find that variations from binary population synthesis and \SFRD assumptions affect the rates of order  $\sim \mathcal{O}(10)$. In the models that we varied we find that the  binary population synthesis variations impact the rate with a factor 2--4 more compared to the variations in \SFRD.  Figure~\ref{fig:IntrinsicRates} and~\ref{fig:ObservedRates} also show that the uncertainties are somewhat independent of each other: our \SFRD assumptions introduce uncertainties of $\mathcal{O}(10)$ for all our binary population synthesis models, and vice versa: varying our binary population synthesis model assumptions introduces uncertainties of order $\mathcal{O}(10)$ for all our \SFRD models. This is because in  all our population synthesis models, the \bhnsSingle rate behaves overall similarly as a function of metallicity: the rate is suppressed both at extremely low and high metallicities for all our binary population synthesis variations. This is shown for our fiducial model in Figure~\ref{fig:BHNS_rate_per_metallicity}.

\subsubsection{Effect from binary population synthesis variations}
 The highest predicted \bhnsSingle rates in Figure~\ref{fig:IntrinsicRates} and~\ref{fig:ObservedRates} are found in the binary population synthesis models G, H, I, K, M, N and O,  which have $\gtrsim 50\%$ of the \SFRD variations typically predict  \bhnsSingle merger rates $\rate^0 \gtrsim 100$\GpcminThree \yearmin and predict for most  \SFRD models $\rate_{\rm{det}} \gtrsim 10$\yearmin.
 
 In model G the \ac{CE} efficiency parameter is increased to $\alpha=2$ compared to $\alpha=1$ for our fiducial model. This means that we assume more binary orbital energy is converted into unbinding the \ac{CE} during the unstable mass transfer phase. This leads to more binaries surviving the \ac{CE} phase as systems that don't successfully eject the envelope are assumed to merge as stars. The result is that the rate of \bhnsSingle mergers forming from formation channels involving a \ac{CE}, increases in our simulations for model G, as also visible for model G in Figure~\ref{fig:ObservedRates-formation-channels}.
 The optimistic \ac{CE} model (H) is the variation on our fiducial model  where we allow Hertzsprung gap donors that engage in a \ac{CE} event to survive.  In this model the \bhnsSingle merger rate is, therefore, enhanced compared to the fiducial model as a significant number of \bhnsSingle systems forms through a \ac{CE} with a Hertzsprung gap donor. Figure~\ref{fig:ObservedRatesNSBH} shows that model H is especially important for forming a large fraction of NS--BH systems. 
 In the rapid \ac{SN} model (I) the rate of \bhnsSingle merger formation is also enhanced compared to our fiducial model that assumes the delayed \ac{SN} remnant mass prescription. This is because most \bhnsSingle mergers form from binaries with secondary stars that have $10 \lesssim \mtwoi / \Msun \lesssim 30$ (Section~\ref{subsec:bhns-BPS-ZAMSm1m2} and top panels Figure~\ref{fig:BHNS_ZAMSmasses}). The rapid \ac{SN} remnant mass function typically allows a larger fraction of secondary masses in this range to form a \ac{NS} whereas in the delayed prescription these form a \ac{BH} instead. This is visible in the bottom panel in Figure~12 of \citet{2012ApJ...749...91F}. 
 This effect causes the \bhnsSingle rate to increase in our model I simulation.
 In model K, which assumes a maximum \ac{NS} mass of $3$\Msun, the \bhnsSingle merger rate is also enhanced as a fraction of the systems that in our fiducial model form  \ac{BHBH} mergers with one \ac{BH} $\lesssim 3\Msun$ in this model form a \bhnsSingle merger instead. 
 Models  M, N and O all have lower \ac{BH} and/or \ac{NS} natal kicks compared to our fiducial model, which increases the fraction of systems that remain bound during the \acp{SN} and therefore increases the \bhnsSingle merger rate compared to our fiducial model. 

In  Figure~\ref{fig:IntrinsicRates} and~\ref{fig:ObservedRates} show that when the mass transfer efficiency is changed to respectively $\beta = 0.25, 0.5 $ and $0.75$ (models B, C and D) that this leads to a decreasing \bhnsSingle rate.  This seems counter-intuitive since higher values for $\beta$ correspond to more mass accretion by a star during mass transfer, which intuitively leads to the formation of more \acp{BH} and higher \bhnsSingle merger rates. However, there is another effect: the detectable  systems are highly biased towards tight binaries that merge within a Hubble time. As almost all our \bhnsSingle mergers in our fiducial model go through the classic formation channel (I) (Figure~\ref{fig:ObservedRates-formation-channels}) which involves a \ac{CE} phase, a more massive companion leads to a more massive shared envelope  that needs to be successfully ejected to avoid a stellar merger from the \ac{CE} event. This leads to  fewer \bhnsSingle systems as in agreement with findings by e.g.  \citet{2018MNRAS.481.1908K}.  Figure~\ref{fig:ObservedRatesNSBH} shows that the fraction of  NS--BH mergers, on the other hand, increases for increasing $\beta$. This is because in this case the contribution of systems that merge as a result from a failed  \ac{CE} ejection is relatively low as these binaries typically form from  initially lower mass stars and  will typically have a lower mass envelope in the \ac{CE} compared to BH--NS binaries.

\subsubsection{Effect from \ac{SFRD}{\ensuremath{(Z_{\rm{i}},z)}\xspace} variations  }
\label{subsec:effect-from-MSSFR-variations}

For a given binary population synthesis model, the highest predicted \bhnsSingle rates are from the \ac{SFRD}{\ensuremath{(Z_{\rm{i}},z)}\xspace} models with the  \citet{2006ApJ...638L..63L}   \ac{MZR} (the models $\rm{xy}1$). For a fixed galaxy mass, this   \ac{MZR} relation results in lower average birth metallicities compared to the other \ac{MZR}  variations as can be seen in Figure~\ref{fig:MSSFR-MZRs} (see  Appendix A of  \citealt{2019MNRAS.490.3740N} for more details). 
This leads to a higher fraction  of stars that form \acp{BH} and therefore to a higher yield of \bhnsSingle mergers. 
The other two \acp{MZR},    \citet{2006ApJ...638L..63L} $+$ offset and \citet{2016MNRAS.456.2140M}, corresponding to $\rm{xy}2$ and $\rm{xy}3$ respectively,   lead to lower \bhnsSingle  yields.  
We find that the \ac{MZR} dominates the uncertainty in the predicted rates for our model variations, consistent with findings by e.g.  \citet{2019MNRAS.487.1675A, 2019MNRAS.488.5300C}.

For the \ac{SFRD}  we find that the  \citet{2004ApJ...613..200S} \ac{SFRD}  ($2\rm{yz}$)  typically has the highest yields, followed by the {\citet{2014ARA&A..52..415M}}  ($1\rm{yz}$) and  \citet{2017ApJ...840...39M}  ($3\rm{yz}$) \ac{SFRD}  assumptions. This is corresponds to each \SFRD prescription having a different total star forming yield,  as can be seen in Figure~\ref{fig:MSSFR-SFRs}.

For the GSMFs we find the highest yield is given by the  \citet{2015MNRAS.450.4486F} functions, either single or double Schechter,  which both give almost identical yields. 
On the other hand, the \citet{2004MNRAS.355..764P}  \ac{GSMF}  ($\rm{x}1\rm{z}$) leads to  relatively lower \bhnsSingle merger  yields. This is because the  \citet{2015MNRAS.450.4486F} \ac{GSMF} prescriptions have relatively more low mass galaxies compared to the \citet{2004MNRAS.355..764P}  \ac{GSMF} as can be seen in Figure~\ref{fig:MSSFR-GSMFs}, which map to lower \Zi and typically to higher \bhnsSingle yields.  

In total this leads to the \ac{SFRD}{\ensuremath{(Z_{\rm{i}},z)}\xspace} models   $\rm{xyz}=231$ and $\rm{xyz}=312$  having (one of)  the highest and lowest \bhnsSingle merger yields for each binary population synthesis model,    as can be seen in Figures~\ref{fig:IntrinsicRates} and \ref{fig:ObservedRates}. Our fiducial \ac{SFRD}{\ensuremath{(Z_{\rm{i}},z)}\xspace} model (000) also produces one of the lowest yields.   These three \SFRD models are highlighted in Figures~\ref{fig:IntrinsicRates}--\ref{fig:ObservedRatesNSdisrupted}.

\begin{figure*}
    \centering
\includegraphics[width=1\textwidth]{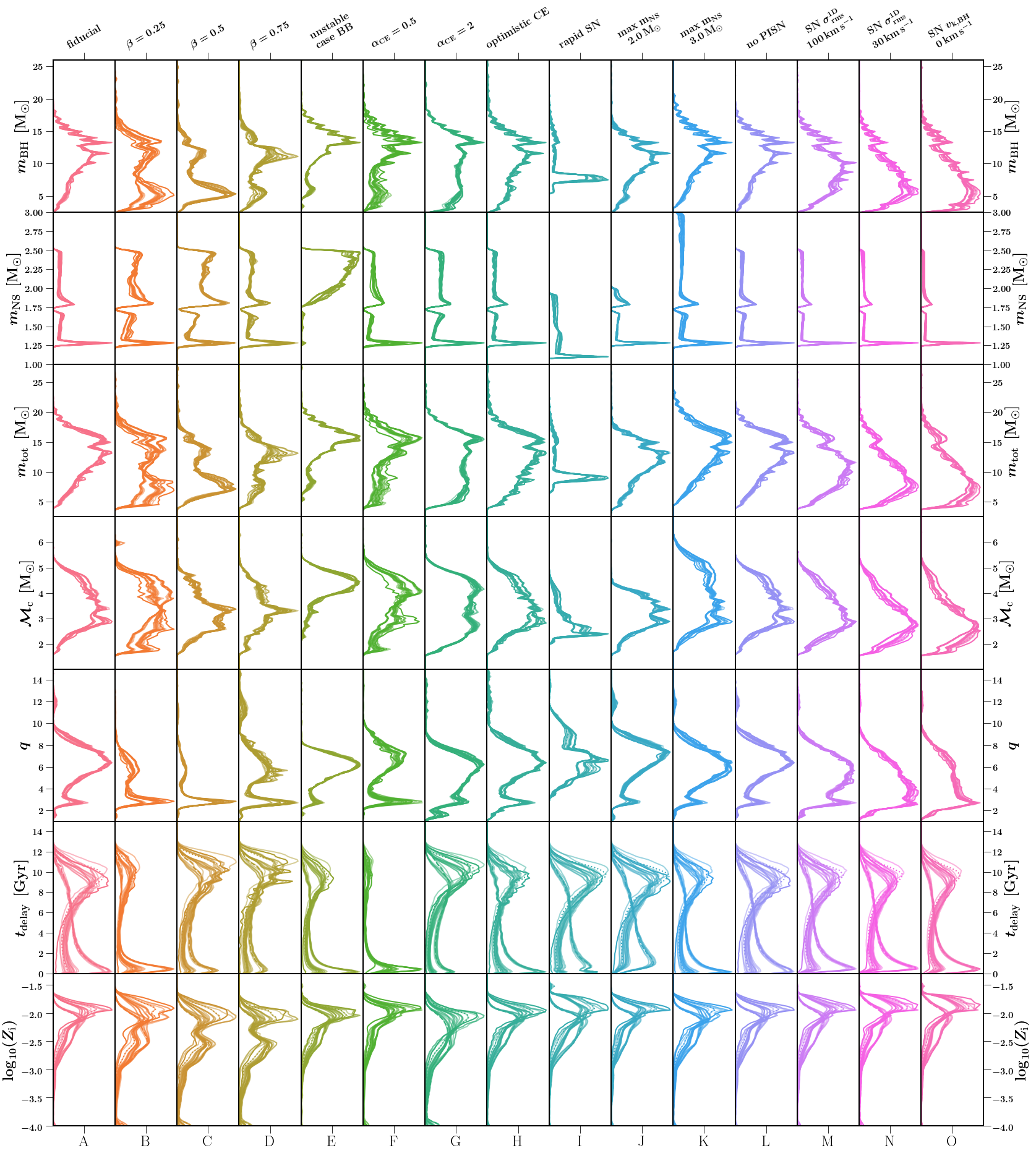} %
    \caption{Predicted normalized probability distribution functions (PDFs) for \bhnsSingle merger characteristics  for our \Nmodels model variations of  binary population synthesis and \SFRD assumptions (Tables~\ref{tab:variations-BPS} and~\ref{tab:MSSFR-variations-labels}). Each row shows, for the \NmodelsBPS  binary population synthesis variations (denoted with letters A, B, ..., O, and different colors), the PDFs  for the \NmodelsMSSFR \SFRD models in a subfigure using a   kernel density estimator. The PDF axis is plotted in linear scale. 
    All distributions are weighted by the detection probability for an observatory equivalent to the LVK network at design sensitivity. 
    For the kernel density functions we use the dimensionless  bandwidth. More details are given in \url{https://github.com/FloorBroekgaarden/BlackHole-NeutronStar}.  }
    \label{fig:Distributions_BHNS_kde}
\end{figure*}

\subsection{Predicted \bhnsSingle distribution functions for varying model assumptions}

\begin{figure*}
    \centering
\includegraphics[width=1\textwidth]{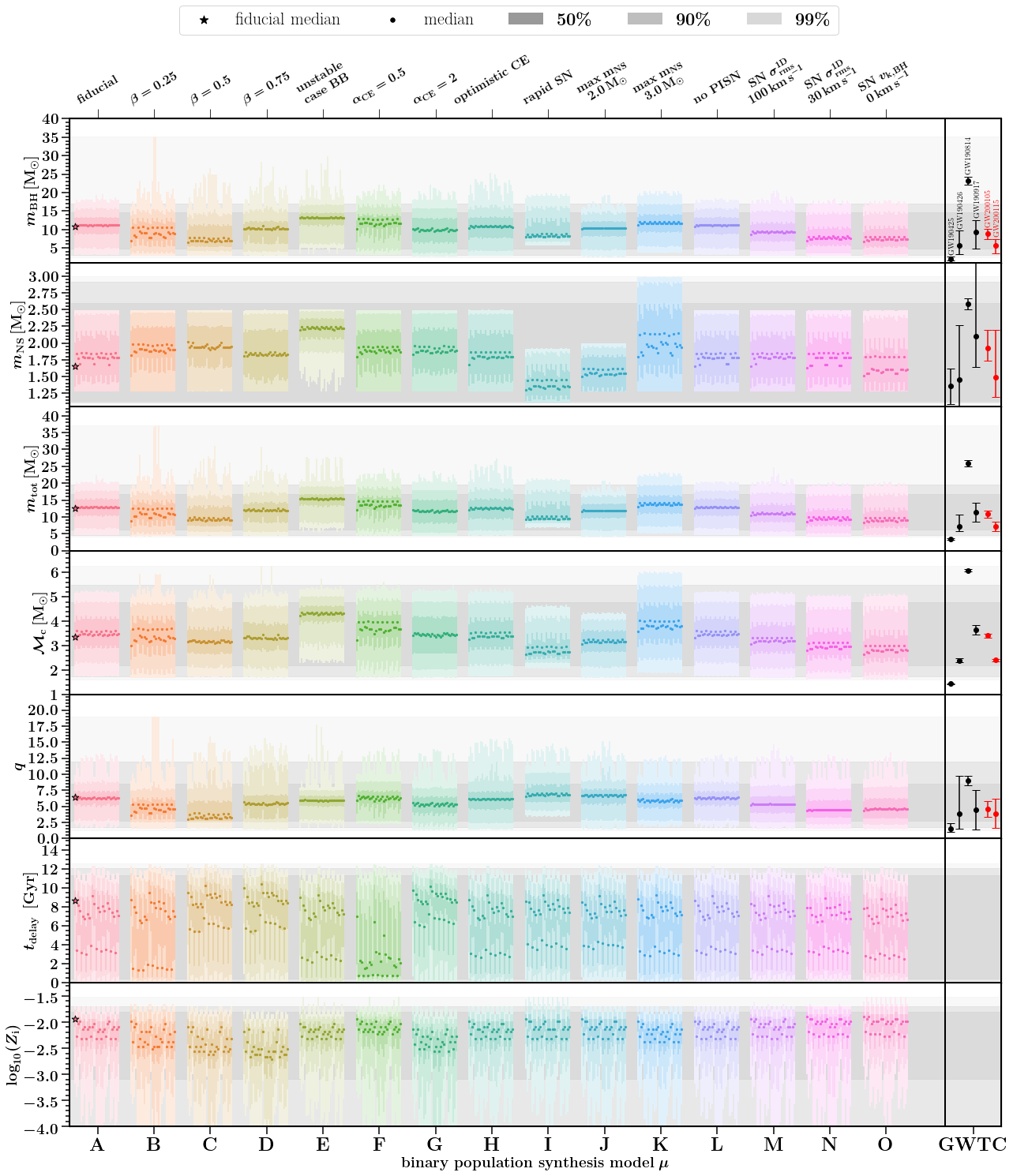} %
    \caption{Distribution quantiles for the predicted \bhnsSingle distributions for our \Nmodels model variations of  binary population synthesis and \SFRD assumptions (Table~\ref{tab:variations-BPS} and~\ref{tab:MSSFR-variations-labels}). Panels, colors and labels are as in Figure~\ref{fig:Distributions_BHNS_kde}. Each colored block, which corresponds to one sub-figure in Figure~\ref{fig:Distributions_BHNS_kde},  shows 28 bars, indicating the 28 \SFRD model variations.  The \SFRD labels of these bars are shown in   Figure~\ref{fig:BHNS-quantiles-ZOOM}.
     Each individual vertical bar representing one \SFRD model shows the median (scatter points), $50\%$ , $90\%$  and $99\%$  (see three shades) quantiles for the normalized probability distribution functions.    The fiducial model (\mAzero) median values are shown with star symbols.   The gray areas  in the background show the minimum and maximum values for the $50\%$, 90$\%$ and 99$\%$ distribution quantiles for all \Nmodels models in each panel.
     All distributions are weighted by the detection probability for an observatory equivalent to the LVK network at design sensitivity.
The rightmost columns show the median and $90\%$ credible intervals for the component masses, total mass, chirp mass and mass ratio of the confident \bhnsSingle detections GW200105 and GW200115 in red \citep{Abbott:2021-first-NSBH} and the possible  detected \ac{GW} sources GW190425, GW190426,  GW190814  and GW190917 in black (where a \bhnsSingle origin has not been ruled out yet;  \citealt{2020arXiv201014527A,GWTC2point1}).}
    \label{fig:ConfidenceINtervals_BHNS}
\end{figure*}

Besides the predicted rates (discussed in Section~\ref{subsec:results-variations-rates}), the shapes of the predicted \bhnsSingle distributions are also sensitive to the binary population synthesis  and \SFRD assumptions. To summarize and compare these effects, we discuss in this section the  shape of the \bhnsSingle merger distributions for our   \Nmodels model variations.  We show the  \ac{BH} and \ac{NS} masses (\mbhf and \mnsf), mass ratio ($\qf = \mbhf / \mnsf$), total mass ($\mtotf =\mbhf + \mnsf$) and chirp mass ($\mchirpf$) distributions as these are properties of the \bhnsSingle mergers that are generally inferred from \ac{GW} observations. In addition, we show the delay times, \tdelay and  birth metallicity, \Zi,  of the detected mergers.  The birth metallicity is highly correlated with the inspiral times, shown in previous sections, as \bhnsSingle mergers with longer inspiral times typically formed from stars with lower \Zi. Throughout this section we show normalized probability distribution functions of these \bhnsSingle characteristics,   to compare their shapes (the rates are given in Figure~\ref{fig:IntrinsicRates} and ~\ref{fig:ObservedRates}). The distributions are weighted for the  detection probability  obtained from Equation~\ref{eq:rate_detector} for a \ac{GW} detector equivalent to the LVK network at design sensitivity.

To summarize and compare the shape of the \bhnsSingle merger characteristics for \Nmodels model variations we  focus on two figures. First, Figure~\ref{fig:Distributions_BHNS_kde} shows the predicted \bhnsSingle probability distribution functions for all \Nmodels model variations. This figure visualizes the overall shape of the distributions.    
Second, Figure~\ref{fig:ConfidenceINtervals_BHNS}, shows  the median and the $50\%$, $90\%$ and $99\%$ distribution quantiles for all \Nmodels model variations. 

\subsubsection{BH mass}
\label{subsec:BH-mass-prediction}
 The top panel in Figure~\ref{fig:ConfidenceINtervals_BHNS} shows that for all our \Nmodels model variations  $90\%$ of the predicted observable \bhnsSingle mergers have  $\mbhf$ in the range $ 2.5$--$ 18\Msun$. We find that in all model combinations less than $5\%$ of the \bhnsSingle mergers are predicted to have \mbhf exceeding $18$\Msun. Moreover, most of the \Nmodels model variations predict $<1\%$ of  \ac{GW} detectable \bhnsSingle mergers have $\mbhf \gtrsim 20\Msun$, and only in models B, C, D and H this fraction is slightly above $1\%$ for some of the \SFRD models. This is especially clear in the probability distribution functions (PDFs) for the \ac{BH} mass shown in Figure~\ref{fig:Distributions_BHNS_kde}, where the region above $\mbhf \gtrsim 20\Msun$ is visibly suppressed for all models. Even the models that do not assume (pulsational)-\ac{PISN} (all models with $\mu= \rm{L}$), or that have \SFRD models that prefer low \Zi, lack a population of \bhnsSingle mergers with $\mbhf \gtrsim 20\Msun$. All in all, we conclude that such massive \acp{BH}  are  extremely rare  in \bhnsSingle mergers  in all our \Nmodels model variations. 
 Detecting more than 1$\%$ of   \bhnsSingle mergers with $\mbhf \gtrsim 20$\Msun   would therefore suggest that a large fraction of the \bhnsSingle systems  form through a different  formation channel than the isolated binary evolution channel, such as in AGN disks or through dynamical formation, which predict a significant fraction of  \bhnsSingle mergers with  \acp{BH} exceeding 20\Msun \citep[e.g.][]{2020MNRAS.497.1563R}.   Alternatively, it could constrain binary population synthesis and \SFRD models as basically all our models would be excluded as they do not support such observations.
 In \PII{} we show that this is very different compared to our results for \ac{BHBH} mergers, where typically our models predict that  $\gtrsim 50\%$ of  \ac{GW} detectable \ac{BHBH} mergers contain two \acp{BH} that both  have masses \mbhf exceeding  $18\Msun$.
 
 The \mbhf distributions for models with the  rapid \ac{SN} remnant mass model (I) sharply peak around $\mbhf \approx 8\Msun$ in Figure~\ref{fig:Distributions_BHNS_kde}. This is because the rapid prescription maps a large range of ZAMS masses to this \ac{BH} mass range \citep[see Figure 12 of][]{2012ApJ...749...91F}
 
The top panel  in Figure~\ref{fig:Distributions_BHNS_kde} and~\ref{fig:ConfidenceINtervals_BHNS}  also show that \bhnsSingle mergers with  $\mbhf \lesssim$5\Msun are  rare  in most of our \Nmodels models. In all models, except most of the \SFRD models in combination with  E (unstable case BB mass transfer) and all models with I (rapid SN remnant mass model) about $5\%$ of \bhnsSingle  are expected to have \ac{BH} masses below $5\Msun$. The sharp boundary in the minimum \ac{BH} mass for model I is caused by the rapid \ac{SN} remnant mass prescription, which assumes no \acp{BH} form with $\mbhf \leq 6\Msun$. Figure~\ref{fig:Distributions_BHNS_kde} and ~\ref{fig:ConfidenceINtervals_BHNS}  show that models including B, C, G, H, N and O have the largest fraction of \bhnsSingle mergers with $\mbhf \lesssim 5$\Msun.

\subsubsection{NS mass}
\label{subsec:NS-masses-predictions}
The $90\%$ confidence intervals of the \bhnsSingle \ac{NS} masses in the second row of Figure~\ref{fig:ConfidenceINtervals_BHNS} typically span the  range  $\mnsf \approx 1.25$--$2.5$\Msun. The exceptions of models where the $90\%$ quantiles shift significantly are   model E (assuming unstable case BB) where the range is  $\mnsf \approx 1.8$--$2.5$\Msun, model  I (the rapid SN remnant mass model) where it shifts to  $\mnsf \sim 1.1$--$1.9$, model J (assuming max $\mnsf=2\Msun$) where it shifts to $1.25$--$2$\Msun  and model K (assuming max $\mnsf=3\Msun$)  where it shifts to $1.25$--$3$\Msun.   Especially in Figure~\ref{fig:Distributions_BHNS_kde} it can be seen that the unstable case BB model highly favors \acp{NS} with $\mnsf \gtrsim 2\Msun$, which is a result from that the low mass \acp{NS} are suppressed as their low mass progenitor helium stars typically undergo a case BB mass transfer phase and merge before forming a \bhnsSingle system  in this model. In addition, it can  be seen that for the rapid \ac{SN} remnant mass model (I) the most probable \ac{NS} mass  in \bhnsSingle mergers is predicted to be $\lesssim 1.25$\Msun. 
In most model variations we find  that the median \ac{NS} mass  is  $\gtrsim 1.8\Msun$ and that except for models I and J, in all other models typically $\gtrsim 25\%$ of  \ac{GW} detectable \bhnsSingle mergers have \ac{NS} with $\mnsf \gtrsim 2\Msun$.  Such massive \acp{NS} are thus predicted to be common in \bhnsSingle mergers as can also be seen in the Figure~\ref{fig:Distributions_BHNS_kde} and~\ref{fig:ConfidenceINtervals_BHNS}.

\subsubsection{Total mass and chirp mass}
\label{subsec:total-and-chirp-masses-predictions}
The \bhnsSingle total mass distributions  follow the shape of the \ac{BH} mass distributions as the \ac{BH} typically dominates the mass of the \bhnsSingle. Figure~\ref{fig:ConfidenceINtervals_BHNS}  shows that in all \Nmodels model variations $90\%$ of the predicted \ac{GW} detectable \bhnsSingle mergers will have $\mtotf $ in the range  $\sim 5$--$20\Msun$ and hence that less than $5\%$ of \bhnsSingle mergers are predicted to have $\mtotf \gtrsim 20\Msun$. The total mass  peaks around $\mtotf \approx 15\Msun$ for models including our fiducial model (A), and has the highest average in the unstable case BB model (E),  as shown in Figure~\ref{fig:Distributions_BHNS_kde}, as in this model many of the binaries with low \ac{BH} mass progenitors merge during unstable case BB mass transfer. 
The total mass peaks as low as  $\mtotf \approx 7\Msun$ for models including model C ($\beta=0.5$) and models N and O with low \ac{SN} natal kicks. The latter is because low mass \acp{BH} can get moderate kicks in our delayed remnant mass prescription, disrupting the binary, whereas in models N and O  such low mass \acp{BH} are given low or zero natal kicks.

Figure~\ref{fig:ConfidenceINtervals_BHNS} shows that  $90\%$ of the predicted \ac{GW} detectable \bhnsSingle mergers will have  $\mchirpf  \approx 1.7$--$5.5\Msun$ and hence that less than $5\%$ of \bhnsSingle mergers are predicted to have $\mchirpf \gtrsim 5.5\Msun$.  The most probable chirp masses, as shown in  Figure~\ref{fig:Distributions_BHNS_kde}, are typically  in the range $\mchirpf \approx 2$--$5$\Msun.

\subsubsection{Mass ratio distribution}
\label{subsec:mass-ratio-predictions}
The $90\%$ quantiles for the \bhnsSingle mass ratio distributions lies for all models in  $\qf \approx 2$--$ 12$ as shown in Figure~\ref{fig:ConfidenceINtervals_BHNS}. The median and most probable \bhnsSingle \qf for our fiducial model (A), and most models, is $\qf \approx 6$ as shown in Figure~\ref{fig:Distributions_BHNS_kde}. For models B, C and D ($\beta = 0.25, 0.5, 0.75)$, however, this changes to a sharper peak around   $\qf \approx 3$, as in these models low mass \acp{BH} are more common.  Figure~\ref{fig:ConfidenceINtervals_BHNS} also shows  that in none of the models more than 5$\%$ of the \bhnsSingle mergers are predicted to have \qf $\gtrsim 12$.  In addition, Figure~\ref{fig:Distributions_BHNS_kde} and ~\ref{fig:ConfidenceINtervals_BHNS}  show that \bhnsSingle mergers with $\qf \lesssim 2$ are predicted to be  rare in all our \Nmodels model variations (less than $5\%$ of all \bhnsSingle mergers).

\subsubsection{Delay times}
Figures~\ref{fig:Distributions_BHNS_kde} and ~\ref{fig:ConfidenceINtervals_BHNS} show that the predicted delay time distributions have a wide spread of \tdelay, which typically peaks around 9\Gyr or 1\Gyr, where the peak depends mostly on the \SFRD model that is used. Figure~\ref{fig:ConfidenceINtervals_BHNS} shows that \SFRD models with relatively shorter median delay times (e.g. models with xy1), correspond to \SFRD models with low median metallicities.  This is because the  delay time of the detected \bhnsSingle mergers is a convolution between the intrinsic delay time distribution and the \SFRD. Intrinsically, most \bhnsSingle mergers have short delay times $\lesssim 1\Gyr$  (e.g. Figure~\ref{fig:BHNS_ObservableDistributions_per_metallicity}), and so the majority of \bhnsSingle systems has already merged before redshift $z=0$. In addition,  the   yield of \bhnsSingle mergers is higher at lower metallicities (i.e., around $\log_{10}(\Zi)\approx -2.5$), as shown in Figure~\ref{fig:BHNS_rate_per_metallicity}. As a result, \SFRD models with high median metallicities only form stars with low \Zi early in the Universe. Because of the high \bhnsSingle formation yield, these low metallicities dominate the rate, leading to the detected population having longer delay times (since the \bhnsSingle with short delay times from low metallicities will already have merged at high redshifts). On the other hand, \SFRD models with low median metallicities still form low \Zi binaries at lower redshifts. All in all, this results in \SFRD models with low median metallicities corresponding to shorter \tdelay values.

\subsubsection{Birth metallicities}
Figures~\ref{fig:Distributions_BHNS_kde} and ~\ref{fig:ConfidenceINtervals_BHNS} show that the predicted range and shape of the birth metallicity distribution of the detected \bhnsSingle mergers is strongly impacted by the \SFRD prescriptions. The scatter is mostly dominated by the  \ac{MZR} prescriptions  in combination with the  inspiral time distributions from the \bhnsSingle mergers. Models with relatively higher metallicities originate from \SFRD prescriptions that have the  \citet{2006ApJ...638L..63L} $+$ offset or \citet{2016MNRAS.456.2140M}  \ac{MZR} relation that, for a given galaxy mass, map to higher average initial metallicities.  
Figure~\ref{fig:Distributions_BHNS_kde} shows that typically the metallicities have values in the range $\Zi \approx  0.001$--$0.03$. The  highest values for the median of \Zi in Figure~\ref{fig:ConfidenceINtervals_BHNS} are \Zi $\sim 0.016$, whereas the lowest median values are \Zi  $\sim 0.0018$.

\subsubsection{Effect from variations in the binary population synthesis  and \SFRD  model assumptions on the shape of the \bhnsSingle distribution functions}
\label{subsec:Effect-from-variations-in-the-BPSmodels-on-PDFs}
We find that the variation in the predicted \bhnsSingle distribution functions is typically dominated by the binary population synthesis model assumptions when comparing with the effect from variations in our \SFRD models. This can  be seen in Figures~\ref{fig:Distributions_BHNS_kde} and \ref{fig:ConfidenceINtervals_BHNS}, that show that typically variations between models that use the same binary population synthesis model but different \SFRD are smaller compared to variations in the predicted \bhnsSingle merger distributions between different binary population synthesis models. The main exception are the \Zi distributions, which are mostly impacted by variations in \SFRD.  
Models where some of the probability density function values change with factors 2 or more over variations of \SFRD within the same binary population synthesis model
include $\mu = \rm{B}$ ($\beta=0.25$), $\mu = \rm{F}$ ($\alpha_{\rm{CE}}=0.5$), $\mu =  \rm{N}$ (reduced SN natal kicks of $\sigma_{\rm{rms}}^{\rm{1D}}=30\kms$) and $\mu =  \rm{O}$ (zero BH natal kicks).
This is particularly also visible in Figure~\ref{fig:ConfidenceINtervals_BHNS}, where for those binary population synthesis models the median changes the most over \SFRD variations. In all the figures the largest impact by \SFRD models comes from the \ac{MZR} assumption, which seems to be an outlier in our \SFRD models.

\subsubsection{Comparison with GW observations}
\label{sec:comparison-with-GW-Observations}
The right-most panels of Figure~\ref{fig:ConfidenceINtervals_BHNS} show the median and  $90\%$ credible intervals for the component masses, total mass, chirp mass and mass ratio of the two confident and four possible \bhnsSingle mergers from the third LIGO-Virgo-KAGRA observing run \citep{Abbott:2021-first-NSBH, 2020arXiv201014527A, GWTC2point1}.

GW200105 and GW200115 are the first two confident \bhnsSingle mergers recently detected by LIGO-Virgo \citep{Abbott:2021-first-NSBH}. The $90\%$ credible intervals of the properties of GW200105 are inferred to be $\mchirpf = 3.41^{+0.08}_{-0.07}$, $\mtotf = 10.9_{-1.2}^{+1.1}$\Msun, $\mbhf = 8.{9}_{-1.5}^{+1.2}$ $\Msun$, $\mnsf = 1.{9}_{-0.2}^{+0.3}$ $\Msun$ and $\qf = 0.22_{-0.04}^{+0.08}$. For GW200115 the properties are $\mchirpf = 2.42_{-0.07}^{+0.05}$\Msun,  $\mtotf = 7.1_{-1.4}^{+1.5}$\Msun,  $\mbhf = 5.{7}_{-2.1}^{+1.8}$ $\Msun$ and $\mnsf = 1.{5}_{-0.3}^{+0.7}$\Msun and $\qf \equiv (\mnsf / \mbhf) = 0.26_{-0.10}^{+0.35}$. Both GW200105 and GW200115 are consistent with the predicted distributions from our population synthesis models as shown in Figure~\ref{fig:ConfidenceINtervals_BHNS}. However, \bhnsSingle mergers with lower black hole masses, such as in GW200115, are more common in a subset of our models such as the simulations with $\beta = 0.25$ or $\beta = 0.5$ (models B and C) and/or  low \ac{SN} natal kicks (models N and O). 
In \citet{BroekgaardenBerger2021} we discuss the formation of GW200105 and GW200115 from isolated binary evolution in more detail and show that our models with low \ac{SN} kick can potentially explain their formation as well as simultaneously  match the inferred \bhnsSingle, \ac{BHBH} and \ac{NSNS} merger rate densities.

GW190425 is the  system of the possible \bhnsSingle detections with the lowest total mass and is most likely a \ac{NSNS} merger. However,  a \bhnsSingle has not been ruled out. If it is a \bhnsSingle system, the component masses are $\mbhf=2.0_{-0.3}^{+0.6}$ and $\mnsf = {1.4}_{-0.3}^{+0.3}$, and \mchirpf, \mtotf and \qf are as shown with errorbars in Figure~\ref{fig:ConfidenceINtervals_BHNS}. The inferred  credible intervals of GW190425, particularly those for the \mbhf, \mtotf and \mchirpf, fall outside of the typical parameter values predicted by our models, which might indicate this is likely not a (typical)  \bhnsSingle system. 

GW190426 (short for GW190426$\_$152155) is a candidate event with a relatively high FAR of 1.4 \yearmin \citep{2020arXiv201014527A}, making it a less significant detection compared to the two events mentioned above. If this event is real, we find that its inferred parameter credible intervals match well with the predicted \bhnsSingle distributions for all shown parameters in  Figure~\ref{fig:ConfidenceINtervals_BHNS}.

GW190814 has been interpreted as a likely \ac{BHBH} merger since its lower component mass is $2.59_{-0.009}^{+0.008}$\Msun. However, a \bhnsSingle interpretation cannot be ruled out \citep{2020ApJ...896L..44A}. We find that although our model K (max $\mnsf = 3.0\Msun$) can produce such a massive \ac{NS}, the other inferred credible intervals for GW190814, such as \mtotf, \mchirpf and \mbhf,  do not match well with typical \bhnsSingle distributions predicted by any of our models. These findings are conform \citet{2020ApJ...899L...1Z} who also discuss the challenges in forming this system through isolated binary evolution.

GW190917 is a marginal \bhnsSingle event from the GWTC-2.1 catalog \citep{GWTC2point1}. If real, this \bhnsSingle merger has inferred source parameters of $\mchirpf = 3.7^{+0.2}_{-0.2}$\Msun, $\mbhf =  9.3^{+3.4}_{-4.4}$\Msun, $\mnsf = 2.1^{+1.5}_{-0.5}$\Msun and $\qf = 0.23^{+0.52}_{-0.09}$, very similar to GW200105, that are consistent with our predicted populations.


\section{Discussion }
\label{sec:discussion}

\subsection{Uncertainties beyond our models}
We have modeled the rate and properties of \bhnsSingle mergers and demonstrated that uncertainties from both the massive (binary) star evolution and \SFRD significantly impact the predicted results. However, even with our large set of \Nmodels model variations, there are still additional uncertainties in the modelling that should be added and further explored in future work. We discuss the most important ones below.

\subsubsection{SN remnant mass functions}
\label{subsec:discussion-delayed-vs-rapid-SN-remnant-mass}
Figure~\ref{fig:Distributions_BHNS_kde} showed that model I, which uses the \textit{rapid} SN remnant mass prescription from \citet{2012ApJ...749...91F}, is the model variation that resulted in one of the most drastic changes in the predicted  \bhnsSingle distributions. This makes the remnant mass prescription  one of the key assumptions in modelling \bhnsSingle merger properties. Many binary population synthesis studies use the \textit{rapid} SN prescription instead of our default delayed SN remnant prescription \citep[e.g.][]{2019IAUS..346..417K,2020A&A...636A.104B}. The {rapid} prescription assumes \ac{SN}  explosions occur within 250 ms (compared to longer timescales assumed for the delayed prescription) and reproduces, by construction,  the  lower remnant mass gap between NSs and BHs that has been inferred from  X-ray binary observations 
\citep{1998ApJ...499..367B, 2010ApJ...725.1918O, 2011APS..APRH11002F} by a lack of BHs in the mass range between about the heaviest neutron stars  $\sim$ 3\Msun \citep{2016ARA&A..54..401O, 2017ApJ...850L..19M, 2020NatAs...4...72C,2020CQGra..37d5006A,,2020arXiv200106102S} and  BHs of $\sim$ 6\Msun.

	It is still under debate whether this lower \ac{BH} mass gap exists. Models of \acp{SN}  predict a gap \citep{2014ApJ...785...28K,2015ApJ...801...90P} and no  gap (e.g. \citealt{1995ApJS..101..181W,2001ApJ...554..548F,2020ApJ...890...51E, 2020MNRAS.495.3751C});  see also \citet{2020MNRAS.491.2715B} who point out that it is  challenging to model remnant masses, and that the compactness measure that is often used in these \ac{SN}  remnant mass studies is not a good metric for \ac{SN} explodability. 
	Moreover, \citet{2012ApJ...757...36K}   argue that the apparent  mass gap  in X-ray binaries could be caused by systematic offsets in the \ac{BH} mass measurements.  
	More recently, \citet{2019Sci...366..637T} found evidence for a $3.3_{-0.7}^{+2.8}$\Msun mass \ac{BH} observed in a non-interacting binary with a red giant, adding to earlier speculation of a  $\sim 2.44\pm0.27$\Msun \ac{BH} \citep{2002A&A...392..909C,2010AAS...21541905R,2011arXiv1107.1537D}. 
	In addition,  \citet{2016MNRAS.458.3012W,2020A&A...636A..20W} studied microlensing events of BHs based on  observations from OGLE-III and Gaia data release 2.  
	They find evidence for a continuous distribution of stellar remnant masses and disfavor a mass gap between NSs and BHs (unless BHs receive natal kicks above 20-80\kms).   
	In addition, the LIGO-Virgo Collaboration reported in the second \ac{GW} catalog three \ac{GW} events,  GW190425, GW190814 and GW190426,  with at least one component possibly in the lower mass gap \citep{2020arXiv201014527A}.	\citep{2020ApJ...899L...1Z} show that particularly the formation of GW190814 might require the absence of the lower mass gap between \acp{NS} and \acp{BH}. 
	Moreover, \citet{2018MNRAS.481.4009V} show that the delayed prescription better matches the  observed distribution of masses of Galactic double \ac{NS} systems.

In practice, both the delayed and rapid \ac{SN} models might not represent the \ac{SN} remnant mass distribution and future work should, therefore, explore other alternatives \citep[cf.][]{2020arXiv201202274R,2021arXiv210612381V,2021MNRAS.500.1380M}. Examples include the \citet{2016MNRAS.460..742M} prescription used in \citet{2018MNRAS.481.4009V}, the remnant mass function studied in \citet{2020arXiv201202274R} and the  probabilistic remnant mass function from  \citet{2020MNRAS.499.3214M}.

\subsubsection{Other binary population synthesis variations}
Besides the remnant mass prescription, there are many additional uncertainties from binary population synthesis prescriptions that should be explored in future work. Examples which we think could significantly impact the rate and characteristics of \bhnsSingle mergers include the wind prescription,      exploring different models and assumptions for \ac{ECSN} and \ac{USSN}, exploring alternative models for the \ac{SN} natal kick magnitude and using different prescriptions for the \ac{CE} evolution \citep[e.g.][]{2020A&A...638A..55K, 2020PASA...37...38V, 2021arXiv210205649O}. 
Beyond this, we also note that instead of varying one population parameter at a time, future work should do a more robust exploration of the full parameter space that includes changing parameters simultaneously \citep[e.g.][]{2018MNRAS.477.4685B}.
In addition, future studies should incorporate more detailed stellar evolution tracks such as discussed in e.g.  \citet[][]{2020MNRAS.497.4549A, 2018MNRAS.481.1908K} and \citet{2020arXiv201016333B}.

\subsubsection{Metallicity-specific star formation rate density}
\label{subsec:disc-MSSFR-assumptions}
As discussed in Section~\ref{subsec:MSSFR-variations}, in this work we used analytical prescriptions, commonly applied in binary population synthesis studies, to construct \NmodelsMSSFR \SFRD models. This allows us to discuss the variation in the modeled properties of BHNS population caused by the different literature \SFRD assumptions.
We note that to properly evaluate the uncertainty of the modelled properties of the BHNS population, one should cover the full range of \SFRD assumptions allowed by observations. Establishing this range presents a challenge in itself \citep{2019MNRAS.488.5300C} and is beyond the scope of this study.
Furthermore, prescriptions introduced in Section~\ref{subsec:MSSFR-variations} rely on important simplifications and do not reproduce all characteristics of the \SFRD expected from observations (see e.g. discussion in \citealt{2019MNRAS.488.5300C}). For instance, our fiducial \SFRD (following the phenomenological model from  \citet{2019MNRAS.490.3740N} by construction produces a log-normal metallicity distribution, which is not supported by observations \citep[e.g.][]{2019MNRAS.488.5300C,2020arXiv201202800B} and does not reproduce the extended low metallicity tail of the distribution.
Those issues could be circumvented with more detailed approaches, for instance employing cosmological simulations to construct a \SFRD \citep[e.g.][]{2010ApJ...716..615O, 2017MNRAS.472.2422M,2018MNRAS.479.4391M,2018MNRAS.480.2704L, 2019MNRAS.487.1675A} or combining a wide range of recent observational results describing the properties of star forming galaxies at different redshifts    \citep{2019MNRAS.488.5300C,2019ApJ...881..157B,2020arXiv201202800B}.

\subsubsection{Initial conditions}
We  assumed fixed initial distributions  for, e.g.  the initial masses and separations. In practice these distributions are uncertain \citep[e.g.][]{2018Sci...359...69S} and may  be metallicity or redshift dependent \citep[e.g.][]{2018ApJ...855...20G}, which is not taken into account here. 
Varying the uncertain initial conditions can impact the results for the predicted rates and shapes of \bhnsSingle merger distributions. Future work should explore this impact similar to studies such as \citet{2015ApJ...814...58D}, and use correlated initial distribution function as described in work including  \citet{2015ApJ...814...58D, 2017ApJS..230...15M} and \citet{2018A&A...619A..77K}.

\section{Conclusions}
\label{sec:conclusions}
In this study we made predictions for the intrinsic and \ac{GW} detectable rate and characteristics of  \bhnsSingle mergers for ground-based \ac{GW} detectors equivalent to the LVK network at design sensitivity. We accomplish this by simulating populations of binaries over a grid of 53 metallicities and convolving this with a  \SFRD  and detection probability. We explore uncertainties arising from both assumptions in massive binary star evolution and the \SFRD, and present the \bhnsSingle rates and characteristics for a total of \Nmodels variations (\NmodelsBPS binary population synthesis and \NmodelsMSSFR \SFRD variations). Our main findings are summarized below.

\subsection{Predictions for \bhnsSingle mergers from our fiducial model}

\begin{itemize}
\item Our fiducial model \mAzero predicts an intrinsic  \bhnsSingle merger rate at redshift zero of $\RateIntrinsicZero \approx \RateIntrinsicAzeroBHNS $\GpcminThree \yearmin, consistent with the inferred merger rate from the first two \bhnsSingle observations by the LIGO/Virgo  network \citep{Abbott:2021-first-NSBH}. We find a \ac{GW} detection rate of $\RateObserved \approx \RateObservedAzeroBHNS $\yearmin for an LVK network at design sensitivity. \\

\item In Section~\ref{subsec:fiducial-formation-channels-GWs} we show that our fiducial model predicts that \PercentageClassicLVK of  the \ac{GW} detectable \bhnsSingle mergers formed through the \textit{classic} formation channel within isolated binary evolution (shown in Figure~\ref{fig:formation-channels-sketch}). This channel involves  first a dynamically stable mass transfer episode, and eventually a reverse dynamically unstable (common-envelope) mass transfer phase. \ac{GW} observations of \bhnsSingle mergers will therefore provide a clean probe of the classic formation channel (whereas \ac{BHBH} and \ac{NSNS} detections might be dominated by different formation channels within isolated binary evolution). \\

\item In Figure~\ref{fig:BHNS_DCO_observed} we present the characteristic distributions for detectable \bhnsSingle mergers for our fiducial model \mAzero. This model predicts that \bhnsSingle mergers observable with the LVK network at design sensitivity typically have \ac{BH} masses of $2.5\lesssim \mbhf / \Msun \lesssim 16$,  with fewer than $5\%$ of \bhnsSingle mergers having $\mbhf \gtrsim 15\Msun$. We also find that $\approx 60\%$ of the \ac{NS} masses are $\mnsf \gtrsim 1.5\Msun$. In addition, we find for \mAzero that the mass ratio is typically in the range $2 \lesssim \qf \lesssim 10$ ($\qf \gtrsim 2$ for $\gtrsim 98\%$ of detectable \bhnsSingle mergers). The \bhnsSingle merger total mass and chirp mass for our fiducial model are predicted to lie in the ranges $5 \lesssim \mtotf / \Msun \lesssim 20$ and $1.5 \lesssim \mchirpf / \Msun \lesssim 5.5$.  The inspiral times for the majority of detected \bhnsSingle mergers are predicted to fall in the range $2 \lesssim \tinspiral / \Gyr \lesssim 12$ in model \mAzero. \\

\item Our fiducial model predicts that $f_{\rm{EM}} \approx 1-28\%$ of the \ac{GW} detected \bhnsSingle mergers will have a disruption of the \ac{NS} outside of the \ac{BH}'s innermost stable circular orbit (Table~\ref{tab:fiducial_BHNS_EM_ejecta_observed}). Such systems are expected to produce electromagnetic  counterparts. 
\end{itemize}

\subsection{Varying binary population synthesis and metallicity-specific star formation rate models}

In Section~\ref{sec:results-variations} we investigate how the predictions for the \bhnsSingle characteristics change over our \Nmodels model variations in binary population synthesis  and \SFRD prescriptions (summarized in Table~\ref{tab:variations-BPS} and~\ref{tab:MSSFR-variations-labels}). 

\subsubsection{Predicted rates}

\begin{itemize}
\item The predicted merger rate is uncertain, spanning two orders of magnitude across our range of assumptions, where both binary evolution model assumptions and \SFRD variations impact the predicted merger rates at the order-of-magnitude level.  We show our predicted intrinsic \bhnsSingle merger rates in Figure~\ref{fig:IntrinsicRates}; these \Nmodels rates span the range $\rate_{\rm{m}}^0 \approx$ \RateIntrinsicAzeroBHNSmin--\RateIntrinsicAzeroBHNSmax \GpcminThree \yearmin.
In Figure~\ref{fig:ObservedRates} we show the rates of detectable events, which are in the range  $\RateObserved \approx  \RateObservedAzeroBHNSmin$--$\RateObservedAzeroBHNSmax$\yearmin for an LVK network \\

\item In Figure~\ref{fig:ObservedRates-formation-channels} we show that in the majority of our models $\gtrsim 50\%$ of the \ac{GW} detectable \bhnsSingle mergers forms through the classic formation channel. However, we also show that, particularly when changing the mass transfer efficiency in models B, C and D, the contribution of the classic channel is reduced to a few percent and that other channels dominate the formation of \bhnsSingle mergers, including the only stable mass transfer channel (II) and single-\ac{CE} as first mass transfer channel (III). We find that the double-core CE channel, which has been found to be important for the formation of NSNS mergers, does not significantly contribute to the formation of \bhnsSingle mergers in any of our \Nmodels models.  \\

\item We present in Figure~\ref{fig:ObservedRatesNSBH} the predicted fraction of \ac{GW} detectable \bhnsSingle mergers where the NS forms first. We find that this fraction lies in the range $f_{\text{NS--BH}} \approx 0-20\%$, and that the fraction is mostly impacted by our mass transfer and \ac{CE} assumptions. This fraction could be possibly inferred from the spin distribution of \ac{GW} detections and the rate of occurrence of pulsar--BH systems. \\

\item We present for our \Nmodels model variations  the predicted fraction of \ac{GW} observable \bhnsSingle mergers where the \ac{NS} is disrupted outside of the \ac{BH} innermost-stable orbit in Figure~\ref{fig:ObservedRatesNSdisrupted}, and find that this fraction lies in the range $f_{\rm{EM}} \approx 0-70\%$. The lowest fractions are from models assuming small \ac{NS} radii and low \ac{BH} spins and from models assuming unstable case BB mass transfer or using the rapid \ac{SN} remnant mass prescription. We find that in most of our \Nmodels models the fraction lies above $1\%$ even for our most pessimistic assumptions for the \ac{NS} radii and \ac{BH} spins.  \\

\end{itemize}

\subsubsection{Predicted shapes of the \bhnsSingle property distributions}
In Figure~\ref{fig:Distributions_BHNS_kde} and \ref{fig:ConfidenceINtervals_BHNS} we present the predicted shapes of the probability distribution functions for the population of \bhnsSingle mergers that will be detected with design-sensitivity advanced \ac{GW} instruments. We present results for \bhnsSingle merger characteristics including the \ac{BH} and \ac{NS} masses (\mbhf, \mnsf), the total mass (\mtotf), the chirp mass (\mchirpf), the mass ratio ($\qf = \mbhf / \mnsf$), delay time (\tdelay) and the initial metallicity (\Zi).  We find that, except for the delay time and initial metallicity distributions, the shapes are predominantly impacted by the binary evolution variations and that the impact from variations in \SFRD are typically minor. 

\begin{itemize}
    \item In all our \Nmodels models, fewer than $5\%$ of the \acp{BH} in \bhnsSingle mergers are predicted to have $\mbhf \gtrsim 18\Msun$, and typically $\lesssim 1\%$ of the \bhnsSingle merges are predicted to have $\mbhf \gtrsim 20\Msun$ (Section~\ref{subsec:BH-mass-prediction}). Thus, our model variations do not commonly form \ac{GW} events similar to GW190814 which is inferred to have $\mbhf \gtrsim 20\Msun$. \\
    \item  In Section~\ref{subsec:NS-masses-predictions} we show that all models except I and J (which assume the `rapid' \ac{SN} remnant mass function and a maximum \ac{NS} mass of $\mnsf=2\Msun$, respectively) predict median \ac{NS} masses of $\gtrsim 1.8\Msun$, with $\gtrsim 25\%$ of the detectable \bhnsSingle mergers having $\mnsf > 2\Msun$, making such massive \acp{NS} common in observations of \bhnsSingle mergers. \\
    
    \item  We find in Section~\ref{subsec:total-and-chirp-masses-predictions} that in all \Nmodels models, $\gtrsim 90\%$ of the detectable \bhnsSingle mergers are predicted to have total masses in the range $\mtotf \approx 5-20\Msun$ and chirp masses in the range $\mchirpf  \approx 1.7-5.5\Msun$. Less than $5\%$ of \bhnsSingle mergers are predicted to have $\mtotf \gtrsim 20\Msun$ or $\mchirpf \gtrsim 5.5\Msun$ in any of our models. \\
    
    \item In Section~\ref{subsec:mass-ratio-predictions}  we show that in all \Nmodels models,  $\gtrsim 90\%$ of the detectable \bhnsSingle mergers are predicted to have mass ratios in the range $\qf \approx 2-12$ and that less than $5\%$ of \ac{GW} detected \bhnsSingle mergers are predicted to have $\qf \gtrsim 12$. We also find that \bhnsSingle mergers with $\qf \lesssim 2$ are predicted to be  rare (less than $5\%$ of all \bhnsSingle mergers) in all our simulations.  \\
\end{itemize}

To summarize, we find that the rate and the shape of the distributions of \ac{GW} detectable  \bhnsSingle mergers are significantly impacted by variations in binary evolution and \SFRD.  Future \bhnsSingle observations can, therefore, help to constrain our models. On the other hand, several predictions above are robust among our \Nmodels variations. If future \ac{GW} observations of \bhnsSingle mergers support distributions that violate these  predictions, it will mean that either all our \Nmodels models miss some of the underlying physics necessary to correctly model \bhnsSingle mergers (as discussed in Section~\ref{sec:discussion}) or that \bhnsSingle mergers predominantly form through a different formation channel than the isolated binary evolution studied in this paper. Future \ac{GW} observations and modelling, particularly simultaneously with constraints from NSNS and BHBH observations and electromagnetic observations, will aid in exploring massive binary star evolution and the metallicity specific star formation history over cosmic time.

\section*{Acknowledgements}
The authors thank Gus Beane, Christopher Berry, Charlie Conroy, Rosanne Di Stefano,  Douglas Finkbeiner,  Chris Fryer, Sebastian Gomez, Griffin Hosseinzadeh,  Lokesh Khandawal,  Floris Kummer, Joel Leja, Mathieu Renzo, Carl Rodriguez, Lieke van Son, Mario Spera,  Ashley Villar, Serena Vinciguerra, Thomas Wagg and  Reinhold Willcox  for helpful discussions and input on this manuscript. FSB thanks Munazza Alam, Marisa Borreggine, Victoria diTomasso, Miranda Eiben,  Claire Lamman, Locke Patton and Christina Richey for crucial input during this project.  The authors also thank everyone in the COMPAS collaboration and Berger Time-Domain Group for help. In addition, the authors thank David Stops and the Harvard FAS research computing group for technical support on the simulations and high performance computing part of the research.   Lastly, FSB wants to acknowledge the  amount of serendipity and privilege that was involved to end up pursuing this astronomy research. The Berger Time-Domain Group is supported in part by NSF grant AST-1714498 and NASA grant NNX15AE50G. Some of the authors are supported by the Australian Research Council Centre of Excellence for Gravitational Wave Discovery (OzGrav), through project number CE170100004.
FSB is supported in part by the Prins Bernard Cultuurfonds studiebeurs 2020. 
IM is a recipient of the Australian Research Council Future Fellowship FT190100574.
A.V-G. acknowledges funding support by the Danish National Research Foundation (DNRF132). SJ and SdM acknowledge funding from the Netherlands Organisation for Scientific Research (NWO), as part of the Vidi research program BinWaves (project number 639.042.728).

\section*{Software} 
Simulations in this paper made use of the COMPAS rapid binary population synthesis code, which is freely available at \url{http://github.com/TeamCOMPAS/COMPAS} \citep{stevenson2017formation, 2018MNRAS.477.4685B, 2018MNRAS.481.4009V,2019MNRAS.490.5228B}. The simulations performed in this work were simulated with a COMPAS version that predates the publicly available code. Our version of the code is most similar to version 02.13.01 of the publicly available COMPAS code. Requests for the original code can be made to the lead author. The authors used {\sc{STROOPWAFEL}} from \citep{2019MNRAS.490.5228B}, publicly available at \url{https://github.com/FloorBroekgaarden/STROOPWAFEL}\footnote{For the latest pip installable version of STROOPWAFEL please contact Floor Broekgaarden.}.
The authors also made use of Python from the  Python Software Foundation. Python Language Reference, version 3.6. Available at \url{http://www.python.org} \citep{CS-R9526}. In addition the following Python packages were used: matplotlib \citep{2007CSE.....9...90H},  {NumPy} \citep{2020NumPy-Array}, SciPy \citep{2020SciPy-NMeth}, \texttt{ipython$/$jupyter} \citep{2007CSE.....9c..21P, kluyver2016jupyter}, pandas \citep{mckinney-proc-scipy-2010}, seaborn \citep{waskom2020seaborn},  \citep{2018AJ....156..123A}. This paper also made use of  the hdf5 library for Python, available at  \url{https://docs.h5py.org/en/stable/}  \citep{collette_python_hdf5_2014,}. 
This research has made use of NASA’s Astrophysics Data System Bibliographic Services. We also made use of the computational facilities from the FAS Research Computing and Birmingham computer cluster. 
This research has made use of data, software and/or web tools obtained from the \ac{GW} Open Science Center (\url{https://www.gw-openscience.org}), a service of LIGO Laboratory, the LIGO Scientific Collaboration and the Virgo Collaboration. LIGO is funded by the U.S. National Science Foundation. Virgo is funded by the French Centre National de Recherche Scientifique (CNRS), the Italian Istituto Nazionale della Fisica Nucleare (INFN) and the Dutch Nikhef, with contributions by Polish and Hungarian institutes.

%
%

\section*{Data Availability}
All data produced in this study are publicly available on Zenodo at doi \url{10.5281/zenodo.4574727} through the link \url{https://doi.org/10.5281/zenodo.4574727}.
 All code, scripts and files to reproduce all the figures, results and supplementary material from the paper are publicly available through the main author's Github repository \url{https://github.com/FloorBroekgaarden/BlackHole-NeutronStar} or at doi \url{https://doi.org/10.5281/zenodo.5555970}.


\bibliographystyle{mnras}
\bibliography{BHNS-MSSFR} 



\appendix

\section{Calculating the cosmological merger rates}
\label{sec:app-calculating-merger-distributionsMSSFR}

In practice we approximate the integral in Equation~\ref{eq:rate_detector}  with the Monte Carlo estimate
%
\begin{align}
	%
	&\rate_{\rm{det}}  =  
	\sum_{z_{{\rm{m}}}^j} \sum_{\Zi^k} \, 
	 \left( 
	 	\int_0^{\tmerger^j} \, 
	 	 \diff \tdelay \, 
		 \rate_{{\rm{form}}}(\Zi^k, \tdelay)  \, {\rm{SFRD}}(\Zi^k, z_{\rm{form}})   
	 \right) \times \notag \\ 
	  &\hspace{1cm}\frac{\Dc^2(z_{{\rm{m}}}^j) }{E(z_{{\rm{m}}}^j) }
	 \frac{1}{1 + z_{{\rm{m}}}^j}
 	 \frac{4 \pi c}{\Hubble} \
	  P_{{\rm{det}} }(\monef, \mtwof,z_{{\rm{m}}}^j) \,
	  \Delta z_{{\rm{m}}}^j \,
	  \Delta  Z^k, 
    \label{eq:full-equation-detectable-rate}
\end{align}

where we sum over 100 equally divided  redshift bins $z_{{\rm{m}}}^j \in [0, z_{\rm{max}}]$ where we choose  $z_{\rm{max}} = 6$, which is a  conservative upper limit for the maximum redshift out to which \acp{DCO} are detectable with a LVK \ac{GW} network at design sensitivity\footnote{$z=0.50$ is equal to a luminosity distance of $D_{\rm{L}}\approx 3 \cdot 10^3$\Mpc.}, see Figure~3 of \citealt[][]{2016PhRvD..93k2004M}, the last columns of Table 3 and 4 of \citet{2020A&A...636A.104B} and \citealt{2018LRR....21....3A}. 
We also sum over our 53 metallicity bins $\Zi^k$  distributed approximately log-uniform in the range 0.0001--0.03, corresponding to the \Zi range in \citet{2000MNRAS.315..543H}. The metallicity grid points are shown with black scatter points in Figure~\ref{fig:BHNS_rate_per_metallicity}. 
 We use $\tmerger^j$ as short hand notation for $\tmerger(z_{{\rm{m}}}^j)$, the merger time at redshift $z_{{\rm{m}}}^j$, and also write 	the short hand notation  $ \Delta z_{{\rm{m}}}^j  = (z_{{\rm{m}}}^{j+1} - z_{{\rm{m}}}^j)$ and  $\Delta  \Zi^k = (\Zi^{k+1} - \Zi^{k})$. 
We also use $z_{\rm{form}}$ as a short hand notation for $z(\tform = \tdelay - \tmerger^j)$.  
\Dc and $E$ are functions of redshift given in \citet{1999astro.ph..5116H}  and  $\Hubble$ is the Hubble constant at redshift $z=0$.

\section{metallicity-specific star formation rate prescriptions}
\label{sec:app-SFRD-figures}

\begin{figure*}
    \centering
        \subfloat[][The four different star formation rate density (\ac{SFRD}) prescriptions  studied in this work.\label{fig:MSSFR-SFRs}]{{\includegraphics[width=1\columnwidth]{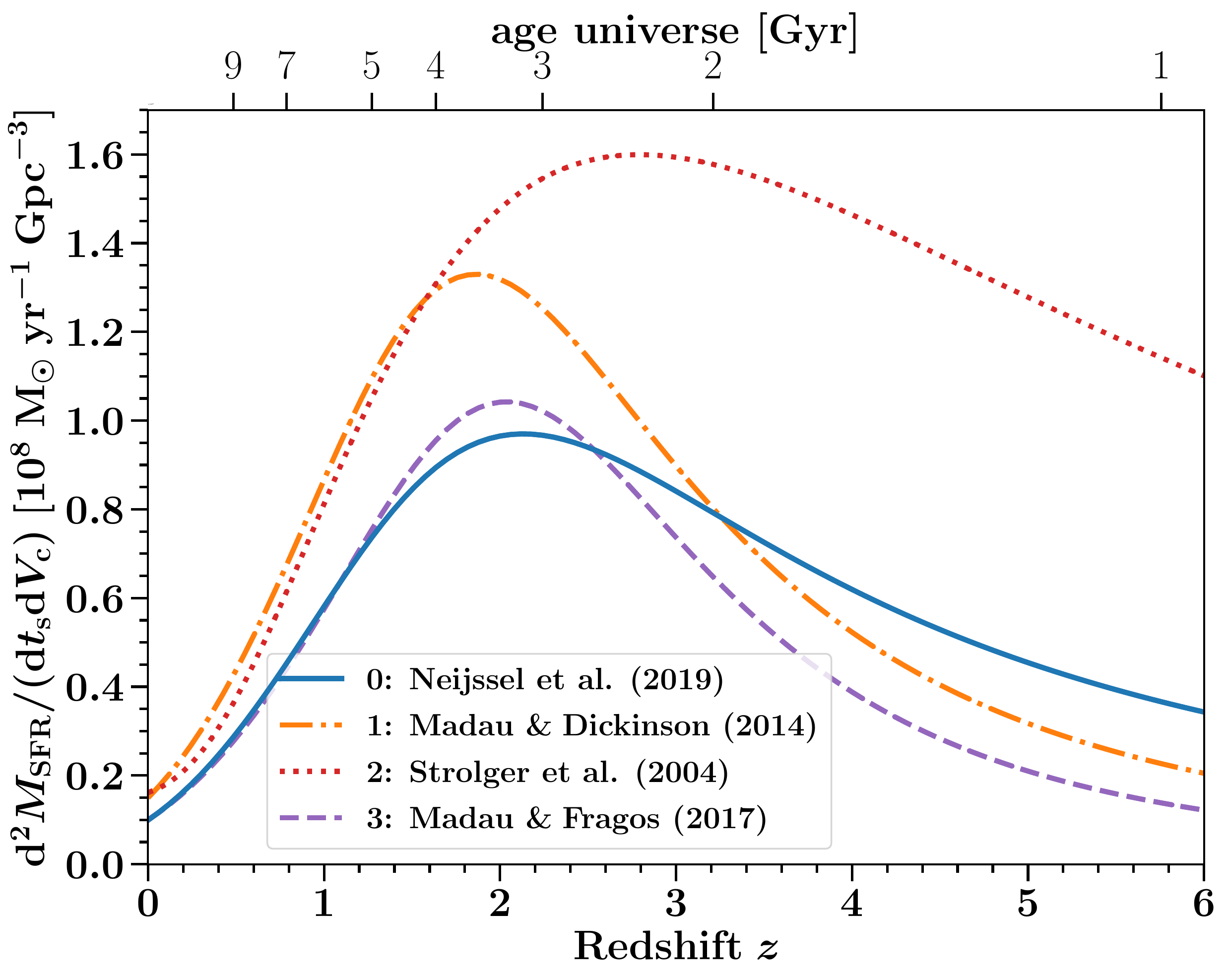} }}%
        \qquad
        \subfloat[][{Three examples of  metallicity probability distribution functions  ($\diff P / \diff \Zi $) used in this work.  For each of the prescriptions the metallicity distribution is shown for four different redshifts. }\label{fig:MSSFR-Z-PDFs}]{{\includegraphics[width=1\columnwidth]{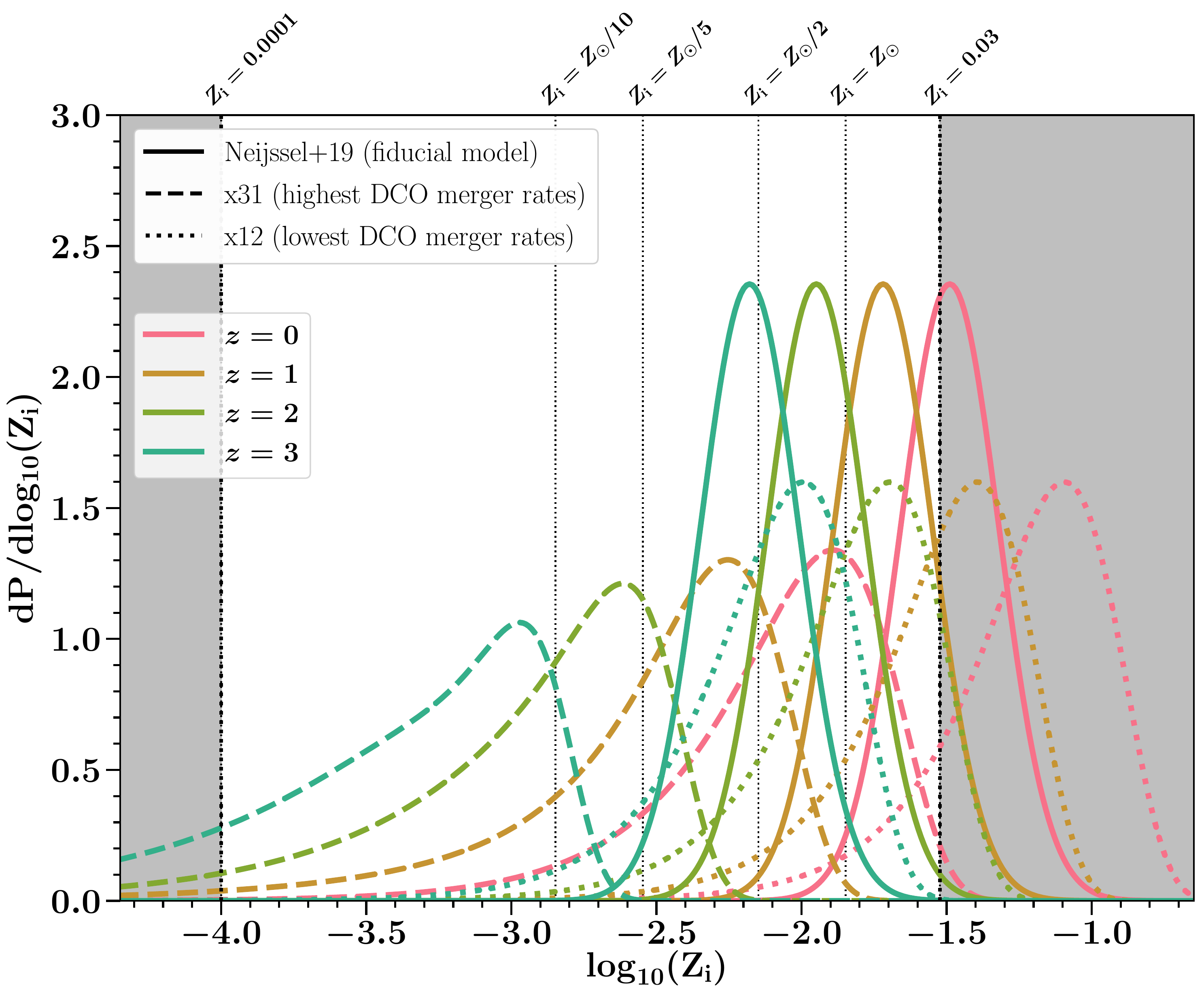} }}%
      \qquad
        \subfloat[][{The three different galaxy stellar mass function ({GSMF}) prescriptions studied in this work shown at seven different redshifts. }\label{fig:MSSFR-GSMFs}]{{\includegraphics[width=1\columnwidth]{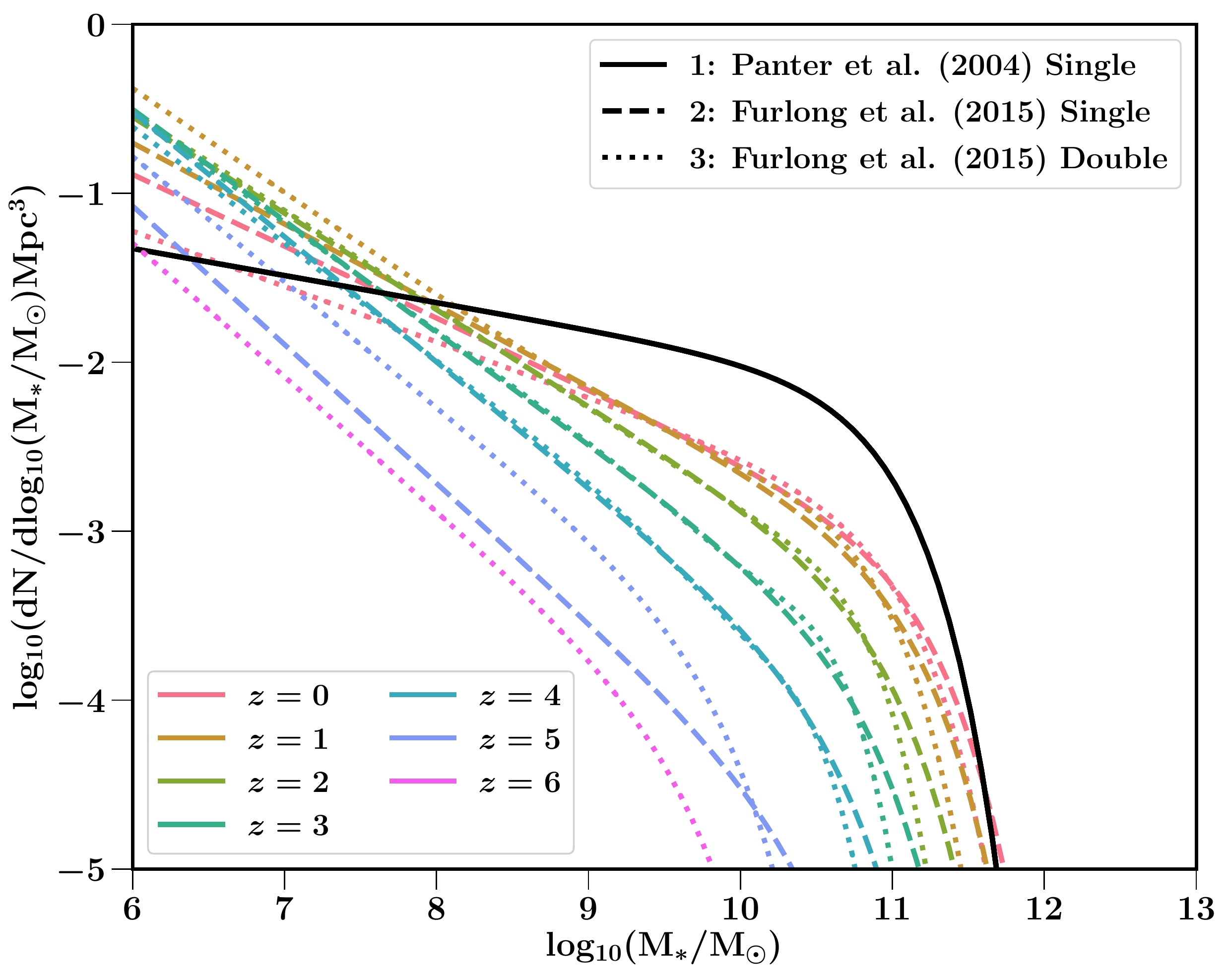} }}%
      \qquad
        \subfloat[][{The three different Mass-metallicity relation (\ac{MZR}) prescriptions studied in this work  shown at four different redshifts. }\label{fig:MSSFR-MZRs}]{{\includegraphics[width=1\columnwidth]{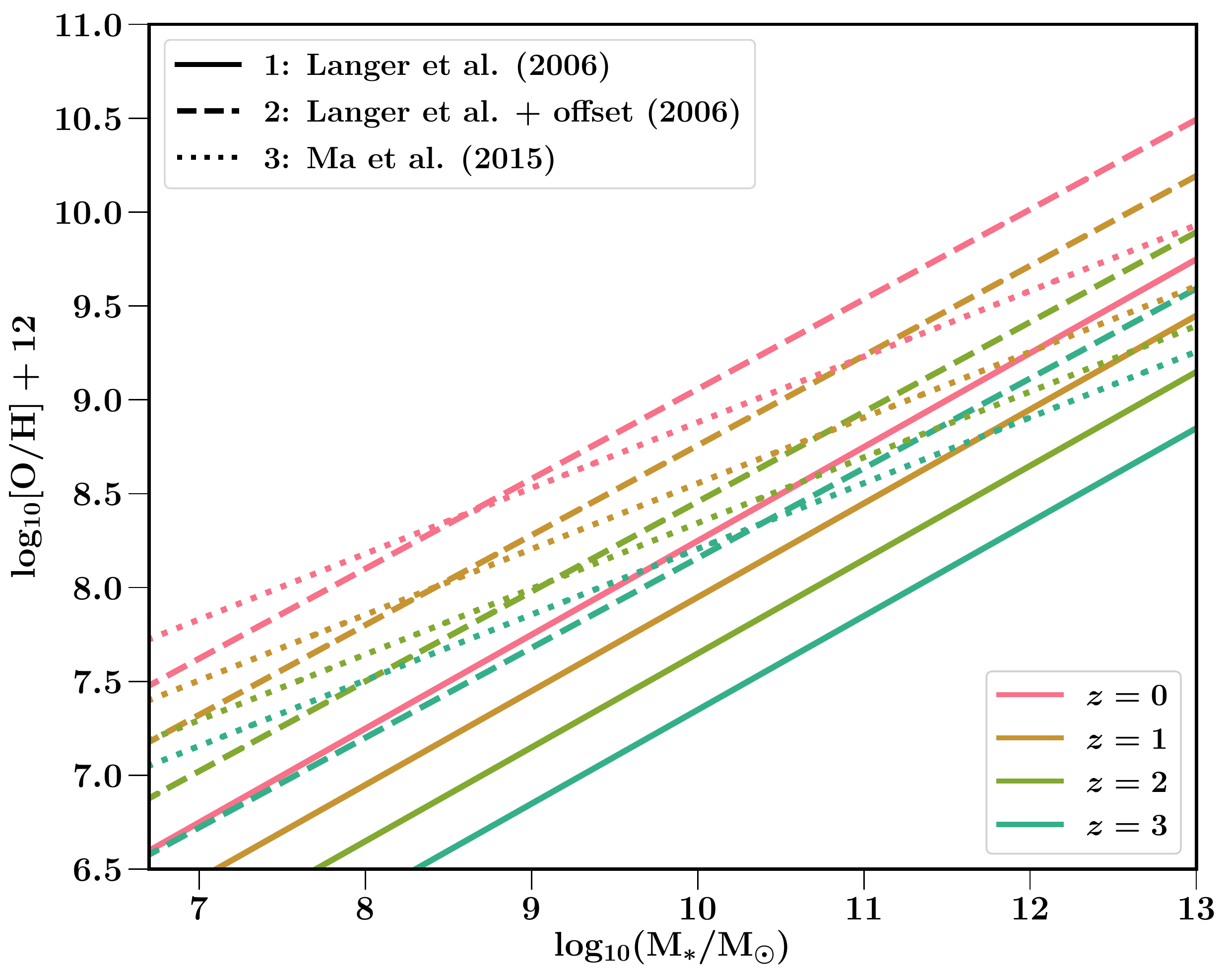} }}%
      \qquad
  \caption{Figures showing the components that  construct the metallicity-specific star formation rate densities (\SFRD) that are used in this study. From left to right and top to bottom we show: (a) the \acp{SFRD}, (b) the metallicity distributions $\diff P / \diff \Zi$, (c) the \acp{GSMF} and (d) the \acp{MZR}.  See for more details  Table~\ref{tab:MSSFR-variations-labels} and Section~\ref{subsec:method-MSSFR}.}
  \label{fig:4panels_examples_SFRD}
\end{figure*}

In Figure~\ref{fig:4panels_examples_SFRD} we show the distributions of the prescriptions that build up our \SFRD models in more detail. 
As discussed in Section~\ref{subsec:method-MSSFR} (and schematically shown in Figure~\ref{fig:MSSFR-sketch}) we  construct our \NmodelsMSSFR \SFRD models from a \ac{SFRD} and metallicity distribution function $\diff P / \diff \Zi$. The \NmodelsMSSFR ($3\times3 +1$) combinations  are summarized in Table~\ref{tab:MSSFR-variations-labels}.

Figure~\ref{fig:MSSFR-SFRs} shows the used  \ac{SFRD} prescriptions. It can be seen that the  \citet{2017ApJ...840...39M}  and \citet{2019MNRAS.490.3740N}  \acp{SFRD} prescriptions give lower yields, particularly for redshifts $z\lesssim 3$, compared to the {\citet{2014ARA&A..52..415M}}  and \citet{2004ApJ...613..200S} \ac{SFRD}. These lower \ac{SFRD} yields  lead to lower predicted merger rates as discussed in Section~\ref{sec:results-variations}. 

Figure~\ref{fig:MSSFR-Z-PDFs} shows three examples of metallicity distributions $\diff P / \diff \Zi$.  In all models except our \SFRD model $\rm{xyz}=000$, the metallicity distribution function is created from convolving a \ac{GSMF} with a \ac{MZR}.  In model $\rm{xyz}=000$ the  $\diff P / \diff \Zi$ is constructed from a phenomenological model as presented in \citet{2019MNRAS.490.3740N}.   The \citet{2019MNRAS.490.3740N} phenomenological metallicity distribution function is shown in solid lines. %
The metallicity prescriptions $\rm{yz}=31$ (dashed lines) and $\rm{yz}=12$ (dotted lines) are shown as example. These prescriptions correspond to the metallicity distribution models with our \citet{2015MNRAS.450.4486F} \ac{GSMF} and \citet{2006ApJ...638L..63L} \ac{MZR} and our  \citet{2004MNRAS.355..764P} \ac{GSMF} and \citet{2006ApJ...638L..63L} + offset \ac{MZR}. Section~\ref{sec:results-variations} shows that these result in the highest and lowest \bhnsSingle merger rates, respectively. The gray areas in the figure show \Zi values outside our modelled metallicity grid (see Table~\ref{tab:population-synthesis-settings}). Parts of the metallicity distribution function that fall outside of the \Zi parameter range are added to the edge bins when integrating over metallicities in Equation~\ref{eq:full-equation-detectable-rate}. 
Figure~\ref{fig:MSSFR-GSMFs} shows our explored \acp{GSMF}, whereas Figure~\ref{fig:MSSFR-MZRs} shows our three explored \acp{MZR}. See for more details \citet{2019MNRAS.490.3740N}.

\section{Zoom in on Figure~15.}
\begin{figure}
\centering
\includegraphics[width=1\columnwidth]{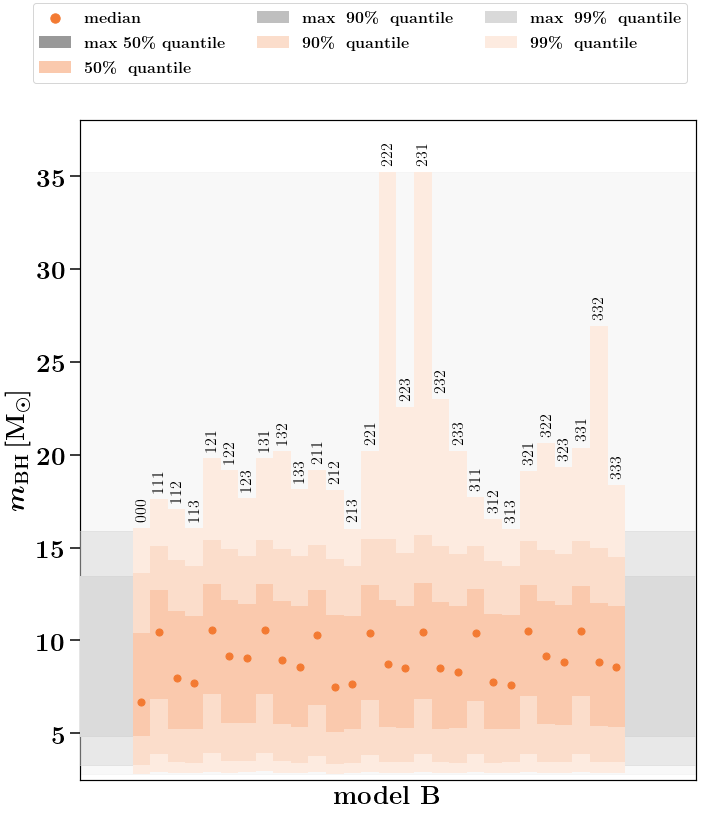} %
    \caption{A zoom in of Figure~\ref{fig:ConfidenceINtervals_BHNS} showing the distribution quantiles for the BH mass distribution for all \NmodelsMSSFR \SFRD model variations in combination with the binary population synthesis model B. The figure highlights the \SFRD bars, adding the \SFRD label denoted by $\rm{xyz}$ (see Table~\ref{tab:MSSFR-variations-labels}) on top of the corresponding bar.  
    Colored bars from dark to light denote the distribution quantiles for $50\%$, $90\%$ and $99\%$ of the distribution. Scatter points show the median values of the distribution. In the background the maximum range of the $50\%$, $90\%$ and $99\%$  quantiles over all \Nmodels model variations is shown.  }  
    \label{fig:BHNS-quantiles-ZOOM}
\end{figure}
In Figure~\ref{fig:BHNS-quantiles-ZOOM} we give a zoom-in of Figure~\ref{fig:ConfidenceINtervals_BHNS} to show the \SFRD models that correspond with each bar. The order of the bars, and hence \SFRD models, is the same throughout all model blocks in Figure~\ref{fig:ConfidenceINtervals_BHNS}.




\bsp	
\label{lastpage}
\end{document}